# Synthesis, Characterization, and Biological Evaluation of Gelatin-based Scaffolds

## Dissertation

zur Erlangung des akademischen Grades
"doctor rerum naturalium"
(Dr. rer. nat.)
in der Wissenschaftsdisziplin
„Materialien in den Lebenswissenschaften"

eingereicht an der
Mathematisch-Naturwissenschaftlichen Fakultät
der Universität Potsdam

von

## Giuseppe Tronci

aus Lecce, Italien

Potsdam, 2010

Gutachter:    Prof. A. Lendlein, Universität Potsdam

Prof. R. v. Klintzing, Technische Universität Berlin

Prof. M. Maskos, Bundesanstalt für Materialforschung und -prüfung

Tag der Annahme der Dissertation: 07.07.2010

Tag der Disputation: 15.12.2010

This work was financed by the German Research Foundation (Deutsche Forschungsgemeinschaft (DFG)) through the Collaborative Research Center 760 (CRC – 760, Sonderfoschungsbereich 760 (SFB 760)), subproject B5.

*"Research is to see what everybody else has seen, and to think what nobody else has thought."*

Albert von Szent-Györgyi de Nagyrápolt

# Statement of Originality

I, Giuseppe Tronci, formally submit the dissertation entitled "Synthesis, Characterization and Biological Evaluation of Gelatin-based Scaffolds" to the University of Potsdam, Faculty of Mathematics and Natural Sciences, Germany, for the acquirement of the academic degree of Doctor of natural sciences (Dr. rer. nat.) in Materials for Life Sciences. The work presented here was carried out from March 2006 to July 2010 at the Center for Biomaterial Development, Institute for Polymer Research, GKSS Research Center, campus Teltow.

I hereby certify that this submission is entirely my own original work and that, to the best of my knowledge and belief, it contains no material previously published or written by another person, except where due reference is made in the thesis itself. Neither the dissertation, nor any sections thereof, has been previously submitted for a degree or other qualification to any other University or Institution. Any contribution made to the research by others, with whom I have worked at GKSS or elsewhere, is explicitly acknowledged in the thesis.

Giuseppe Tronci

Potsdam, 07.07.2010

# Contents

# List of abbreviations

| | |
|---|---|
| AFM | Atomic Force Microscopy |
| a.u. | Arbitrary unit |
| BMP | Bone Morphogenetic Protein |
| ca. | Approximately |
| CHI | Cyclohexylisocyanate |
| CLSM | Confocal Laser Scanning Microscopy |
| CO | Carbonyl |
| $d_{dry}$ | Diameter of dry interconnection |
| $D_{dry}$ | Diameter of dry pore |
| DDW | Doubly Distilled Water |
| DMSO | Dimethylsulfoxide |
| DMTA | Dynamic Mechanical Analysis at varied Temperature |
| DNA | Deoxyribonucleic acid |
| DSC | Dynamic scanning calorimetry |
| $D_{wet}$ | Diameter of wet pore |
| $\Delta\sigma/\Delta\varepsilon$ | Slope of the plateau region |
| $E$ | Young's modulus |
| $E'$ | Storage modulus |
| $E''$ | Loss modulus |
| $E_c$ | Compression modulus |
| ECM | Extracellular Matrix |
| ER | Endoplasmic Reticulum |
| ESI | Electrospray ionization |
| EtO | Ethylene Oxide |
| EtOH | Ethanol |
| $\varepsilon_b$ | Elongation at break |
| $\varepsilon_{el}^*$ | Compressive plateau strain |
| FDA | Food and Drug Administration |
| FTIR | Fourier Transform Infrared Spectroscopy |
| g | Gram |
| GAG | Glycosaminoglycan |
| gel-sol | Gelation-solution |
| Gly | Glycine |
| $g \cdot mol^{-1}$ | Gram per mole |
| h | Hour |
| $H$ | Water Uptake |
| $H^+$ | Proton |
| $h_0$ | Height of a sample before compression test |
| $H_2O$ | Water |
| $h_c$ | Height of a compressed sample |
| HDI | Hexamethylene diisocyanate |
| HFIB-D | Human dermal fibroblasts |
| HLB | Hydrophilic-lipophilic balance |
| $h_r$ | Height of a compressed sample after removal of loads |
| Hyp | Hydroxyproline |
| Hz | Hertz |
| kPa | Kilopascal |
| L | Liter |
| LAL | Limulus Amebocyte Lysate |
| LDH | Lactate Dehydrogenase |
| LDI | Ethyl ester lysine diisoyanate |
| LPS | Lipopolysaccharide |
| Lys | Lysine |
| $M$ | Molecular weight of the surfactant molecule |
| MEM | Minimal Essential Medium |
| $M_h$ | Molecular weight of the surfactant hydrophilic segment |
| min | Minute |
| mm | Millimeter |
| $M_n$ | Number-averaged molecular weight |
| MPa | Megapascal |
| MS | Mass spectrometry |
| MTS | Mitochondrial Dehydrogenase |
| MW | Molecular weight |
| m/z | Mass-to-charge ratio |
| μCT | Micro-computed Tomography |
| μm | Micrometer |
| NCO | Isocyanate group |
| n.d. | Not determined |
| $NH_2$ | Amino |
| nN | Nanonewton |
| n.o. | Not observed |
| PBS | Phosphate Buffer Solution |
| PEG | Poly(ethylene glycol) |
| PEO | Poly(ethylene oxide) |
| PHEMA | Poly(2-hydroxyethylmethacrylate) |
| PPO | Poly(propylene oxide) |
| Pro | Proline |
| PVA | Poly(vinyl alcohol) |
| $Q$ | Degree of Swelling |
| $R_a$ | Averaged roughness |
| rel. mol-% | Relative mole percent |
| $R_q$ | Root mean square roughness |
| $\rho_1$ | Density of water at 25 °C |
| $\rho_2$ | Density of the dry gelatin sample |
| s | Second |
| SEM | Scanning Electron Microscopy |
| SFB | Sondeforschungsbereich |
| SME | Shape-Memory Effect |
| SR | Shape recovery |
| $\sigma_{el}^{*}$ | Compressive plateau stress |
| $\sigma_{max}$ | Maximum tensile strength |
| $T$ | Temperature |
| $tan\,\delta$ | Loss tangent |
| TE | Tissue Engineering |
| $T_g$ | Glass transition temperature |
| TGA | Thermal gravimetric analysis |
| TNBS | Trinitrobenzene sulphonate |
| TNF-α | Tumor necrosis factor-alpha |
| vol.-% | Volume percent |
| $Wall_{dry}$ | Wall thickness in the dry state |
| $Wall_{wet}$ | Wall thickness in the wet state |
| WAXS | Wide Angle X-ray Scattering |
| $W_d$ | Weight of a dry sample |
| WHO | World Health Organization |
| $W_s$ | Weight of a swollen sample |
| wt.-% | Weight percent |
| 2-D | Two-Dimensional |
| 3-D | Three-dimensional |
| °C | Degrees Celsius |

# Abstract

This work presents the development of entropy-elastic gelatin-based networks in the form of films or scaffolds. The materials have good prospects for biomedical applications, especially in the context of bone regeneration. Entropy-elastic gelatin-based hydrogel films with varying crosslinking densities were prepared with tailored mechanical properties. Gelatin was covalently crosslinked above its sol-gel transition, which suppressed the gelatin chain helicity. Hexamethylene diisocyanate (HDI) or ethyl ester lysine diisocyanate (LDI) were applied as chemical crosslinkers, and the reaction was conducted either in dimethyl sulfoxide (DMSO) or water. Amorphous films were prepared as measured by Wide Angle X-ray Scattering (WAXS), with tailorable degrees of swelling ($Q$: 300-800 vol.-%) and wet-state Young's modulus ($E$: 70-740 kPa). Model reactions showed that the crosslinking reaction resulted in a combination of direct crosslinks (3-13 mol.-%), grafting (5-40 mol.-%), and blending of oligoureas (16-67 mol.-%).

The knowledge gained with this bulk material was transferred to the integrated process of foaming and crosslinking to obtain porous 3-D gelatin-based scaffolds. For this purpose, a gelatin solution was foamed in the presence of a surfactant, Saponin, and the resulting foam was fixed by chemical crosslinking with a diisocyanate. The amorphous crosslinked scaffolds were synthesized with varied gelatin and HDI concentrations, and analyzed in the dry state by micro computed tomography (µCT, porosity: 65±11–73±14 vol.-%), and scanning electron microscopy (SEM, pore-size: 117±28–166±32 µm).

Subsequently, the work focused on the characterization of the gelatin scaffolds in conditions relevant to biomedical applications. Scaffolds showed high water uptake ($H$: 630-1680 wt.-%) with minimal changes in outer dimension. Since a decreased scaffold pore size ($115\pm47$–$130\pm49$ µm) was revealed using confocal laser scanning microscopy (CLSM) upon wetting, the form-stability could be explained. Shape recoverability was observed after removal of stress when compressing wet scaffolds, while dry scaffolds maintained the compressed shape. This was explained by a reduction of the glass transition temperature upon equilibration with water (dynamic mechanical analysis at varied temperature (DMTA)). The composition-dependent compression moduli ($E_c$: 10-50 kPa) were comparable to the bulk micromechanical Young's moduli, which were measured by atomic force microscopy (AFM). The hydrolytic degradation profile could be adjusted, and a controlled decrease of mechanical properties was observed. Partially-degraded scaffolds displayed an increase of pore size. This was likely due to the pore wall disintegration during degradation, which caused the pores to merge.

The scaffold cytotoxicity and immunologic responses were analyzed. The porous scaffolds enabled proliferation of human dermal fibroblasts within the implants (up to 90 µm depth). Furthermore, indirect eluate tests were carried out with L929 cells to quantify the material cytotoxic response. Here, the effect of the sterilization method (Ethylene oxide sterilization), crosslinker, and surfactant were analyzed. Fully cytocompatible scaffolds were obtained by using LDI as crosslinker and $PEO_{40}$-$PPO_{20}$-$PEO_{40}$ as surfactant. These investigations were accompanied by a study of the endotoxin material contamination. The formation of medical-grade materials was successfully obtained (<0.5 EU/mL) by using low-endotoxin gelatin and performing all synthetic steps in a laminar flow hood.

# Zusammenfassung

Diese Arbeit beschreibt die Entwicklung Entropie-elastischer Gelatine-basierter Netzwerke als Filme und Scaffolds. Mögliche Anwendungen für die entwickelten Materialien liegen im biomedizinischen Bereich, insbesondere der Knochenregeneration.

Im ersten Schritt der Arbeit wurden Entropie-elastische, Gelatine-basierte Hydrogel-Filme entwickelt, deren mechanische Eigenschaften durch die Veränderung der Quervernetzungsdichte eingestellt werden konnten. Dazu wurde Gelatine in Lösung oberhalb der Gel-Sol-Übergangstemperatur kovalent quervernetzt, wodurch die Ausbildung helikaler Konformationen unterdrückt wurde. Als Quervernetzer wurden Hexamethylendiisocyanat (HDI) oder Lysindiisocyanat ethylester (LDI) verwendet, und die Reaktionen wurden in Dimethylsulfoxid (DMSO) oder Wasser durchgeführt. Weitwinkel Röntgenstreuungs Spektroskopie (WAXS) zeigte, dass die Netzwerke amorph waren. Der Quellungsgrad ($Q$: 300-800 vol.-%) und der Elastizitätsmodul ($E$: 70-740 kPa) konnten dabei durch die systematische Veränderung der Quervernetzungsdichte eingestellt werden. Die Analyse der Quervernetzungsreaktion durch Modellreaktionen zeigte, dass die Stabilisierung der Hydrogele sowohl auf kovalente Quervernetzungen (3-13 mol.-%) als auch auf Grafting von (5-40 mol.-%) und Verblendung mit Oligoharnstoffen (16-67 mol.-%) zurückgeführt werden kann.

Die Erkenntnisse aus dem Umgang mit dem Bulk-Material wurden dann auf einen integrierten Prozess der Verschäumung und chemischen Quervernetzung transferiert, so dass poröse, dreidimensionale Scaffolds erhalten wurden. Dafür wurde eine wässrige Gelatinelösung in Gegenwart eines Tensids, Saponin, verschäumt, und durch chemische Quervernetzung mit einem Diisocyanat zu einem Scaffold fixiert. Die Scaffolds hergestellt mit unterschiedlichen Mengen HDI und Gelatine, wurden im trockenen Zustand mittels Mikro Computertomographie (μCT, Porosität: 65±11–73±14 vol.-%) und Rasterelektronenmikroskopie (SEM, Porengröße: 117±28–166±32) charakterisiert.

Anschließend wurden die Scaffolds unter Bedingungen charakterisiert, die für biomedizinische Anwendungen relevant sind. Die Scaffolds nahmen große Mengen Wasser auf ($H$: 630-1680 wt.-%) bei nur minimalen Änderungen der äußeren Dimensionen. Konfokale Laser Scanning Mikroskopie zeigte, dass die Wasseraufnahme zu einer verminderten Porengröße führte (115±47–130±49 μm), wodurch die Formstabilität erklärbar ist. Eine Formrückstellung der Scaffolds wurde beobachtet, wenn Scaffolds im nassen Zustand komprimiert wurden und dann entlastet wurden, während trockene Proben in der komprimierten Formen blieben (kalte Deformation). Dieses Entropie-elastische Verhalten der nassen Scaffolds konnte durch die Verminderung der Glasübergangstemperatur des Netzwerks nach Wasseraufnahme erklärt werden (DMTA). Die zusammensetzungsabhängigen Kompressionsmoduli ($E_c$: 10-50 kPa) waren mit den mikromechanischen Young's moduli vergleichbar, die mittels Rasterkraftmikroskopie (AFM) gemessen wurden. Das hydrolytische Degradationsprofil konnte variiert werden, und während des Abbaus kam es nur zu kontrolliert-graduellen Änderungen der mechanischen Eigenschaften. Während der Degradation konnte ein Anstieg der mittleren Porengröße beobachtet werden, was durch das Verschmelzen von Poren durch den Abbau der Wände erklärt werden kann.

Die Endotoxinbelastung und die Zytotoxizität der Scaffolds wurden untersucht. Humane Haut-Fibroblasten wuchsen auf und innerhalb der Scaffolds (bis zu einer Tiefe von 90 μm). Indirekte Eluat-Tests mit L929 Mausfibroblasten wurden genutzt, um die Zytotoxizität der Materialien, insbesondere den Einfluss des Quervernetzertyps und des Tensids, zu bestimmen. Vollständig biokompatible Materialien wurden erzielt, wenn LDI als Quervernetzer und $PEO_{40}-PPO_{20}-PEO_{40}$ als Tensid verwendet wurden. Durch den Einsatz von

Gelatine mit geringem Endotoxin-Gehalt, und die Synthese in einer Sterilarbeitsblank konnten Materialien für medizinische Anwendungen (Endotoxin-Gehalt < 0.5 EU/mL) hergestellt werden.

# Acknowledgements

I am very grateful to my supervisor, Prof. A. Lendlein, for accepting me at the GKSS Research Center, for introducing me into the multidisciplinary biomaterial research and for his guidance during the progress of my work. I would like to thank my group leader, Dr. A. Neffe, for the constructive discussions along the years and for reviewing my thesis.

A special thank goes to Ben Pierce for his helpful suggestions about the work and the writing, and to Martin Roessle for his kind supervision at the beginning of my thesis. I am very thankful to Dr. W. Albrecht for his guidance during the start of my Ph. D., and for his important advice throughout my thesis. I take this opportunity to convey my sincere thanks to Prof. D. Hofmann for the first discussions about my project and his important suggestions later on.

I would like to thank all the people who supported me with the different methods during these years: thank you to M. Schossig for the SEM investigation on the scaffolds, Dr. T. Weigel for the μCT measurements and his nice availability, Dr. H. Kosmella for his help with the mechanical tests, Dr. W. Wagermaier and Dr. U. Noechel for the WAXS measurements. I want to thank R. Apostel for her assistance in the lab during the last period of my work, and M. Rettschlag for his help in setting up the clean room.

I am very grateful to Dr. G. Boese, Dr. B. Hiebl and the Biocompatibility department, for the cell tests, Prof. H. Volk (*Charitè, Berlin*) and his group for the endotoxin tests, Dr. T. Neumann (*JPK Instruments AG, Berlin*) for the AFM measurements on the gelatin scaffolds, and Dr. Jana Falkenhagen (*BAM, Berlin*) for the mass spectra. I also want to thank Dr. J. Lineau (*Julius Wolff Institut, Berlin*) and her group for the nice cooperation in context of the SFB-760.

I owe my deepest gratitude to all members of the PBB department as well as the Ph. D. students, and the Italian community (Giuseppe Tripodo, Alessandro Zaupa, Susanna Piluso and Stefania Federico), for the nice atmosphere and their help during the final stage of my thesis. A special thank goes to Alessandro, with whom I have started this experience and lived together all the challenges along the way. I am grateful to Dieter Hofmann, Karola Luetzow, Wolfgang Wagermaier, Jutarat Pimthon, Matthias Heuchel, Barbara Seifert, Ulrich Noechel, Maria Entrialgo-Castaño, and Silke Pelzer for the nice time spent together and their important advice. In particular, I thank Karola and Barbara for their valuable help and important suggestion concerning the completion of my thesis, as well as Dr. M. Schroeter. I also take this opportunity to thank my officemates, Wolfgang and Ulrich, with whom it was a great pleasure to share the office.

I am very grateful to my parents and to my brother, Cesare, for their support throughout my thesis and for their help during the challenging time. My final thank goes to Susanna for sharing with love this experience with me.

# 1. Introduction

The change of demographic structure and life style, the longer life expectancy, and the advances in medicine have and will continue to have a strong influence on the demand for and the development of novel technologies in the field of medicine. Especially, the treatment of degenerative diseases and injuries is challenging. Natural healing processes often lead to the formation of scar tissue due to the limited capacity of our body parts to regenerate both structurally and functionally.[1] Also, the number of donors is too low to fulfil the pressing demand of organ and tissue replacement. The dramatic limitations inherent in the current clinical approaches have motivated the development of Regenerative Medicine as a new research field. The vision is to accomplish a full functional regeneration of the respective tissue, avoiding transplantations, pure technical solutions, or permanent pharmacotherapy.[2]

## 1.1 Clinical need of bone regeneration

Bone is a prime tissue target for Regenerative Therapies, with more than two million bone replacement procedures carried out annually worldwide,[3] and about $ 849 billions spent for bone and joint health in the US in the year 2004.[4] Aiming at the development of new

clinical treatments, the World Health Organisation (WHO) has proclaimed the current decade as the "Bone and Joint Decade".

Bone regeneration is a complex and long lasting process. Once the bone fracture occurs, a callus is initially formed between the fracture sites by the periosteal cells differentiation in chondroblasts and osteoblasts.[5,6] This is followed by slow remodeling, ultimately resulting in trabecular bone. This process can only be successful in case of small, simple fractures. In contrast, critical sized defects do not self-heal and their complete regeneration can only be ensured by the implantation of materials, i. e. bone grafts or bone substitutes.[7]

Surgical treatments of critical sized bone defects are limited to graft transplantations, such as auto-, allo-, and xeno-grafts.[8,9,10] The best clinical outcome is so far obtained through autologous bone transplantation from an area of the patient's body, e. g. the iliac crest in the hip, to the critical defect. The graft integrates reliably with host bone, in contrast to the immune- and disease-related complications resulting from the application of allogeneic bone (i.e. bone from a human cadaver) or xenogeneic bone (i.e. bone from an animal source). However, the use of autografts is severely hampered by the short supply and the considerable donor site morbidity associated with the harvest.

In view of developing successful Regenerative Medicine strategies, a multidisciplinary approach is required to understand and stimulate the natural bone regeneration process. Besides the knowledge from cell biology and genetics, the bone structure has to be investigated with respect to its hierarchical organization, which spans over several length scales, from the macro- (centimeter) scale to the nanostructured (extracellular matrix, ECM) components.[10,11] Starting from the macroscopic structural level, bone tissue can be arranged in either a compact pattern (cortical bone) or a trabecular pattern (cancellous bone). Moving from the macroscopic to the lower levels, bone is a composite material, consisting of an organic phase, i. e. cells and the ECM, in which mineral nanoparticles (based on hydroxyapatite carbonate) are integrated (Figure 1.1). Especially, the ECM plays a key role in the localization and presentation of biomolecular signals which are vital for neo-tissue morphogenesis.[1]

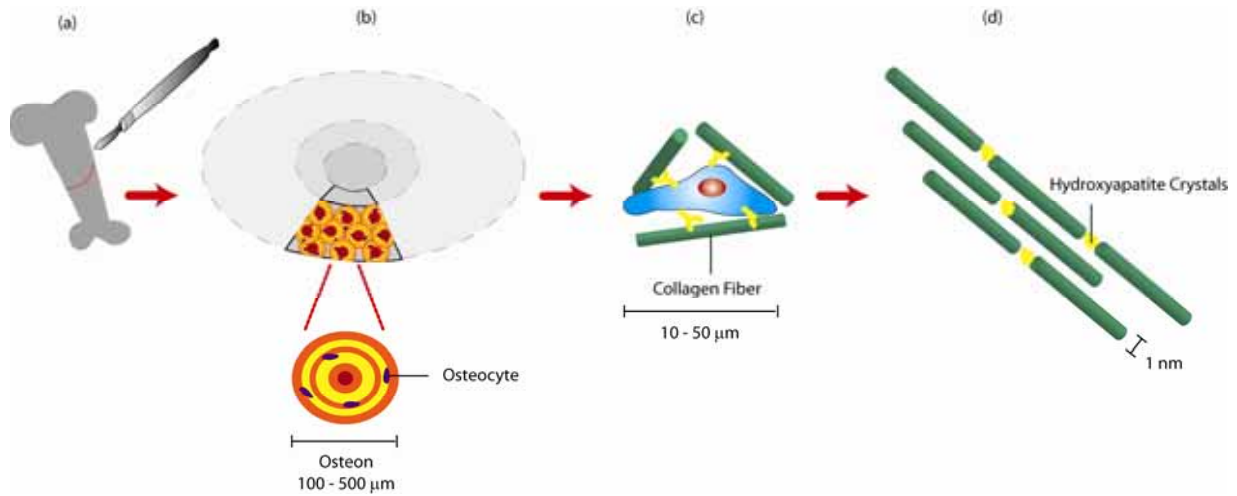

**Figure 1.1. Hierarchical organization of bone over different length scales. Bone consists of a strong calcified outer compact layer (a), including many cylindrical Haversian systems, or osteons (b); cells are immersed in the ECM (c), which is mainly based of collagen fibrils mineralized with hydroxypapatite crystals (d).**

Understanding of the healing process forms the basis for the development of biomaterials supporting bone regeneration. These different fields are covered in the Sonderforschungsbereich (SFB) 760 "*Biomechanics and Biology of Musculoskeletal Regeneration – From Functional Assessment to Guided Tissue Formation*", in which this dissertation was conducted (project B-5). In view of developing biomaterials supporting the bone regeneration process, mimicking the structure and functions of the ECM is essential. In this context, next section will provide an overview of the ECM and respective components.

# 1.2 The extracellular matrix

The extracellular matrix (ECM) is a complex structural entity surrounding and supporting cells that are found within mammalian tissues. It regulates intercellular communication and water content of tissue, segregates tissues from one another, and is directly related to processes like tissue growth or healing. As described in Figure 1.2, the ECM is composed of an interlocking mesh of fibrous proteins and glycosaminoglycans (GAGs). Collagen is the most abundant protein of the ECM, and is found in tendons, ligaments, and in the connective tissue of skin, blood vessels, and lungs. Elastin is another structural protein of the ECM, mainly occuring in the artery walls, lungs, intestines, and skin. Collagen and elastin work in partnership, so that collagen gives rigidity to connective tissues and organs, while elastin is responsible for the tissue elasticity, enabling tissue stretching and recovery.

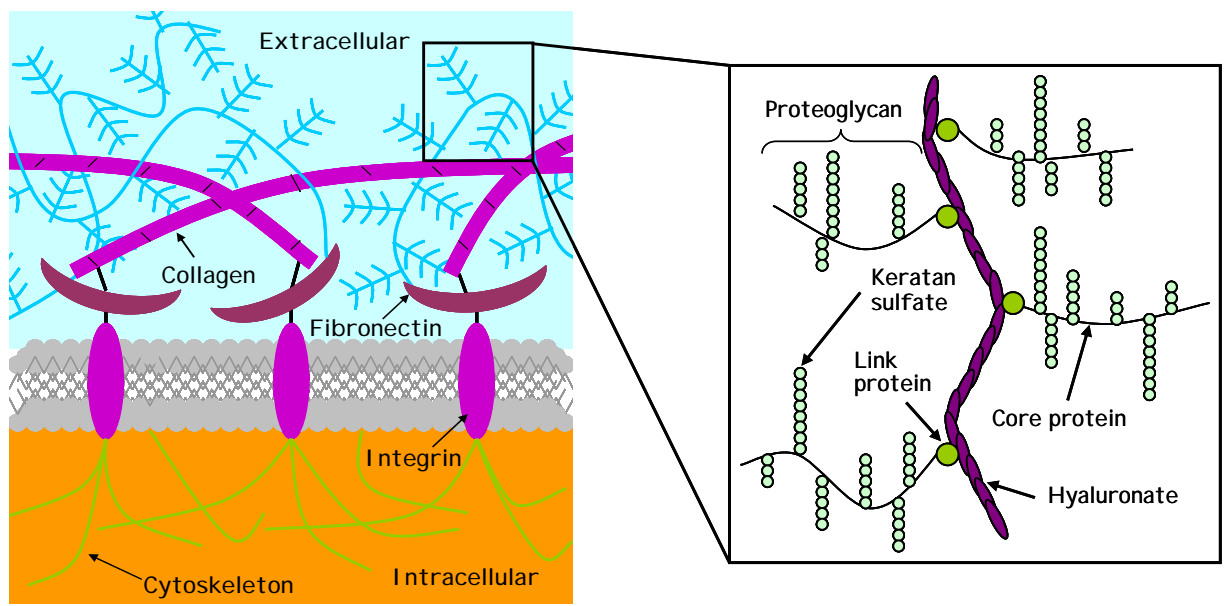

**Figure 1.2. Macromolecular organization of the tissue ECM (left) and the proteoglycan complex (right).**

Proteoglycans are also present in the ECM. They are composed of a core protein and one or more covalently attached sulfated glycosaminoglycan (GAG) chains (Figure 1.2, right). The GAGs are linear polymers of repeated disaccharidic units mostly of hexosamine and hexuronic acid. Due to the occurrence of sulfate and carboxylic groups, GAGs interact with water, resulting in the formation of an ECM hydrogel. Hyaluronic acid is the only non-sulfated, non-attached to protein GAG of the ECM, and is composed of D-glucuronic acid and D-glucosamine. Ultimately, specialized protein, e. g. fibronectin, are also present in the ECM. Fibronectin connects the ECM collagen fibers with the cell surface integrins, thereby enabling the reorganization of the cell's cytoskeleton.

# 1.3 Regenerative Medicine

Since the 1970s, several independent attempts were undertaken to create tissue substitutes using cultured cell sheets, or natural components of the ECM.[12] Despite the long time period and the intensive scientific research pursued, biomaterial-based clinical procedures have demonstrated only partial success.

In the field of Regenerative Medicine, there are several strategies relying on the use of biomaterials, two of which are Tissue Engineering (TE) and Induced Autoregeneration.[1] TE aims at the formation of a specific tissue through the selection and manipulation of cells, matrices, and biologic stimuli.[9] Once the tissue construct is created *in vitro*, it will be transferred *in vivo* at the tissue defect site. Other than TE, induced autoregeneration is based on the application of a degradable biomaterial *in vivo*, and is especially promising for the reconstruction of critical sized, non self-healing defects.[13] Once applied at the defect site, the biomaterial should act as a temporary mechanical and biological support system. *In vivo*, cells attracted from the surrounding tissues should colonize the implant,[14,15] so that formation of functional neo-tissue takes place while the implant completely degrades.[16,17] This approach is convenient for the patient and cost-effective compared to cell-based therapies.[13]

Complex biomaterial requirements need to be fulfilled in order to successfully stimulate the endogenous regeneration. In its basic form, the biomaterial should be non-toxic and non-immunogenic, in order to promote cell colonization and growth in the implant, thereby guiding the formation of the neo-tissue. Important steps in such a testing scheme are: (i) the sterilization of the material (typical sterilization techniques are ethylene oxide, water vapor, and γ-irradiation); (ii) the determination of the endotoxin content; (iii) eluate and direct contact test with fibroblasts, e. g. L929 mouse fibroblasts; (iv) testing with application specific cells. In the case of polymer matrices, the potential material toxicity needs to be investigated with respect to any presence of specific reactive groups or non-reacted moieties in the polymer. Geometrically, the biomaterial should act as a porous three-dimensional (3-D) scaffold, ensuring structural integrity in physiological conditions. The scaffold should also perform a time-limited function, leaving the space to the new functional biological system.[18] For this purpose, the scaffold needs to be biodegradable, ensuring a controlled change of material properties during degradation.[19] Degradable materials are especially suitable if drug delivery functionality is intended, which is important to enhance the regenerative potential of tissue.[20] For this purpose, not only classical small molecule drugs, but also novel protein-based bioactives, e. g. BMPs in bone regeneration, play an important role. In this context, BMP-2 is highly acclaimed for its osteoinductivity in orthopaedic surgery,[21] although mimicking of the protein (e. g. BMP-2) release regulation *in vivo* is a challenging task.[22]

In order to fulfil the above-mentioned biomaterial requirements, hydrogels have received a lot of attention[23,24] since they can be specifically synthesized in order to mimic the ECM structure and composition.[25] As a result, they have shown promising results for cellular proliferation and differentiation.[26] The following sections initially focus on hydrogel materials, in terms of network architecture and morphology. Afterwards, biological hydrogels, e. g. based on collagen and gelatin, will be presented.

# 1.4 Hydrogels for Regenerative Medicine

Hydrogels are 3-D networks formed from hydrophilic homopolymers, copolymers, or macromers, which are crosslinked to form insoluble polymer matrices,[27,28] while maintaining large fractions of water (> 20 wt.-%). In physiological conditions, these gels are typically soft and elastic, as explained by the thermodynamic compatibility of the dry polymer with water, and the low glass transition temperature ($T_g$) in the wet state. In view of these properties, the importance of hydrogels in biomedical applications was first realized in the early 1960s, with the development of poly(2-hydroxyethyl methacrylate) (PHEMA) gels as a contact lens material.[29] Subsequently, hydrogels were developed based on other synthetic polymers, such as poly(ethylene glycol) (PEG) and poly (vinyl alcohol) (PVA).

The crosslinked structure of hydrogels is characterized by junctions or net-points between polymer chains. The net-points can result from chemical crosslinks, permanent or temporary physical entanglements, or physical crosslinks, determined from intermolecular interactions between polymer chains, e. g. hydrogen bonds or Van der Waals interactions (Figure 1.3). Several synthetic routes enable the formation of covalently crosslinked hydrogels. Free-radical polymerization is a widely used method where a polymer chain propagates through the consumption of vinyl monomers. This reaction scheme was recently applied for the formation of PEG hydrogels with defined surface topographies, and tunable chemical and physical properties.[30] In order to achieve high chemical selectivity, alternative synthetic methods include Michael conjugate addition, e. g. between an acrylated and a thiolated macromer,[31] and click chemistry, between azides and terminal acetylenes.[32] Alternatively, hydrogels can also be prepared as blends or interpenetrating networks. Based on the synthetic route, and the starting monomers and macromers used, hydrogels can be obtained from equilibration of an amorphous or semi-crystalline polymer network in water.

An amorphous polymer network is especially important in view of accomplishing a homogeneous profile during degradation.

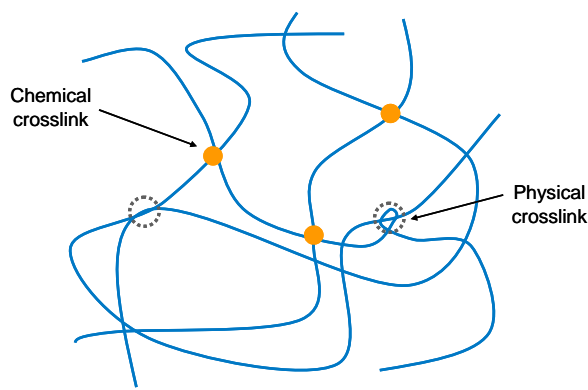

**Figure 1.3. Schematic representation of the molecular architecture of a hydrogel network. Junctions can occur in form of chemical and physical crosslinks, e. g. as entanglement.**

During equilibration in aqueous solution, the dry networks will not dissolve, as due to the presence of the netpoints between polymer chains. In contrast, the dry networks will swell and absorb a certain amount of water, as based on hydrophilic building blocks in the network chemical structure. As a result, a two-component system is obtained. Two opposing forces are thereby responsible for the swelling of a dry polymer network:[33] on the one hand, swelling is driven by the spontaneous mixing of the network chains with water; on the other hand, covalent crosslinks prevent the dissolution of the network. During swelling, the network chains assume an elongated configuration, associated with an elastic retractile entropy-driven force. The mixing tendency of the solvent and the network chains is expressed by the entropy of dilution. As long as the swelling proceeds, the elastic retractile force increases, while the diluting force decreases. When these two forces are balanced, the swelling equilibrium is reached between the solvent within the hydrogel and the surrounding solution. Additionally, the ionic contribution needs to be considered for the equilibrium swelling in the case of ionic gels. The degree of swelling ($Q$) is generally used to elucidate the network hydrogel architecture, as $Q$ is inversely related to the network crosslinking density. $Q$ can be measured by the ratio between the macroscopic volume of the hydrogel and the macroscopic volume of the dry network. Details for the determination of $Q$ can be found in chapter 9 (section 9.7).

Once the swelling equilibrium is achieved, the network chains are in a randomly coiled conformation and are oriented, so that the elasticity gets partially lost. The resulting hydrogels can exhibit different responses when subjected to mechanical stress. They can display fully elastic recovery following an applied stress, or a time-dependent recovery, which is typical of a viscous behaviour. The hydrogel mechanical behaviour is directly related to the $T_g$ of the swollen network, which depends on the $T_g$ of the dry polymer network and its degree of swelling. At temperatures below the $T_g$, the network is in the glassy state.[34] Upon swelling, the water tends to plasticize the polymer chains, so that a reduction of $T_g$ is expected in the swollen network compared to the dry network. As a result, a thermal transition from the glassy to the rubbery state, or entropy-elastic state, can be expected at e. g. room or body temperatures.

Once that the chemical and physical properties of synthetic hydrogels have been presented, the next section gives an overview about the hydrogels based on the components of the ECM. Therefore, collagen will be introduced as the main component of the ECM. Afterwards, the attention focuses on gelatin, which is the product of the partial collagen denaturation, and on gelatin hydrogels as potential biomaterials for application in Regenerative Medicine.

# 1.5 Collagen

The defining feature of the collagen molecule is the triple helix, which consists of three parallel left-handed helical polypeptides coiled about each other to form a right-handed triple helix. Biochemically, the collagen triple helix is synthesized in the endoplasmic reticulum (ER) lumen, starting from three peptide chains. These peptide precursors, known as preprocollagen, have registration peptides on each end and a signal peptide. The peptide chains are sent into the ER lumen, where the signal peptides are cleaved and triple helical structure is formed (procollagen). Outside the cell, the registration peptides are cleaved and the collagen molecule (tropocollagen) is formed by procollagen peptidase. Multiple tropocollagen molecules form collagen fibrils, via covalent cross-linking by lysyl oxidase which links hydroxylysine and lysine residues. Multiple collagen fibrils assemble into collagen fibers. These hierarchically organized networks are responsible for the tissue elasticity.

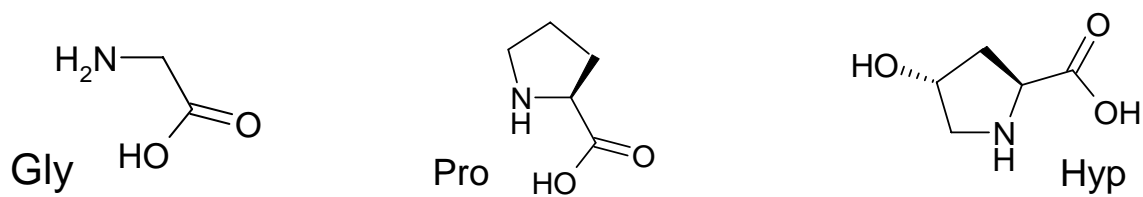

**Figure 1.4. Chemical structure of glycine (Gly), proline (Pro) and hydroxyproline (Hyp).**

The amino-acidic sequence (Gly-X-Y) represents the primary structure of the collagen chain, whereby Gly is glycine, X or Y are often proline (Pro) or Y hydroxyproline (Hyp) (Figure 1.4). Either Pro or Hyp display a lateral group which loops back to re-attach the main chain, resulting in a ring. The presence of these rings confers enhanced localised rigidity to the polypeptide backbone, and is essential to explain the stabilization of the collagen triple helix.[35] As a result, each strand in the triple helix is twisted in a left-handed helix (pitch ~ 0.9 nm), and the three strands are stabilized by wrapping into a right-handed helix (pitch ~ 8.6 nm, Figure 1.4).[36] The interchain hydrogen bonding occurs perpendicular to the chain axis and takes place between carbonyl (CO) and amino (NH) groups belonging to two adjacent backbones, eventually mediated by water molecules.

# 1.6 Gelatin

Gelatin is obtained from collagen by partial hydrolysis and denaturation, during which the regular triple helix structure is broken down to form random gelatin coils. Despite the collagen macromolecular organization is lost, the chemical composition is closely maintained in resulting gelatin.[37] Two basic processes are involved in the denaturation process from the regular structure of collagen to the more random gelatin, i. e. thermal treatment and hydrolytic breakdown of covalent bonds.[38] The thermal treatment ($T \sim 40$ °C) occurs in the presence of water and is necessary in order to destroy both hydrogen and electrostatic interactions. Subsequently, hydrolytic degradation of collagen takes place either in acidic or basic conditions, leading to the formation of gelatin type A or B, respectively. In the acidic process, collagen is soaked in dilute acid and then extracted at about pH 4. At this pH, non-collagenous tissue proteins are slightly soluble.[39] In contrast, many of the impurities are soluble in alkaline conditions and can be removed during extraction, so that gelatin type B is considered purer than gelatin type A. However, gelatin type B normally displays an isoelectric point in the acidic pH.[38] Consequently, a higher molar excess of the carboxyl groups (compared to the amino groups) is expected in gelatin type B with respect to gelatin type A. This observation is important in perspective of gelatin functionalization, since carboxyl groups are challenging to functionalize compared to the highly reactive amino groups.

The positions of the bond breaks during collagen hydrolytic degradation determine the molecular weight, the number of polypeptide chains, the number of each kind of amino acid residue, and their position with respect to the chain ends of the released gelatin molecule. However, no one bond is known to be so labile to specifically break during collagen hydrolysis. Several bond positions are at risk and can break on a probability basis, depending on the pH and temperature. It is therefore not surprising that the random characteristic of bond hydrolysis is the main cause of the gelatin molecular heterogeneity. Additionally, the collagen source also counts for this observation, since the animal species, tissue, age, and sex are all variables leading to variability in the collagen amino acid composition. Consequently, controlling the chemical composition and the overall properties of gelatin materials is therefore a challenging task, also in view of the control of gelatin properties in water.

Once collagen crosslinks are broken down, gelatin coils are dissolved in solution ($T$= 40-50 °C). When temperature is lowered below roughly 35 °C, the gelatin chains undergo a progressive conformational change, which is known as the coil-to-helix transition (Figure 1.5).

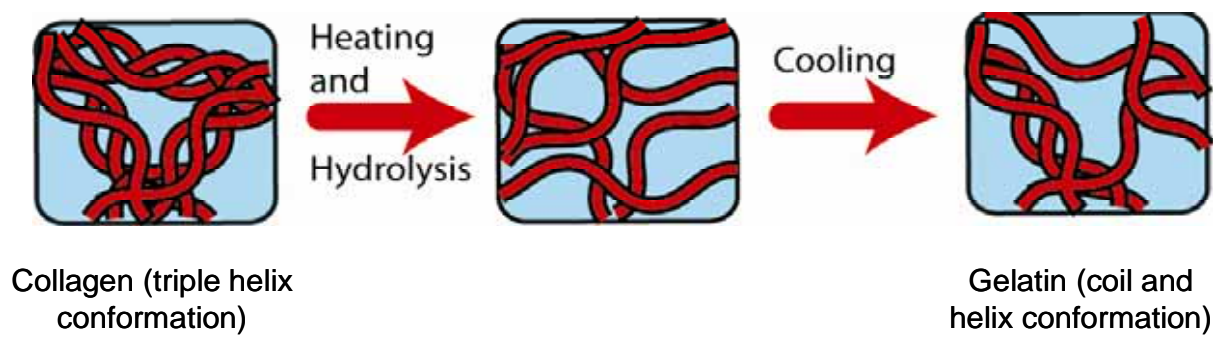

Collagen (triple helix conformation)

Gelatin (coil and helix conformation)

**Figure 1.5. Denaturation process of collagen to obtain gelatin: the thermal treatment and hydrolytic degradation lead to the irreversible breakdown of the triple helix structure. Random gelatin coils are thereby formed, which can partially renaturate on cooling.**

During this process, the solution viscosity is progressively increased, so that a transparent, thermo-reversible physical gel is formed. The basic gelation mechanism implies the triple helix renaturation, with the renatured triple helix being the favored thermodynamic conformation of the gelatin chains below 35 °C. The gelatin coil-to-helix transition is very slow compared to the renaturation rates of other helix-forming biopolymers, i. e. polysaccharides and DNA. This is attributed to the high proportion of peptide bonds, e. g. Gly-Pro and X-Hyp, which must undergo *cis* to *trans* isomerization reactions during gelatin folding.[40,41] The physical crosslinks which are formed by the partial gelatin renaturation are responsible for the mechanical and swelling properties of resulting hydrogels. With that regard, the gelatin concentration and the thermal history of the solution are crucial for the renaturation process, thereby influencing the hydrogel strength.[42]

# 1.7 Gelatin hydrogels: state of the art and challenges

ECM-derived biopolymers, such as collagen and gelatin, represent excellent candidates for the design of cell-compatible and degradable biomaterials.[43] The advantage of such materials lies in mimicking tissue components once implanted *in vivo*, which is not the case for most synthetic hydrogels. However, the clinical use of biopolymer-based materials is challenging in view of the batch to batch variation, since their chemical composition might differ based on the ECM source (section 1.6). Furthermore, the application of ECM-derived biopolymers raises concerns about the material contamination with pathogens or endotoxins. Ultimately, their high interaction with water results in high swelling (gel formation) and weak mechanical properties, so that the accomplishment of elastic materials with controlled dimensions is challenging.

Collagen has been extensively employed in the form of sponges as implant material for orthopedic applications as well as for the design of bioprosthetic vascular devices, i. e. heart valves and vascular grafts.[44,45] However, drawbacks such as form instability (shrinkage)[46,47,48] and immunogenic responses[49] have been reported. Crosslinking techniques attempting to ensure mechanical integrity[50,51,52,53,54] have only led to slight macroscopic variation of mechanical properties. Methods directed to collagen interfibril crosslinking[55,56,57] have been partially successful, but procedures which can easily tune the final properties of collagen materials are still lacking.

Other than collagen, gelatin is proangiogenic[39] and non-immunogenic,[58,59] exhibits low levels of cytotoxicity[60] and has FDA approval as a clotting agent, e. g. in the form of sponges.[61] Although gelatin possesses the ability to form gels by triple helicalization of chains, chemical crosslinking has been usually applied to ensure material structural integrity and controlled, reproducible gel properties *in vivo*.[62] Examples of crosslinking agents include glutaraldehyde,[63,64,65,66] genipin,[67,68,69] carbodiimide,[70] transglutaminase,[71,72] oxidized chondroitin sulfate,[73] oligomeric proanthocyanidins,[74] dextran dialdehyde,[75] bis(vinylsulfonyl)methane,[76] and hexamethylene diisocyanate.[77] Crosslinked gelatin has been investigated as a peripheral nerve guide conduit material,[78] bone substitutes,[79] and for protein-releasing matrices,[20] and has been processed into films,[63-68] foams,[80,81,82,83,84] and other forms.[85] Despite crosslinked gelatin materials have been deeply investigated, reports have so far concentrated on linking mechanical properties to the helical gelatin content. Bigi et al. demonstrated that the material Young's modulus ($E$: 4-26 MPa) for uncrosslinked and crosslinked gelatin films greatly depends on the Bloom index and the renaturation level of gelatin triple helices.[63] McDermott et al. also measured the change in Young's modulus ($E$: 10 – 125 kPa) for transglutaminase-crosslinked gelatin films, showing that $E$ increased with increasing gelatin concentrations and Bloom index.[86]

Although these reports are insightful, systematic structure-property relationships are not yet established between the physical material properties and molecular parameters, i. e. crosslinking density, chain type, segment length, orientation, and conformations of chains. This would enable the formation of novel biomimetic materials with tailored properties and functions, which can only be accomplished when specifically controlling the gelatin chain helicity.

# 1.8 Design of degradable and form-stable hydrogels

The many requirements that a biomaterial needs to fulfil for biomedical applications demand for the development of hydrogel systems with tailorable properties and functions. The synthesis of polymer systems has been demonstrated to be a successful strategy for the adjustment of material properties.[87] Polymer systems are families of related polymers in which small changes in the molecular compositions lead to systematic variations of overall material properties.[88,89]

Other than the inherent material properties, specific functions can be incorporated in the material either in an extra step, i. e during the device manufacturing, or can be recalled per demand under certain environmental conditions.[90] Specific functions for hydrogels to be applied in tissue regeneration are the form-stability and the control over degradation. The form-stability is of primary importance in the context of induced autoregeneration, in view of accomplishing the functional regeneration of critical sized tissue defect, avoiding scar tissue formation. Control over degradation can be incorporated as another functional element, in order to enable the ingress of the neotissue complemented with the structural evolution of the implant. In this context, the fixation of a porous geometry in hydrogel systems, resulting in 3-D scaffolds, is highly demanded, in order to provide an appropriate physiological environment to cells, with the necessary physical and structural properties.[91] The role of the biomaterial should be therefore to facilitate and support existing biological process, and serve as a synthetic mimic of the ECM. Therefore, microenvironment and macroenvironment should be closely investigated in order to guarantee the hierarchical organization of the neo-tissue. Understanding of how these length scales can be defined and controlled during material degradation is also crucial to ensure a successful biomaterial-assisted autoregeneration.

# 2. Aim of the Ph. D. thesis

The goal of this work was to develop a structured hydrogel system based on gelatin for potential application in Regenerative Therapies, such as for biomaterial-induced autoregeneration. The biomaterial should mimic the structure and function of the ECM and should display material properties tailorable to the *in vivo* situation. For this purpose, inherent structure-property relationships needed to be established in order to control and adjust the swelling behaviour and the mechanical properties of the material in physiological conditions. Especially for potential application in bone regeneration, the following biomaterial requirement needed to be fulfilled: (1) a porous geometry should be fixed in the material in order to allow tissue in growth; (2) form-stability must be ensured in contact with physiological environments including full elastic recoverability; (3) the material shall provide a suitable microenvironment for regiospecific cells, so that a local Young's modulus in the kPa range shall support selective cell growth and direct stem cell differentiation; (4) the macroscopic mechanical properties have to be adjusted to the stress applied during surgery and the elasticity of the tissue to be replaced; (5) as a temporary substitute of the ECM, the material should be degradable and the degradation should lead to controlled change of mechanical properties. In this way, material degradation can be complemented with the regeneration process, allowing the formation of the neo-tissue; (6) the material should display full cytocompatibility and minimal endotoxin contamination.

By establishing structure-property/functions relationships between the chemical structure, material properties and the biological material competence, the design of smart structured gelatin-based hydrogels should be developed at both the structural and functional level.

# 3. Strategies and concept

In order to induce the regeneration of a critical defect, a biomaterial scaffold should interact with the surrounding cells and tissues as well as with blood, interstitial liquids, and the immune system. In order to provide an interface to the organism similar to the naturally ECM, a biopolymer derived from the ECM should be the basis for the material development. By transferring the material design strategies used for the synthetic polymers to biopolymers, a structured, entropy-elastic hydrogel system[92] should be formed to achieve defined structure-property relationships for homogeneous films. To accomplish thermo-mechanical properties tailored to the *in vivo* situation, and controlled swelling behavior, chemical crosslinking should lead to the permanent suppression of the intra- and inter-molecular chain organizations into single and triple helices. Once such reference hydrogel films are established, the above-mentioned synthetic strategy should be transferred to the design of a scaffold system.[13] Especially, the embodiment of the gelatin network in a porous architecture would lead to the formation of hierarchically structured hydrogels. Here, the control of material properties should be ruled by the bulk properties of the polymer network as well as by the scaffold geometry.[93] By keeping the scaffold geometry constant, the micro- and macromechanical properties are expected to follow the structure-property relationships defined for the hydrogel reference system. The biological applicability of the scaffolds will have to be investigated according to direct and indirect cell tests and endotoxin contamination.

The synthesis of an entropy-elastic hydrogel shall form the basis to design a structured gelatin-based scaffold. Gelatin was selected as the ECM-derived biopolymer, since it is degradable, pro-angiogenic,[39] non-immuogenic,[58,59] and has been widely employed for biomedical applications.[61] When synthesizing an entropy-elastic hydrogel by crosslinking gelatin above its gel-sol transition ($T > 40$ °C), gelatin should be permanently captured in its randomly coiled state because of the presence of covalent crosslinks. A low molecular weight diisocyanate will be used for the chemical crosslinking of gelatin (chapter 4). Isocyanate chemistry was recently applied for the synthesis of polyester urethane networks, which showed a controlled change of mechanical properties during degradation.[19] Crosslinking of gelatin with diisocyanates leads to the formation of degradable urea bonds, resulting from the nucleophilic addition of the lysine ε-amino terminations and aminotermini of protein chains. Therefore, a fully degradable gelatin network is expected. Besides the reaction with gelatin lysines, isocyanates can react with water by decarboxylation, resulting in a newly-formed amino group. In this context, a complex cascade of complex reaction can occur simultaneously, potentially resulting in a combination of oligourea chains either grafted or crosslinked to gelatin. Therefore, understanding of the crosslinking reaction on a molecular level will be required, e. g. using model reactions. In this way, not only the control of the reaction should be enabled, but the knowledge of the elucidated network internal structure serves as the basis for the development of structure-function relationships. An excess of isocyanate is thereby required to accomplish direct crosslinking rather than simple grafting. For these reasons, crosslinking will be initially carried out in DMSO, under easily controllable reaction conditions, and then transferred to water as the optimal solvent to ensure biomedical applicability of the resulting hydrogels. Hexamethylene diisocyanate (HDI) and ethyl ester lysine diisocyanate (LDI) are used as chemical crosslinker. The presence of the HDI aliphatic chain should lead to a reduction of the gelatin swelling in aqueous medium. Other than HDI, LDI was selected since it is known to result in non-toxic lysine derivative degradation products.[94] Both HDI and LDI are only slightly soluble in water, though they may react with it. To confer the solubility of the diisocyanates in water, a surfactant will have to be employed in order to form micelles that incorporate the hydrophobic crosslinker. The excess of isocyanate groups (NCO) will be varied with respect to the gelatin lysine (Lys) content in the crosslinking reaction. The resulting hydrogels will then be characterized by their thermo-mechanical properties, swelling behavior, as well as the organization of gelatin chains.

Once the hydrogel system is characterized, an integrated foaming crosslinking process will then be employed for the development of three-dimensional scaffolds. Gelatin is known to possess foaming properties in aqueous solution,[95,96] so that the formation of gelatin foams was the basis for the design of defined porous scaffolds. The aqueous solution of gelatin and a surfactant will be foamed above the gelatin sol-gel transition temperature ($T$ = 45 °C). Chemical crosslinking of the gelatin foaming solution is then expected to provide the porous shape fixation, resulting in the formation of an entropy-elastic porous hydrogel. Alternatively, chemical crosslinking could be carried out on the resulting gelatin foam in a separated step.[97] However, this approach could not lead to the formation of an entropy-elastic hydrogel, since the foam is initially stabilized based on the gelatin triple helix renaturation. The hydrophilic-lipophilic balance (HLB) of the surfactant can be adapted as general criterion for the surfactant selection, in order to reach foam formation, homogeneous pore distribution, and HDI miscibility in the aqueous solution. The stability of the resulting foams will be investigated by measuring the steady-state volume of non-foamed solution, after hydrogel formation. Crosslinked scaffolds will be synthesized with different surfactants, HDI excess, and gelatin concentrations. Variation of the HDI excess is expected to affect the network architecture, while variation of gelatin concentration was expected to influence the density of the scaffold bulk phase. SEM and μCT can be used to characterize the porous internal structure, while test reactions and WAXS will be again carried out to confirm the formation of an amorphous gelatin network in the resulting scaffold.

As for application *in vivo*, scaffolds will be characterized in biologically-relevant conditions. Uptake and swelling studies will be carried out on the scaffolds in physiological conditions, and will be complemented by swelling tests on the powdered bulk material. In this way, the effect of the material porous geometry can be considered in order to understand the swelling mechanism of the scaffold. With that regard, evaluation of the pore size in the wet compared to the dry state is crucial to elucidate the mechanism of material form-stability. Wet-state compression tests will be conducted to determine the macroscopic mechanical properties of the scaffolds at the tissue level. Thermal properties will be analyzed in the wet state to elucidate the elastic recoverability of the material through identification of the network morphology. Additionally, atomic force microscopy (AFM) will be carried out on the pore walls in order to evaluate the microscopic stiffness sensed by the cells during adhesion. Then, a full description can be accomplished about the scaffold mechanical properties at both cell (AFM) and tissue (compression) level. The degradation behaviour will furthermore be studied through gravimetric measurements and compression tests, as well as WAXS and morphological analyses.

In order to investigate the scaffold cytocompatibility, direct cell tests will be conducted to understand whether the porous structure enables cell proliferation within the implant. These tests will be complemented with indirect tests, in order to quantify the material cytotoxic response. The endotoxin material contamination will be quantified by the tumor necrosis factor (TNF-α) induction in the whole blood and by the limulus amebocyte lysate (LAL) assay.

In the following, chapters 4-7 describe the results of this thesis and provide detailed discussion as to their context. Chapter 4 includes the synthesis and characterization of entropy-elastic gelatin-based hydrogels, which served as a foundation on which the following work was based. Chapter 5 describes the design of a gelatin-based scaffold system prepared using the knowledge gained from the gelatin-based hydrogels. As the scaffolds are intended for clinical use as biomedical implants, their behavior in aqueous environments was investigated and is discussed in Chapter 6. Finally, chapter 7 discusses the development of medical-grade scaffolds, which were prepared under 'clean' conditions and evaluated biologically.

# 4. Entropy-elastic gelatin-based hydrogels: synthesis and characterization

Films prepared by reacting Gelatin with HDI in DMSO at 45 °C were performed in Petri dishes as the materials were crosslinked within seconds. These films were analyzed using WAXS, and their swelling and mechanical properties were quantified. Consequently, the reaction was transferred to water, which is a preferred solvent for biomedical applicability. Test reactions elucidated the crosslinking mechanism as well as the internal structure of the gelatin hydrogels on the molecular level. As a result, structure-property relationships could be established.

## 4.1 Networks synthesized in DMSO

DMSO is one of the few solvents in which gelatin is soluble and which does not react with HDI.[77] The crosslinking reaction was conducted above the disaggregation temperature, in order to stabilize the gelatin chains in the randomly coiled state. By permanently suppressing the gelatin chain helicity, the material properties were expected to be systematically ruled by the only variation of crosslinking density (Figure 4.1). In this way, the principle of entropy-elastic covalent networks, which is a well known method for tailoring the mechanical properties of synthetic materials, should be transferred to a gelatin-based material. Consequently, this chemical modification aims at producing an amorphous hydrogel rather than a thermoplastic material, requiring large amounts of crystalline hard segments to obtain a wide range of Young's moduli.[98,99]

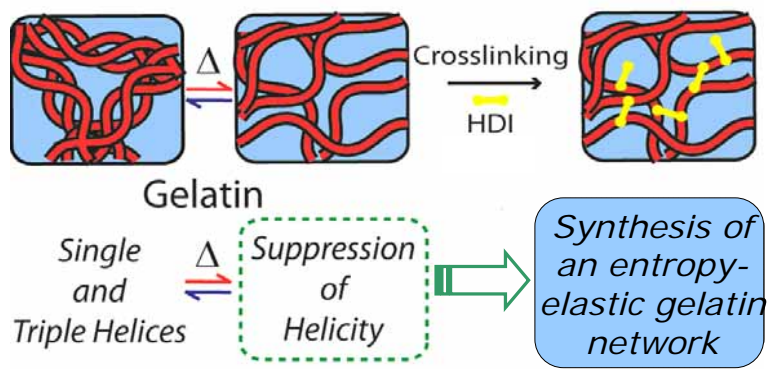

**Figure 4.1. Gelatin is in the coil conformation above the sol-gel transition (T > 40 °C). In these conditions, specific chemical crosslinking could be accomplished with hexamethylene diisocyanate (HDI). The resulting entropy-elastic network is expected to show permanently suppressed triple helices, thereby enabling defined structure-property relationships.[92]**

The reaction of gelatin with four different amounts of HDI by a polymer analogous reaction in solution, followed by washing, drying, and swelling in water, resulted in flexible films. For each sample composition, gelation occurred within seconds. The samples are referred to as GX_HNCOZ*, with X = gelatin concentration in wt.-%, H = HDI (crosslinker), and Z = molar excess of isocyanate groups compared to gelatin lysines (NCO/Lys).

# 4.1.1 Investigation of crystallinity in Networks synthesized in DMSO using WAXS

In order to investigate the effect of crosslinking on the gelatin chain helicity, WAXS spectra of crosslinked films were compared with the one of non-crosslinked gelatin. A typical WAXS spectrum of pure gelatin displays three main peaks, at $2\Theta = 8°$, $20°$, and $31°$. These peaks depict the diameter of the triple helix (repeat distance of 11-12.6 Å), the amorphous halo, and the amino acid contacts along the axis of single helices (repeat distance of 2.9 Å), respectively.[100,101] WAXS spectra of the dried films showed only a peak representative for the amorphous organization of chains ($2\Theta = 20°$, Figure 4.2). In contrast, gelatin peaks indicating single or triple helical regions ($2\Theta = 31°$ or $8°$) were not found in the crosslinked films. Therefore, it was proved that the above-described crosslinking strategy successfully enabled the suppression of gelatin chain helicity, which is crucial for the synthesis of a gelatin network with defined structure-property relationships.

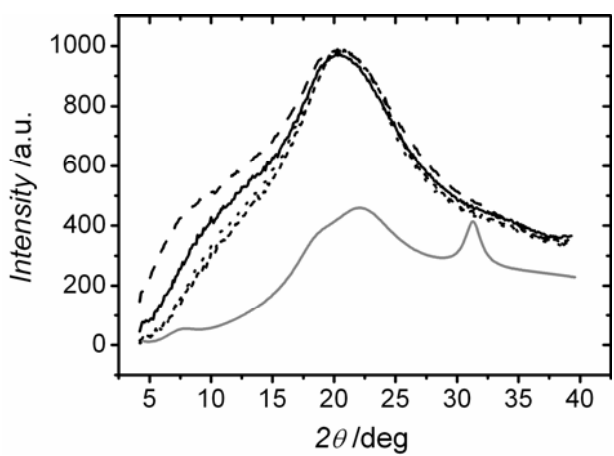

**Figure 4.2. WAXS spectra of dried gelatin films crosslinked with increasing amounts of HDI in DMSO as solvent. (---) G10_HNCO1*, (—) G10_HNCO3*, (·····) G10_HNCO5*, (---) G10_HNCO8*, (—) Gelatin.**

## 4.1.2 Investigation of Swelling behavior of Networks synthesized in DMSO

Gelatin is known to swell in aqueous environment, often resulting in the formation of hydrogels with limited elasticity and mechanical properties. In order to ensure the clinical applicability of hydrogel films *in vivo*, gelatin was crosslinked with varied crosslinker excess (1-8 NCO/Lys molar ratio). The resulting materials were investigated through uptake and swelling tests in aqueous solutions, whereby the following equations are introduced for the determination of the water uptake (*H*, eq. 4.1) and swelling degree (*Q*, eq. 4.2):

$$H = \frac{W_s - W_d}{W_d} \qquad (4.1)$$

$$Q = 1 + \frac{\rho_2}{\rho_1} \frac{W_s - W_d}{W_d} \qquad (4.2)$$

where $\rho_2$ is the density of the gelatin material, here determined by the weight/volume ratio of dry samples, $\rho_1$ is the density of water at 25 °C (here as equal as 0.997 g·mL$^{-1}$), and $W_s$ and $W_d$ are the sample weights in the swollen and dry state, respectively.

All samples displayed an increase of both *H* and *Q* during the first ten hours, though they were stable for several days (no degradation was observed). Furthermore, films were also tested in cell culture medium, in order to study the material behavior in biologically-relevant conditions.

The films crosslinked in DMSO with a 1:1 NCO/Lys molar excess showed significantly higher values of *H* and *Q* than the hydrogels prepared with higher crosslinker excess (3-8 NCO/Lys molar excess). Additionally, more water than cell culture medium was taken up in hydrogels with this composition (Figure 4.3). On the contrary, only small differences in *H* or *Q* were found for hydrogels with 3, 5, or 8 NCO/Lys molar excess, and the values for water or cell culture medium were very similar.

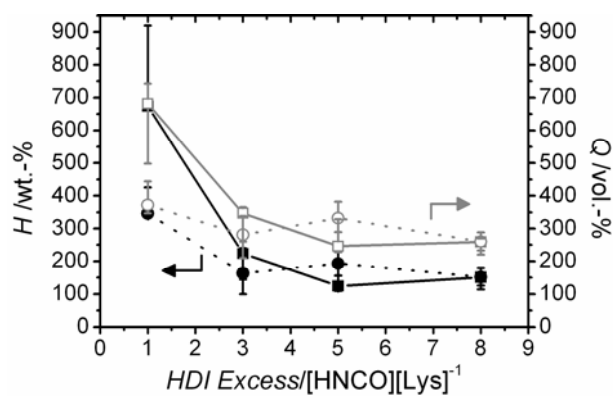

**Figure 4.3: Water uptake *H* (black) and degree of swelling *Q* (grey) of/in water (■□) and cell medium (●○) of the gelatin films crosslinked in DMSO.**

The above-described results proved that it was possible to control the swelling behavior of crosslinked films, though only slight variation of *Q* was observed in films synthesized with different crosslinker excess. Additionally, the swelling behavior of films was only slightly different in cell culture medium compared to water, suggesting that the network architecture ruled the material swelling, irrespective of the ionic contribution resulting from the presence of salts or sugars in the swelling medium.

## 4.1.3 Investigation of Mechanical Properties of Networks synthesized in DMSO using tensile tests

Besides the swelling behavior, the mechanical properties of swollen films were investigated by tensile tests. Therefore, films synthesized with varied HDI excess were analyzed. Figure 4.4 describes the variation of Young's modulus (*E*) in films with different composition. When the HDI excess was raised from 1 to 8 NCO/Lys molar ratio, films displayed an increase of E from roughly 60 kPa to about 450 kPa, which was accompanied by a decrease of elongation at break ($\varepsilon_b$), and an increase of maximum tensile strength ($\sigma_{max}$) (Table 4.1). However, the variation between different replicas of the films G10_HNCO5[*] and G10_HNCO8[*] was quite large. Also, while the mechanical properties of the hydrogels prepared in DMSO could be tailored, H and Q of the crosslinked films were affected to a smaller extent.

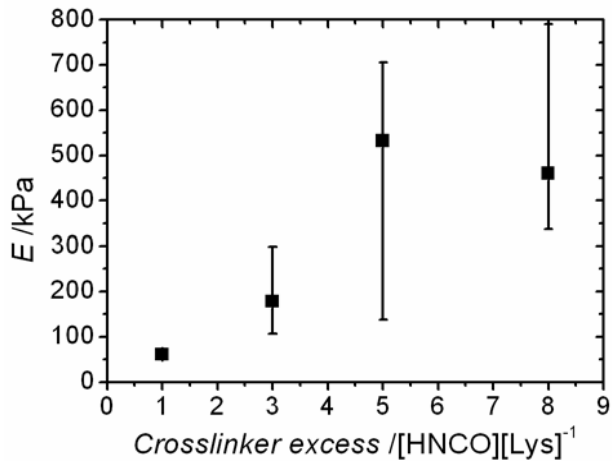

**Figure 4.4. Young's modulus (*E*) of gelatin hydrogel films crosslinked with HDI in DMSO determined in tensile tests depending on the crosslinker concentration. The reported value is the median measurement, and the error represents the highest and lowest values.**

In order to achieve a systematic variation of material properties, the crosslinking reaction was transferred to water, resulting in the formation of gelatin hydrogels. Water is indeed an optimal solvent in perspective of accomplishing cytocompatible biomaterials, since it is the basic natural biological medium and is the main solvent for gelatin. In contrast, DMSO can be hardly extracted from the films after synthesis, thereby potentially leading to a cytotoxic material response.

# 4.2 Networks synthesized in water

Water, if sterile and endotoxin free, is a suitable solvent for preparing gelatin-based implants, especially since gelatin is highly water-soluble. However, the crosslinker HDI is only slightly soluble in water, while the isocyanate groups can react with water resulting in decarboxylation. Therefore, questions addressing this issue were: how to improve on the HDI solubility in water and how to realize the crosslinking reaction in water.

Olde Damink et al. employed a tenside to confer the solubility of HDI for crosslinking of collagen.[54] Here, saponin[102] was employed as surfactant as it has a lower hydrophilic/lipohilic balance than the formerly used tween-80[54] and therefore should lead to an even better solubility of the hydrophobic HDI in water. Furthermore, the reaction was carried out at 45 °C, aiming at the suppression of the gelatin triple helical conformation by the presence of chemical crosslinks. HDI crosslinked films were prepared from 10 wt.-% gelatin solutions with three different crosslinker ratios (3, 5, and 8 NCO/Lys molar ratio) and three different gelatin concentrations (7, 10, and 13 wt.-%) with an 8 NCO/Lys molar excess. Besides HDI, LDI was also used as crosslinker, and resulting films were prepared from 10 wt.-% gelatin solutions with three different LDI ratios (3, 5, and 8 NCO/Lys molar ratio). The kinetics of the reaction with LDI was about 2-fold slower than for the reaction with HDI, i.e. gelling occurred only after about 20 min compared with 8 min. The samples are referred to as GX_YNCOZ, with X = gelatin concentration in wt.-%, Y = type of crosslinker (H = HDI or L=LDI), and Z the molar excess of isocyanate groups.

During crosslinking of gelatin, isocyanate can potentially react with water, resulting in a release of carbon dioxide, e. g. as bubbles. This could potentially cause a non-homogeneity in the material surface, which could lead to artifacts during the material characterization. In order to address this point, the surface and section of was investigated. Dry films showed homogeneous surface when analyzed by SEM (Figure 4.5, a-b). The slightly waved sample surface observed with scanning electron microscopy (SEM) was most likely caused by the drying process needed for the sample preparation. Likewise, optical profilometry showed relatively flat surfaces with an averaged roughness ($R_a$) in the range of 0.1-1 μm (Figure 4.5, c-d). Therefore, the extent of bubble/foaming in the films was successfully minimized during the reaction, so that a homogeneous material was formed.

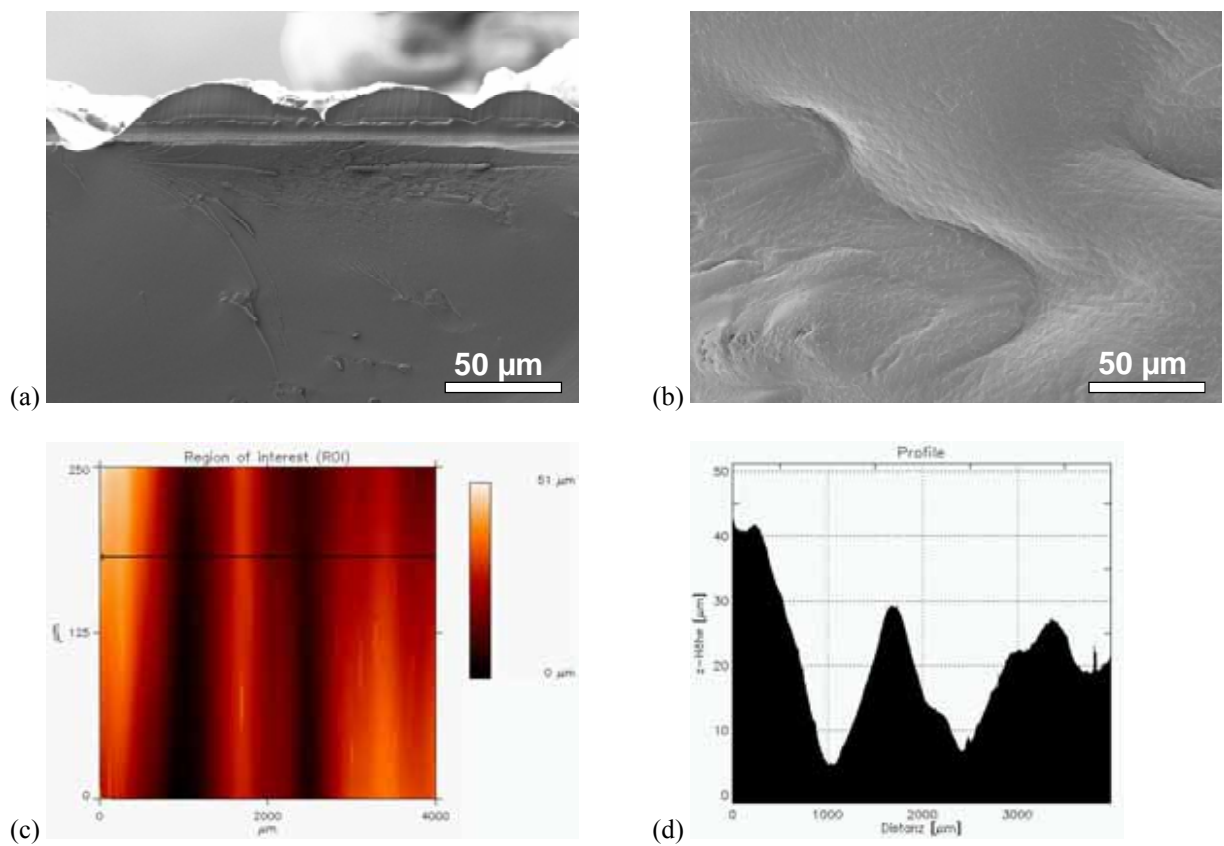

**Figure 4.5. Scanning electron microscopy pictures of lyophilized G10_HNCO8 from the top-surface (a) and cross-section (b). A homogeneous surface on a crosslinked gelatin film (G7_HNCO8) was also observed by optical profilometry (c: overview scan size, d: relative profile), whereby an averaged roughness ($R_a$) and root mean square roughness ($R_q$) of roughly 0.1 and 0.272 μm, respectively, was measured.**

## 4.2.1 Investigation of crystallinity in Networks synthesized in water using WAXS

The crosslinking reaction was conducted above the gelatin sol-gel transition temperature. At $T > 40$ °C with the given concentration, gelatin gels are soluble in water because the fibril bundles become disaggregated. In order to systematically rule material properties by the only variation of crosslinking density, the reactions of gelatin with HDI were necessarily carried out above this temperature threshold, thereby stabilizing the gelatin chains in the randomly coiled state. The dried crosslinked films were investigated using

WAXS to further explore the link between network architecture and macroscopic properties of the resulting material. Specifically, the structural organization of either crosslinked or non-crosslinked gelatin chains as amorphous coils, single helices, or triple helices could be explored (Figure 4.6).

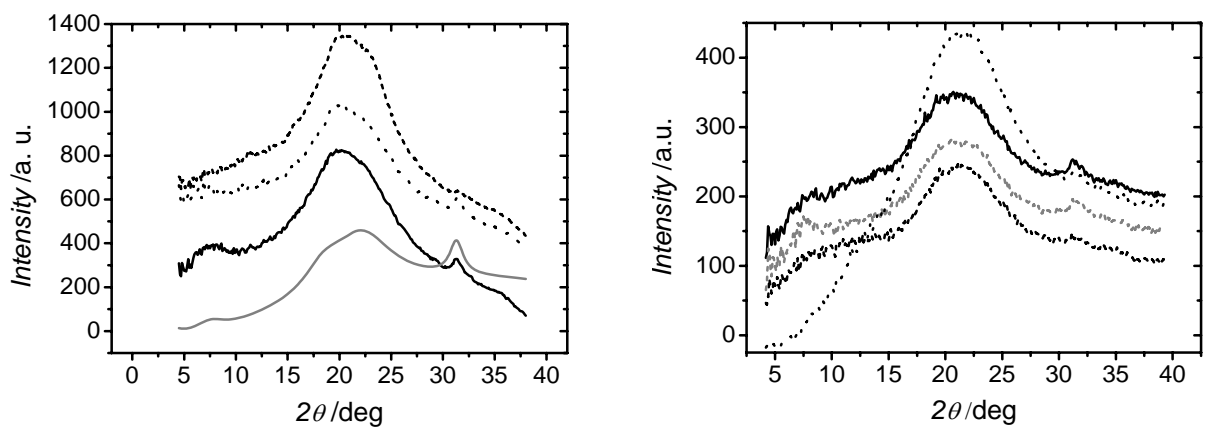

**Figure 4.6. WAXS spectra of dry crosslinked samples at different isocyanate excess: a) HDI films: (—) G10_HNCO3, (·····) G10_HNCO5, (--) G10_HNCO8, (—) Gelatin; b) LDI films: (—) G10_LNCO3, (·····) G10_LNCO5, (—) G10_LNCO8, (—) G10_LNCO8 (2x Saponin).**

Regardless of the composition, the WAXS spectra of the films displayed peaks at 20°, indicating the presence of amorphous regions, as also observed in films crosslinked in DMSO. With increasing HDI content, the 8° peak intensities were suppressed, and the sample with the highest HDI excess (G10_HNCO8) showed only minimal signal intensity near that region. A similar reduction was observed for the peak at 31° in samples synthesized with increasing HDI excess. As the films were treated and casted identically and intensely ordered helical structures were observed in the case of pure gelatin films, it is likely that the incorporation of crosslink junctions inhibited the renaturation of single and triple helical structures.[103,104,105] Such a functionalization presents a novel method of microarchitectural control for biopolymers, where the triple- and single-helical renaturation is systematically hindered according to varying degrees of crosslinking. Samples synthesized with varied LDI excess in the same ratio as HDI similarly showed a suppression of the peaks related to single-and triple-helices with increasing crosslinking density.

## 4.2.2 Swelling behavior of Networks synthesized in water

The swelling behavior was investigated in water and cell culture medium in order to investigate the material performance in biologically-relevant conditions (Figure 4.7). An increase in HDI concentration resulted in a decrease of both water and cell culture medium uptake ($H$). The effect on the degree of swelling ($Q$) in the two media was similar: increasing the LDI crosslinker resulted in reduced Q and H. With the same crosslinker excess compared to HDI crosslinked films, LDI crosslinked films displayed higher $H$ and $Q$ values.

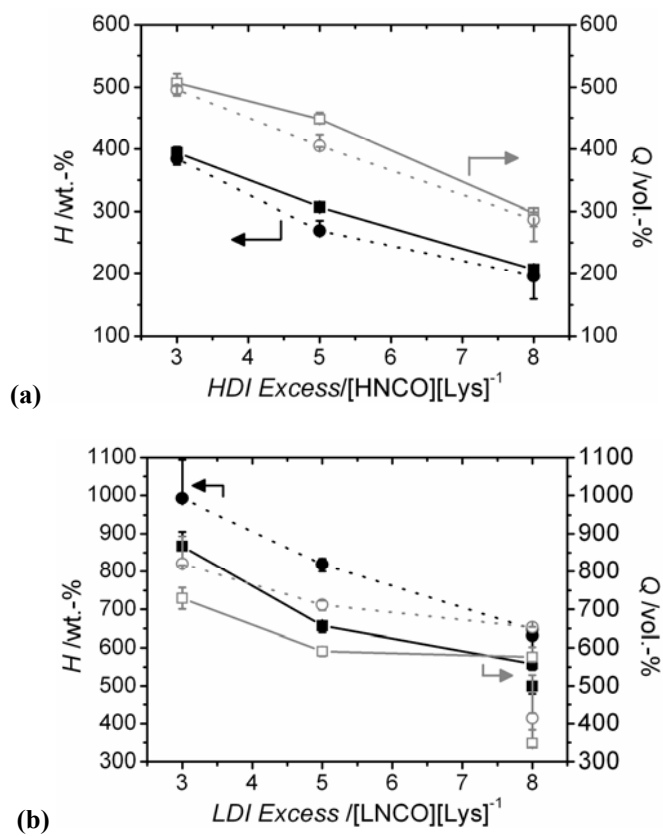

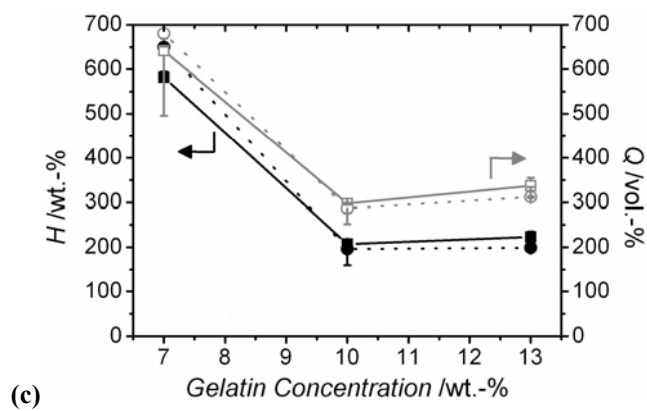

**(c)**

**Figure 4.7: Uptake (black) and swelling (grey) of/in water (■□) and cell culture medium (●○) of gelatin films G10 crosslinked in H$_2$O depending on the HDI concentration (a) or LDI concentration (b)[*] and of gelatin films crosslinked with 8 eq. HDI depending on gelatin concentration (c).[*] Data points which are not connected to the guideline curves refer to samples G10_LNCO10 synthesized with double amount of saponin.**

Comparing samples synthesized with different gelatin concentrations but constant NCO/Lys molar ratio, it was found that G7_HNCO8 hydrogels had much higher $H$ and $Q$ values than the ones corresponding to G10_HNCO8 and G13_HNCO8 films. On the other hand, samples G10_HNCO8 and G13_HNCO8 exhibited similar swelling and uptake behavior.

The above-mentioned observations correspond to the formation of covalently crosslinked networks having an increasing crosslinking density with increasing crosslinker excess during the reaction. Varying the gelatin concentration also had an effect on crosslinking density and/or reaction kinetics, even if the NCO/Lys ratio was kept constant during the reaction. The presence of a chemical network is also supported by the longer (at least 3 days) stability of swollen crosslinked films at 37 °C, in contrast to pure gelatin films, which completely dissolved in 30 °C water after only 16 h.

# 4.2.3 Mechanical properties of Networks synthesized in water

The mechanical properties of the films swollen in water were examined at 25 °C by tensile and compression tests (Figure 4.8, Figure 4.9, and Table 4.1) in air.

In tensile tests, HDI crosslinked films displayed increasing Young's moduli ($E$: 140-740 kPa) with increasing HDI excess (3-5 NCO/Lys molar ratio). On the other hand, increasing the gelatin concentration from 7 to 13 wt.-% did not alter $E$ (ca. 740 kPa), when keeping the HDI excess constant (8 NCO/Lys molar ratio). LDI films likewise showed an increase of $E$ (70-230 kPa) with increasing crosslinker concentration, though $E$ was much lower compared to the HDI crosslinked samples. The elongation at break $\varepsilon_b$ decreased with increasing crosslinker density for both HDI- and LDI-based samples, while $\sigma_{max}$ did not exhibit a clear trend (and large data variation). The mechanical differences between the HDI- and LDI-based materials and the higher swelling and lower Young's moduli for LDI- compared to HDI crosslinked materials indicated that the crosslinkers showed a different reactivity in the reaction mixture.[106] The different reactivities might be due to differing solubilities and / or interactions of the crosslinker with the surfactant. This point will be addressed in section 4.3. The variation of gelatin concentration did not show a clear influence on $\varepsilon_b$ and $\sigma_{max}$, and an influence of the surfactant concentration on the mechanical properties was not observed in the studied samples.

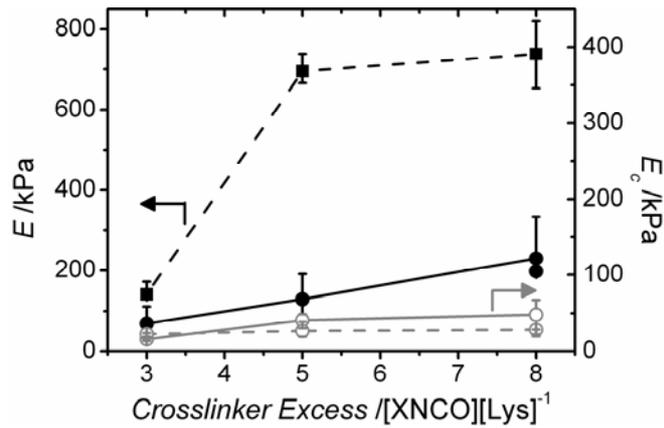

**Figure 4.8: Young's modulus $E$ (black) and compressive modulus $E_c$ (grey) depending on crosslinker type and concentration (■□ = HDI crosslinked films, ●○ = LDI crosslinked films[*]). [*] Data points which are not connected to the guideline curves refer to samples G10_LNCO10 synthesized with double amount of saponin.**

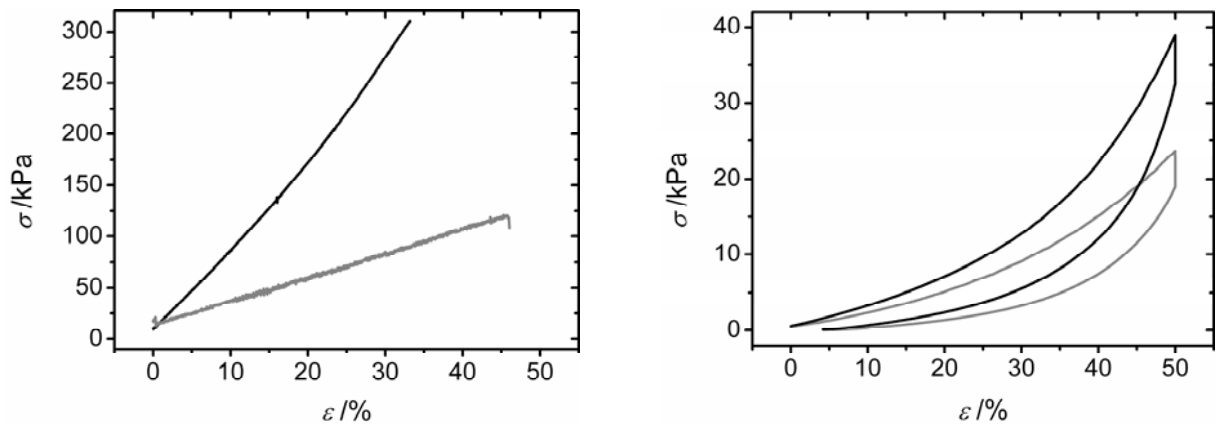

**Figure 4.9. Stress-strain curves measured during tensile (G10_XCNCO8, left) and compression (G10_XNCO5, right) tests of swollen crosslinked gelatin samples at 25 °C. (—): HDI-crosslinked gelatin, (—): LDI-crosslinked sample.**

Some of the tensile experiments were repeated in a water tank as measurements in air might be affected by drying of the samples during the measurements. However, the results for samples pre-swollen in water and measured in air were insignificantly different from those measured in a water tank. Compressive moduli ($E_c$) of 15-50 kPa were measured for the HDI materials, which increased with increasing crosslinker density. LDI materials showed only a very slight increase of $E_c$ from 22 to 28 kPa, with a large variation between replicas. Samples synthesized with constant LDI excess and double amount of surfactant showed a decrease of $E_c$ to ca. 17 kPa, compared to the samples with constant surfactant content. All samples exhibited shape recovery up to 95% when mechanical load was removed, which was not observed in the case of non-functionalized gelatin materials.

**Table 4.1. Summary of mechanical properties of the films: the median value and the lowest and highest values for the three samples are given.**

| Sample ID | $E_c$[a] | $E$[a] | $\varepsilon_b$[a] | $\sigma_{max}$[a] |
|---|---|---|---|---|
| | kPa (range) | kPa (range) | % (range) | kPa (range) |
| G10_HNCO3 | 16 (14-18) | 141 (136-172) | 122 (121-134) | 252 (207-254) |
| G10_HNCO5 | 41 (33-43) | 695 (666-738) | 56 (49-57) | 487 (445-566) |
| G10_HNCO8 | 48 (29-74) | 738 (653-820) | 35 (33-38) | 310 (288-350) |
| G7_HNCO8 | n.d. | 741 (581-773) | 34 (32-35) | 359 (244-425) |
| G13_HNCO8 | n.d. | 730 (485-840) | 41 (29-46) | 338 (276-449) |
| G10_LNCO3 | 22 (21-26) | 69 (61-111) | 170 (144-285) | 364 (267-369) |
| G10_LNCO5 | 27 (23-34) | 129 (83-191) | 50 (27-56) | 74 (56-121) |
| G10_LNCO8 | 28 (28-37) | 230 (191-238) | 46 (25-62) | 121 (70-199) |
| G10_LNCO8[†] | 17 (12-19) | 198 (198-334) | 54 (23-62) | 229 (96-240) |
| G10_HNCO1[*] | n.d. | 61 (50-66) | 48 (34-60) | 30 (23-33) |
| G10_HNCO3[*] | n.d. | 178 (107-299) | 31 (16-45) | 52 (28-74) |
| G10_HNCO5[*] | n.d. | 532 (137-706) | 26 (12-27) | 84 (42-139) |
| G10_HNCO8[*] | n.d. | 460 (337-790) | 21 (20-104) | 146 (89-284) |

[a]: $E_c$ = compressive modulus, $E$ = Young's modulus, $\varepsilon_b$ = elongation at break, $\sigma_{max}$ = tensile strength, [*]: samples synthesized in DMSO, [†] synthesized with doubled amount of saponin, 'n.d.' stands for 'not determined'.

$E_c$ was always much lower than $E$, e. g. for G10_HNCO8 about 15-fold. This relationship has been observed in materials with a negative Poisson's ratio.[107,108] However, the macroscopic observation that the material did not expand when elongated speaks against a negative Poisson's ratio for these gelatin-based hydrogels. The more likely explanation is that during the formation of the shaped bodies for tensile tests (dogbone-shaped) and compression tests (cylinders), the samples do not solidify immediately but allow a setting of polymer chains on the bottom of test pieces, resulting in an anisotropic density. The samples for the tensile tests are punched from the bottom and therefore have a higher density than the more voluminous cylinders used for compression tests, which explains their different behavior in the mechanical tests.[93,109] Other than that, a narrow range of data between sample replicas, either from compression of tensile tests, was obtained. This highlights that the synthesis of the gelatin network and film formation was reproducible. In order to minimize the setting of polymer chains, reduced volumes of crosslinking mixture could be cast for the preparation of bulk samples for compression tests.

Overall, these mechanical results followed classical entropy-elastic mechanical properties. The increase of the crosslinker ratio during the synthetic step directly affected the network crosslinking density, resulting in materials with adjustable mechanical properties. In entropy-elastic polymer networks the segment chain length of the flexible chains directly correlates to $\varepsilon_b$ of the material. Also, the increase in the network crosslinking density leads to a drastic increase of $E$, though this is mainly observed for networks with very low crosslinking ($E$ is relatively constant for higher degrees of crosslinking). The fact that the mechanical properties of this system correspond so well with classical polymer-based networks indicates that the presented strategy successfully yielded tunable biopolymer-based materials.

## 4.2.4 Investigation of Thermal properties of Networks synthesized in water

Thermal transitions of the hydrogels were investigated with DMTA, which is a much more sensitive method than DSC in measuring the thermal transitions of networks (Figure 4.10). Several peaks/shoulders are observed for tan$\delta$, two (20, 45 °C) and three (20, 30, 50 °C) peaks for HDI- and LDI-crosslinked samples, respectively. The temperatures at which maxima are obtained for the loss modulus ($E''$) and the loss tangent (*tan$\delta$*) depict thermal transitions.[110] Here, clear peaks could be observed for both samples in the range 17-20 °C, at which temperatures neither melting nor glass transitions could be measured by DSC.

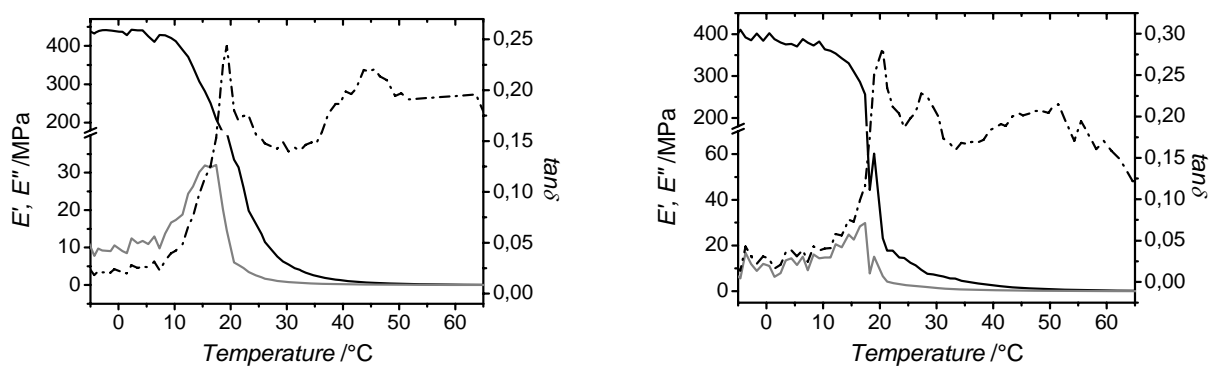

**Figure 4.10. Compressive Dynamical Mechanical Analysis at varied Temperature (DMTA) of HDI- (left) and LDI-crosslinked (right) samples (G10_XNCO5) after swelling in water at 25 °C in air. (—): Storage modulus ($E'$), (—): Loss modulus ($E''$), (---): Loss tangent (*tan$\delta$*).**

In order to understand the above-mentioned DMTA transition, WAXS spectra were considered and indicated the suppression of triple helices by the incorporation of gelatin in the chemical network, after crosslinking with either HDI or LDI (Figure 4.6). Based on DMTA and WAXS observations, it is likely that the peak of *tan$\delta$* at roughly 20 °C indicates a glass transition of the network, suggesting a rubbery-like state of the material at the compression temperature, and therefore explaining the full shape recovery of samples after compression. Besides that, samples synthesized with different crosslinker content did not show remarkable differences in the thermo-mechanical properties by DMTA. The other peaks with lower intensities could not be attributed to distinct thermal transitions and could potentially be due to noise (small storage and loss moduli) or slipping.

# 4.3 Analysis of the crosslinking reaction

The isocyanate groups of HDI or LDI can principally react with all nucleophilic functional groups of gelatin. Therefore, reactions could be expected with amino groups of lysine (and, in case of the collagen/gelatin system, hydroxylysine), aminotermini, aromatic NH-groups (histidine), the alcohols of serine and threonine, phenols (tyrosine), as well as thiols (cysteine), carboxylic acids and amides. However, a detailed study of this reaction revealed the more or less exclusive reaction of amino groups.[111,112,113] Gelatin generally has an approximate lysine/hydroxylysine content of 3 mol.-%,[114] which was verified in the starting materials using a Trinitrobenzene sulfonate (TNBS) colorimetric assay.[115] Ideally, during the crosslinking reaction, the isocyanates groups of HDI or LDI react with the lysine $\varepsilon$-amino functions along two opposing gelatin chains. On the other hand, in aqueous conditions, isocyanate decarboxylation may occur, which leads to the formation of free amino groups that can react with additional isocyanate groups. Therefore, the reaction of gelatin with HDI or LDI is quite complex and the following reactions are possible: i) the crosslinking of gelatin by monomers or oligomers; ii) the grafting of gelatin by monomers or oligomers; iii) the decarboxylation of both isocyanate groups, resulting in a blend of gelatin and diisocyanate-based oligomers; and iv) the decarboxylation of a single isocyanate group, leading to the formation of a monomer or oligomeric ring (Figure 4.11). As all of these could contribute to the changes in mechanical behavior after crosslinking, a model reaction was studied to identify the most predominant compounds formed.

HDI or LDI (2.5 molar excess) was reacted with glycine dissolved in water in the presence of saponin or in DMSO, which corresponded to the synthetic procedure of G10_XNCO5, although the solution viscosity could not be mimicked. Following the reaction, the resulting oligomers were analyzed and quantified by ESI mass spectrometry,[116,117] where the peaks corresponding to saponin were disregarded.

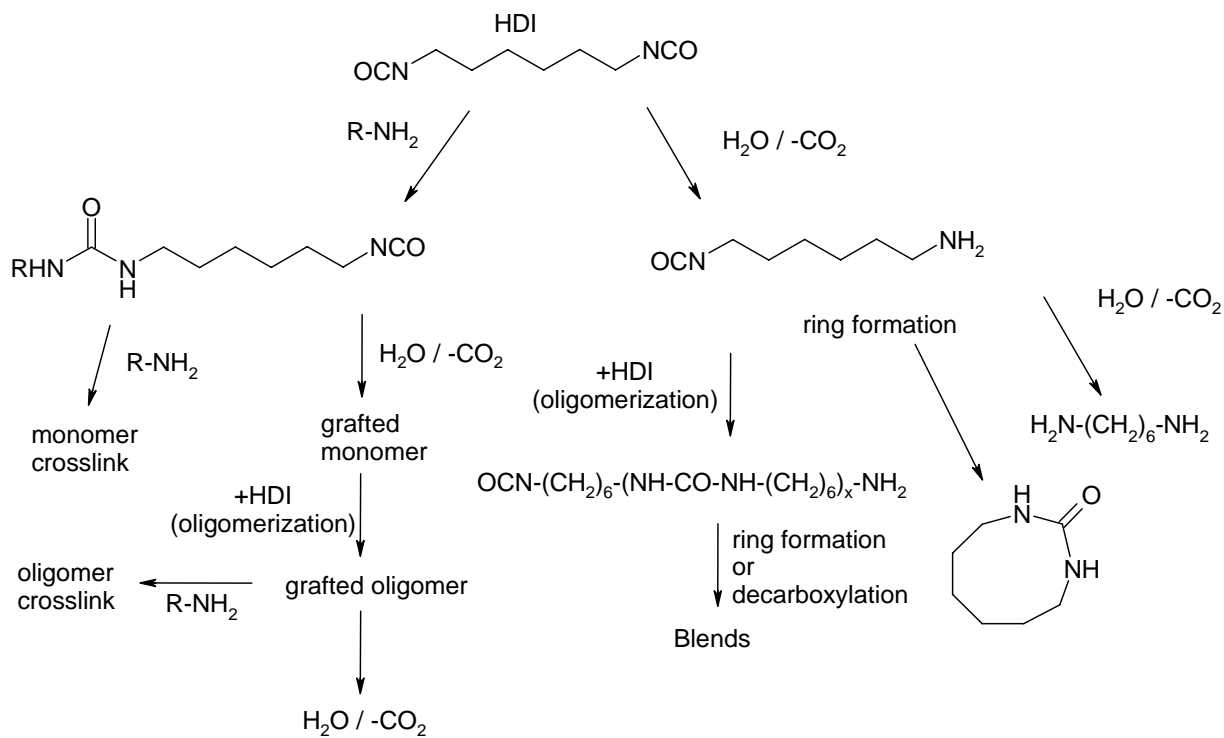

**Figure 4.11. Possible reactions of HDI in an aqueous gelatin solution. Left section: reaction of HDI with an amino function of lysine. The reaction can proceed involving direct crosslinking or oligomer grafting, thereby subsequently terminating with crosslinking or decarboxylation. Right section: initiation of blend formation and ring closure by decarboxylation of the HDI.**

**Table 4.2. Expected and found m/z for the M+H⁺ specimens in the ESI MS analysis of the reaction of glycine with HDI in water. n.o.: not observed.**

| Monomer | Graft | | Link | | Blend | | Ring | |
|---------|-------|------|------|------|-------|------|------|------|
| Units | MW $g \cdot mol^{-1}$ | rel. mol-% | MW $g \cdot mol^{-1}$ | rel. mol-% | MW $g \cdot mol^{-1}$ | rel. mol-% | MW $g \cdot mol^{-1}$ | rel. mol-% |
| 1 | 218 | 23 | 319 | 11 | 117 | 6 | 143 | 26 |
| 2 | 360 | 16 | 461 | 2 | 259 | 5 | 285 | 3 |
| 3 | 502 | 1 | 603 | n.o. | 401 | 4 | 427 | 2 |
| 4 | 644 | n.o. | 745 | n.o. | 543 | 1 | 569 | n.o. |

In the reaction of HDI with glycine in aqueous medium (Table 4.2), interestingly crosslinked species (*i. e.* Gly-HDI-Gly and Gly-HDI₂-Gly) only accounted for 13 mol.-% of the total products, while the amount of grafted Gly-HDI$_x$ species was much higher (50 mol.-% altogether).

**Table 4.3. Expected and found m/z for the M+H⁺ specimens in the ESI MS analysis of the reaction of glycine with HDI in DMSO. n.o.: not observed.**

| Monomer | Graft | | Link | | Blend | | Ring | |
|---|---|---|---|---|---|---|---|---|
| Units | MW g·mol⁻¹ | rel. mol-% | MW g·mol⁻¹ | rel. mol-% | MW g·mol⁻¹ | rel. mol-% | MW g·mol⁻¹ | rel. mol-% |
| 1 | 218 | 23 | 319 | 11 | 117 | 6 | 143 | 26 |
| 2 | 360 | 15 | 461 | 2 | 259 | 5 | 285 | 3 |
| 3 | 502 | 1 | 603 | n.o. | 401 | 4 | 427 | 2 |
| 4 | 644 | n.o. | 745 | n.o. | 543 | 1 | 569 | n.o. |

**Table 4.4. Expected and found m/z for the M+H⁺ specimens in the ESI MS analysis of the reaction of glycine with LDI in water. n.o.: not observed.**

| Monomer | Graft | | Link | | Blend | | Ring | |
|---|---|---|---|---|---|---|---|---|
| Units | MW g·mol⁻¹ | rel. mol-% | MW g·mol⁻¹ | rel. mol-% | MW g·mol⁻¹ | rel. mol-% | MW g·mol⁻¹ | rel. mol-% |
| 1 | 276 | 1 | 377 | 2 | 175 | 30 | 201 | 13 |
| 2 | 476 | 3 | 577 | 1 | 375 | 29 | 401 | 12 |
| 3 | 676 | 1 | 777 | n.o. | 575 | 8 | 601 | 3 |
| 4 | 876 | n.o. | 977 | n.o. | 775 | n.o. | 801 | n.o. |

Additionally, some HDI$_x$ oligomers (16 mol.-%) and cyclic systems (31 mol.-%) were found. Unreacted crosslinker was not detected in the mass spectra. For the reaction with DMSO, basically the same product distribution was observed (Table 4.3). The test reaction of glycine with LDI in water (Table 4.4) gave a significantly different outcome, as grafting (5 mol.-%) and direct crosslinking (3 mol.-%) was much lower than for HDI, while LDI oligomers possibly resulting in blends in the products added up to 77 mol.-%, rings being quantified as 28 mol.-%.

Because of its prominence in the model reaction, simple grafting was investigated as to whether this alone could result in the stabilization of a gelatin scaffold. For this purpose, gelatin (10 wt.-%) was reacted with cyclohexylisocyanate (CHI, 8 NCO/Lys molar ratio), a monofunctionalized molecule, resulting in a material that readily dissolved in water during washing at 25 °C. Therefore, the physical interactions provided by simple grafting were evaluated to be not sufficient to achieve a stable network. Clearly, the crosslinking occurring in the stable gelatin-based films is relatively low but efficient for the formation of entropy-elastic networks. Additionally, grafted chains and linear and cyclic diisocyanate oligomers were also a prominent product of the reaction and might have additionally contributed to the stabilization of the networks. The different reaction products observed for HDI and LDI explain why LDI crosslinked films prepared from the same reagent ratios have a higher degree of swelling and lower Young's moduli than the corresponding HDI crosslinked films.

Ultimately, the solution pH during the reaction has an influence on the reactivity of the amino functions. The used gelatin type A has a pH in water of ca. 5.2 (rising to 5.7 during the reaction due to the decarboxylation of HDI). Some earlier publications report that the fastest and highest yielding reaction of gelatin with HDI takes place at pH 9.6, whereby a gel is formed after roughly five hours.[54] On the other hand, a fast reaction and the formation of a stable gel were here observed within minutes (~ 8 min, starting pH ~ 5.2). In the presented acidic reaction conditions, the fast gel formation might be related to an increased electrophilicity of the isocyanate groups by protonation. In view of the faster reaction kinetics with respect to previous poublications,[54] the reaction of gelatin with HDI was not investigated at other solution pHs.

# 4.4 Summary

A system of gelatin-based entropy-elastic networks was prepared. Model materials were prepared in DMSO to investigate the changing trends in mechanical properties and swelling (section 4.1). Tailorable properties were observed, such as the enhancement of mechanical properties (section 4.1.1) by increasing the crosslinker concentration, although the degree of swelling (section 4.1.2) was less affected by the sample composition.

In order to form hydrogels with adjustable properties that are applicable in Regenerative Therapies applications, the crosslinking reaction was transferred to water (section 4.2). The crosslinking reaction enabled the formation of films with tailorable mechanical ($E$: 141-738 kPa, $E_c$: 16-48 kPa, $\varepsilon_b$: 35-122 %, section 4.2.1) and swelling ($Q$: 300-800 vol.-%, section 4.2.2) properties. Furthermore, permanent fixation of the gelatin chains in the amorphous coil conformation was achieved (section 4.2.3), The swelling behavior was investigated in both water and cell culture medium, and while it was possible to tailor the swelling behavior, no clear preference was found for the water or cell culture medium uptake. Crosslinking with HDI resulted in much higher $E$ moduli (141-738 kPa) than reaction with LDI (69-230 kPa). This could be in principle related to a different yield of crosslinking density between LDI- and HDI-gelatin networks, also in view of the 2-fold slower gelation time of the formed compared to the latter hydrogels.

Test reactions allowed the elucidation of the crosslinking reaction (section 4.3). Only a comparably small number of direct covalent crosslinks stabilize the network (3-13 mol.-%), which is supported by the interaction of hydrophobic grafted and blended diisocyanate oligomers. However, grafting alone is not responsible for the network stability, given that the corresponding reaction with a monofunctional isocyanate yielded materials that readily dissolved in water.

Since ECM-derived materials mimic the biological tissues on a mechanical and molecular level, this gelatin-based hydrogel represents a novel biomimetic system for potential applications in Regenerative Medicine. By suppressing the gelatin chain helicity in the resulting network, defined structure-property relationships were established, regardless of the material thermal history. In this way, gelatin properties can be adjusted in a controlled manner to the complex requirements of specific clinical applications.

# 5. Design of a gelatin-based scaffold system

In this chapter, a strategy is described that translates the chemistry used to create two-dimensional films described in Chapter 4 into a method of preparing three-dimensional porous scaffolds. Porous scaffolds prepared using similar component mixture as the homogeneous hydrogels give an additional structured level that may serve as a valuable geometric construct for applications in Regenerative Medicine.

## 5.1 The integrated foaming-crosslinking process

The strategy used to make crosslinked, foamed gelatine-based scaffolds involves an integrated one-step process. Such integrated processes are known for crosslinked polyolefin foams[118] and are only recently being used for biomaterial-based scaffolds[119,120]. There are few reports in the literature that describe the preparation of gelatin-based scaffolds for bone regeneration; however, the synthetic methods were not integrated processes (i.e. crosslinking was performed following foam formation).[121] Although mechanical properties are increased, methods of crosslinking foams generally augment the material mechanical properties, but they do not ensure that the crosslinking occurs within the pore walls. Integrated foaming and crosslinking processes ensure that the crosslinking step is homogeneous, so the advantages of such a process are controllability, tailorability, and reproducibility.

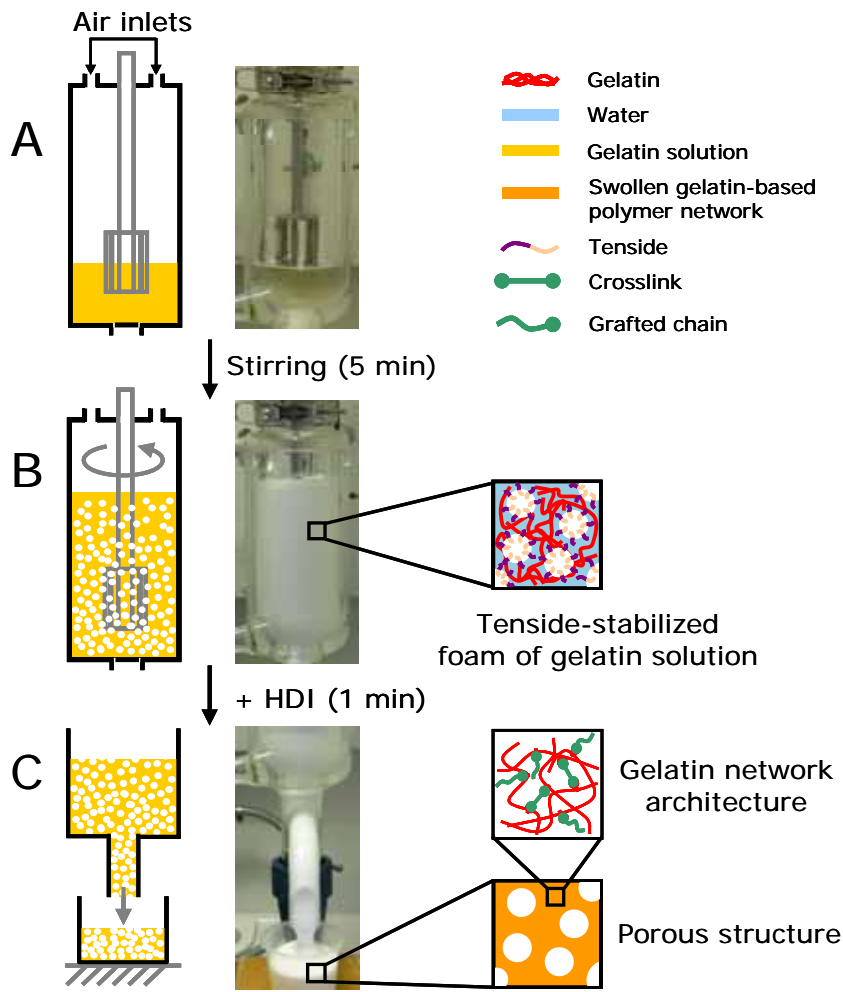

**Figure 5.1. Formation of gelatin-based scaffolds by an integrated process of foaming and crosslinking. A: an aqueous solution of gelatin is achieved at 45 °C. B: addition of saponin as surfactant and stirring led to the formation of a foam. C: stabilization of the scaffolds was achieved by reaction with HDI, resulting in a network of crosslinked and grafted gelatin chains.[13]**

The formation of porous scaffolds envisioned in this work involved the embodiment of an entropy-elastic gelatin hydrogel into a porous scaffold. Gelatin crosslinking above the gel-sol transition successfully enabled a systematic adjustment of material properties by only small changes at the molecular level. Following a similar approach, the goal became the design of structured rather than bulky homogeneous hydrogels. For this purpose, an integrated process was applied combining crosslinking reaction and formation of the porous shape: an aqueous gelatin solution ($T = 45$ °C) was intensively stirred in the presence of a surfactant and subsequently crosslinked using HDI (Figure 5.1). The chemical crosslinking was carried out after solution foaming, in order to ensure both the network formation and foam structure fixation in the resulting material. Therefore, the scaffold architecture resulted from the fixation of the gelatin foam structure. Additionally, the scaffold bulk phase was based on the synthesis of the covalent gelatin network following the crosslinking step. In this context, the bulk material should allow the control of the scaffold local mechanical properties. Test reactions under foaming confirmed the synthesis of a network consisting of both crosslinked and grafted chains, as observed for the synthesis of bulky hydrogels (section 4.3). On the other hand, the mechanical properties of the whole scaffold should also be influenced by the geometry of the porous system. Therefore, by keeping the geometry of the system constant, the influence of the bulk composition on the macromechanical scaffold properties was investigated.

The following sections describe the design of gelatin-based scaffolds and the preliminary characterization of the resulting material. The foam formation is presented with special emphasis on the surfactant parameters, i. e. the type (saponin vs. Polysorbate 20) and the concentration (0.5-3 g surfactant added to 100 g reaction solution) of surfactants. Likewise, the gelatin concentration is also proven to play a central role in the foam stabilization. Scaffolds were prepared with five different starting material compositions. In a first series, the gelatin concentration (7, 10, and 13 wt.-%) was varied in the reaction mixture, in a second series the molar excess of the HDI (3, 5, and 8 NCO/Lys molar ratio). The scaffolds are referred to as SGX_YNCOZ, where S identifies a scaffold (not a film) sample; X describes the gelatin concentration (wt.-%); Y refers to the crosslinker (H: HDI, L: LDI); Z indicates the molar excess of the crosslinker isocyanate groups with respect to gelatin amino functions of lysine.

# 5.2 Foam formation and stabilization

The addition of a surfactant was an important point for as well the foaming as the crosslinking reaction. Indeed, surfactants support the foaming of an aqueous solution by reducing the surface tension at the air-water interface, and therefore stabilize the air-bubbles introduced in the solution during stirring. At the same time, surfactant molecules can also form micelles in the bulk solution, which drive the miscibility of hydrophobic compounds in the aqueous phase. Therefore there were multiple reasons for incorporating a surfactant. The surfactant supported the foam formation and stabilization (1), promoted a homogeneous distribution of pores (2), and mediated the solubility of HDI in the aqueous phase during crosslinking (3).

Saponin or Polysorbate 20 was applied in varied concentrations for the foaming of an aqueous solution of gelatin. Saponin was selected since it is known to promote the formation of stable foams in water.[102] Additionally, it was employed for the synthesis of HDI-crosslinked gelatin hydrogels, and proved to mediate the solubility of HDI in the aqueous phase (chapter 4). The use of Polysorbate 20 was also investigated since it is widely applied for the foaming of aqueous solutions,[122] and for the formation of biocompatible films.[123] Therefore, the attention was first focused on the foaming of different aqueous-gelatin solutions containing different surfactants (without crosslinking). The distribution of the foamed with respect to the non-foamed phase was analyzed by measuring the respective volumes in the resulting foam. Second, several foams prepared with either saponin or Polysorbate 20 were crosslinked and investigated by SEM, to investigate how the resulting pore distribution was affected by the surfactant choice or concentration.

The stability of foams was first investigated on non-chemically crosslinked systems. In these conditions, the renaturation of gelatin chains in collagen-like triple helices should be responsible for the stabilization of the foam structure. Solutions with 7 wt.-% gelatin were analyzed as their corresponding foam included a relatively high amount of non-foamed phase. Therefore, the surfactant concentration was varied aiming at the formation of foams with minimal phase separation. Therefore, the resulting foamed solutions ($T = 45$ °C) were poured in a centrifuge tube and analyzed after equilibration with room temperature. By introducing 1 g saponin to 100 g gelatin solution, stable foams (only 4 vol.-% of non-foamed solution) were observed, while 20 vol.-% non-foamed solution was observed in the case of Polysorbate 20 (1 g Polysorbate 20 to 100 g of solution). The increase of the saponin amount (3 g per 100 g solution) obviously did not result in any detectable change of foam volume, while a reduction of the non-foamed solution from 20 to 15 vol.-% occurred for the same increase in Polysorbate 20 concentration. Likewise, lower concentrations of saponin resulted in increased non-foamed volume. For example, 5 vol.-% of non-foamed phase was measured when 0.5 g saponin was added to 100 g solution gelatin solution. In contrast, a clear non-foamed layer (> 20 vol.-%) was observed when the same conditions (0.5 g surfactant added to 100 g solution) were applied with Polysorbate 20-mediated foaming. These results indicated that saponin was superior to Polysorbate 20 in creating foams with minimal phase separation.

The stability of chemically crosslinked foams was also investigated. Therefore, HDI was added to the foaming solution, so that the resulting crosslinked product was poured in a centrifuge tube and analyzed. In the above-mentioned conditions (1 g saponin added to 100 g solution), aqueous solutions with less than 7 wt.-% gelatin resulted in at least 10 vol.-% of the non-foamed phase and were therefore not further considered. On the other hand, increasing the gelatin concentration from 7 to 10 wt.-% resulted in a decrease of the non-foamed solution from 4 to 1 vol.-% (saponin: 1 g surfactant to 100 g aqueous gelatin, HDI excess: 8 NCO/Lys molar ratio). Additionally, a sample of 13 wt.-% gelatin (8 NCO/Lys molar ratio) was prepared in order to study an extreme system with minimal (< 1 vol.-%) non-foamed solution. The above-mentioned observations suggested that (i) a chemically crosslinked network was successfully synthesized under foaming conditions; and that (ii) the presence of the chemical network enabled the formation of stable foams.

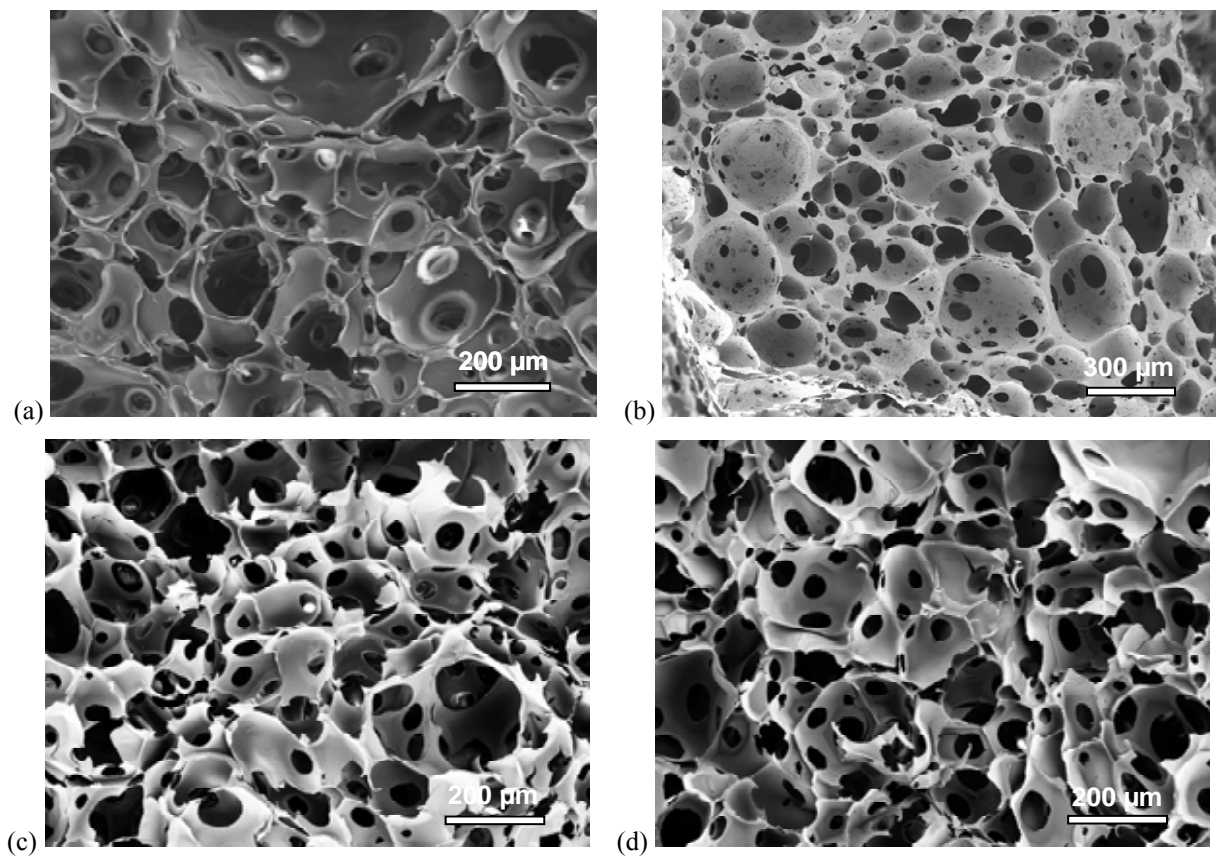

**Figure 5.2. SEM of gelatin-based scaffolds (SG10_HNCO8), synthesized by foaming and crosslinking of an aqueous solution (100 g) of gelatin. (a): 1 g saponin, (b): 1 g Polysorbate 20, (c): 2 g Saponin, and (d): 3 g saponin.**

In order to investigate the effect of the surfactant on the resulting scaffold architecture, 10 wt.-% gelatin-water solution was foamed with either saponin or Polysorbate 20. Once that crosslinking (8 NCO/Lys molar ratio) was completed, resulting scaffolds were analyzed by SEM. A porous architecture was observed in both samples, with a homogeneous distribution of pores. The pore size was only slightly different in foams obtained with Saponin (166 ± 32 μm) compared to foams obtained with Polysorbate 20 (210 ± 83 μm), as shown in Figure 5.2 a-b. Also, no difference of pore size and distribution was detected in scaffolds synthesized with different saponin concentration (Figure 5.2 a, c-d). Therefore, the choice of the surfactant was mainly based on its ability to mediate the crosslinking reaction as well as on the differences observed in the phase volume distributions.

The formation of air-water foams depends on the steady state stability of air bubbles within the solution, and is reversely related to the drainage of liquid between, and the coalescence and disproportionation of, air bubbles.[124,125] The hydrophilic/lipophilic balance (HLB) is one parameter to classify the surfactant foaming properties, and depends on the surfactant chemical structure.[126] Saponin from *Quillaja Saponaria* is a high molecular weight glycoside, with a sugar moiety linked to a triterpene aglycone.[102,127] Under the general assumption of a sugar chain containing 2-5 monosaccharide residues,[102] the saponin HLB[i] value was calculated to be approximately 9-13[128], while the one of Polysorbate 20 is in the range of 16-17.[129] Therefore, saponin is slightly more hydrophobic than Polysorbate 20. This suggests that, during crosslinking reaction, the miscibility of saponin with HDI should be enhanced compared to the miscibility of Polysorbate 20 with HDI. Therefore, a reduced surface tension should be expected at the air-water interface when employing saponin compared to Polysorbate 20. Consequently, foam scaffolds with higher foam stability should be formed in the case of foaming with saponin compared to foaming with Polysorbate 20. For these reasons, saponin was preferred to Polysorbate 20 as foaming agent during the synthesis of the gelatin-based scaffolds.

Besides the variation of surfactant type and concentration, the viscosity of the bulk solution also influences the foam stability, as it is directly related to the drainage of solution from the interior of the air pockets.[97] Correspondingly, the volume of non-foamed solution is directly related to the variation in solution viscosity.[97,130,131] Therefore, the decrease of non-foamed solution observed in foams with increased gelatin concentration can be explained by the resulting increased viscosity in the starting aqueous gelatin solution.

---

[i] HLB = $20 \times M_h/M$, $M_h$: molecular weight of the surfactant hydrophilic segment; $M$: molecular weight of the overall surfactant molecule.

# 5.3 Analysis of the porous structure

Once that stable foams were established with homogeneous pore distributions, crosslinked scaffolds were prepared with different gelatin concentrations and HDI excess. At this point, the aim was to control the system geometry over the whole set of sample compositions. By keeping constant the scaffold geometry, it was hypothesized that the material properties of the whole scaffolds could be tailored by the only variation of the network architecture. The five scaffolds were investigated by SEM and μCT in order to analyze the dry-state porous architectures. The SEM images of porous scaffolds at 50-, 100-, and 500-fold magnification are displayed in Figure 5.3. Besides open pores in the range of 117±28 − 166±32 μm, pore interconnections were also observed (32±11 − 50±12 μm).

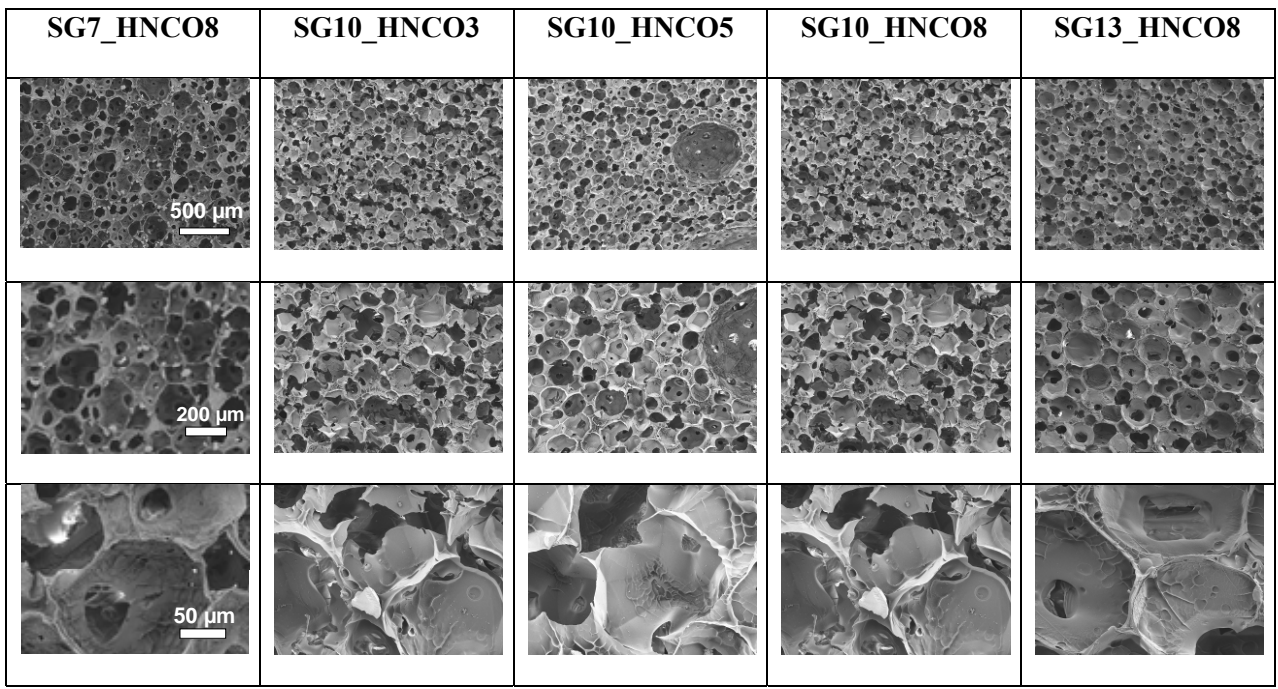

**Figure 5.3. Porous morphology of scaffolds after incubation in 70% aq. EtOH, as investigated by SEM. Pictures of the same raw were obtained with the same magnification and the same scale bar applies.**

The thickness of the pore walls was also measured and varied from 20±10 to 24±11 µm.[ii] Furthermore, a slightly lower pore size was found for samples disinfected in a 70 vol.-% ethanol solution, as for cell culture purposes (Table 5.1).

**Table 5.1. Investigation of the scaffold morphology by SEM and µCT. Samples were washed either in water or ethanol prior to the measurements.**[*]

| Method | SEM | | | | | µCT | | |
|---|---|---|---|---|---|---|---|---|
| *Medium* | H$_2$O | | EtOH | | | H$_2$O | | |
| Sample ID | $D_{dry}$ /µm | $D_{dry}$ /µm | $D_{dry}$ /µm | $D_{dry}$ /µm | $Wall_{dry}$ /µm | $D_{dry}$ /µm | $D_{dry}$ /µm | *Porosity* /vol.-% |
| SG7_HNCO8 | n. d.[*] | 46±14 | 150±35 | 45±13 | 22±11 | 260 | 50 | 73±14 |
| SG10_HNCO3 | n.d.[*] | 41±12 | 146±25 | 50±12 | 22±10 | 300 | 43 | 68±2 |
| SG10_HNCO5 | 156±29 | 32±11 | 127±27 | 48±16 | 20±10 | 350 | 48 | 65±11 |
| SG10_HNCO8 | 166±32 | 45±16 | 117±28 | 48±14 | 21±13 | 280 | 45 | 66±3 |
| SG13_HNCO8 | 148±43 | 40±16 | 130±38 | 49±18 | 24±11 | 280 | 48 | 67±6 |

[*] *Medium*: washing medium, either water (H$_2$O) or 70 vol.-% ethanol solution (EtOH); *n. d.*: not determined; $D_{dry}$: pore size (average ± standard deviation, n= 20-30 (SEM)); $d_{dry}$: interconnection size (average ± standard deviation, n= 23-40 (SEM)); $Wall_{dry}$: wall thickness (average ± standard deviation, n=30-57 (SEM)); porosity (average ± standard deviation, n= 2).

In order to further confirm the results obtained by SEM and evaluate the overall sample volume, the freeze-dried scaffolds were also analyzed by µCT. In contrast to SEM, the pore size detected by µCT was much larger, and was found to range from 260 to 350 µm (Table 5.1). The higher pore size observed by µCT compared to SEM is likely due to the fact that the pore size range detected with µCT depends on the sample-to-X-ray source distance, as shown in Figure 5.4, (left). On the other hand, the size range of pore interconnections corresponded well with the one observed by SEM (Figure 5.4, right). In view of these observations, only the measurements of pore size performed by SEM were further considered.

---

[ii] Measurements in the SEM are subject to methodological and statistical errors due to the sample preparation (drying, cutting, breakline, angle to the observer) and non-statistical choice of data points.

Other than pore and interconnection size, µCT analysis successfully enabled the determination of scaffold porosity. A narrow range of porosity (65±11– 73±14 vol.-%) was observed, with only a slight variation among the different scaffold compositions (Table 5.1).

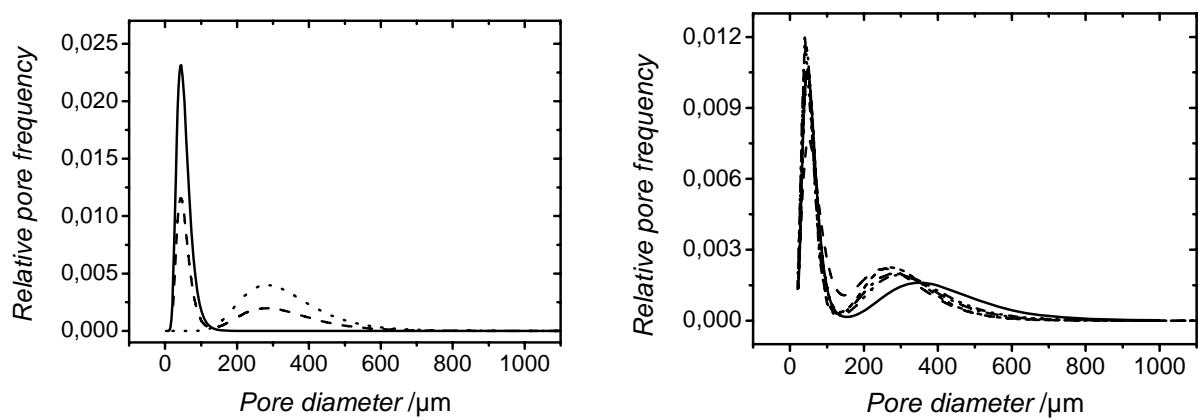

**Figure 5.4. µCT investigation in gelatin scaffolds. Left: pore distribution in sample SG10_HNCO8 as measured with a distance of the sample from the detector of either 21 mm (—) or 143 mm (····), and described as interpolation curve (---). Right: interpolation curves from µCT measurements of the whole set of samples: (–··–) SG10_HNCO8, (- - -) SG13_HNCO8, (– –) SG7_HNCO8, (····) SG10_HNCO3, (—) SG10_HNCO5.**

Finally, µCT and SEM showed negligible pore size differences in scaffolds prepared in different batches. This means that the air bubble distribution was homogeneous and reproducible during the foaming process. The pore interconnection size (SEM), the wall thicknesses (SEM), and the porosity (µCT) were insignificantly different for the tested sample compositions. Importantly, the results of SEM and µCT confirmed the presence of open pores in all sample compositions, which is necessary in situation *in vivo* for promoting cell adhesion and proliferation within the scaffold. For these reasons, the use of an integrated one-step crosslinking foam-formation method of scaffold formation was successful.

# 5.4 Gelatin chain helicity investigated by WAXS

WAXS was carried out on the freeze-dried scaffolds to elucidate details pertaining to the inherent molecular architecture of the scaffold. Two main questions were addressed: at first, it was essential to understand whether the presence of a defined chemical network could again hinder the gelatin chain helicity, leading to an amorphous scaffold material, as observed in the case of the corresponding films. At second, it was investigated whether the material structural organization was permanently defined by the presence of chemical crosslinks. If this was the case, material properties could be varied, regardless of environmental factors *i. e.* sample thermal history or conditioning, which, in contrast, impose serious reproducibility concerns in the case of non-functionalized gelatin.

Figure 5.5 (left) depicts the WAXS spectra of freeze-dried gelatin scaffolds, whereby samples crosslinked with varying HDI excess (3 – 8 NCO/Lys molar ratio) are displayed. As for the gelatin films, the peak describing the amorphous region (20°) was increased, while the two other gelatin peaks (8° and 31°) decreased, according to increasing crosslinker content during the synthesis. In principle, this observation would lead to the conclusion that reaction of gelatin with HDI at 45 °C enabled the fixation of gelatin coil chains, similarly to what was achieved for the crosslinked gelatin films. Films and the scaffolds were treated differently after the synthesis (thermal history), so that the observed chain organization could be related to either a kinetic or thermodynamic effect.[65,132] Indeed, gelatin films were formed by mild drying at room temperature, so that gelatin chains could re-equilibrate over time into the more favorable thermodynamic state. In contrast to that, freeze-drying was applied to the foam scaffolds, involving a -30 °C cooling after the synthesis. In such extreme conditions, coil-to-helix transition of gelatin chains could be hindered.[41]

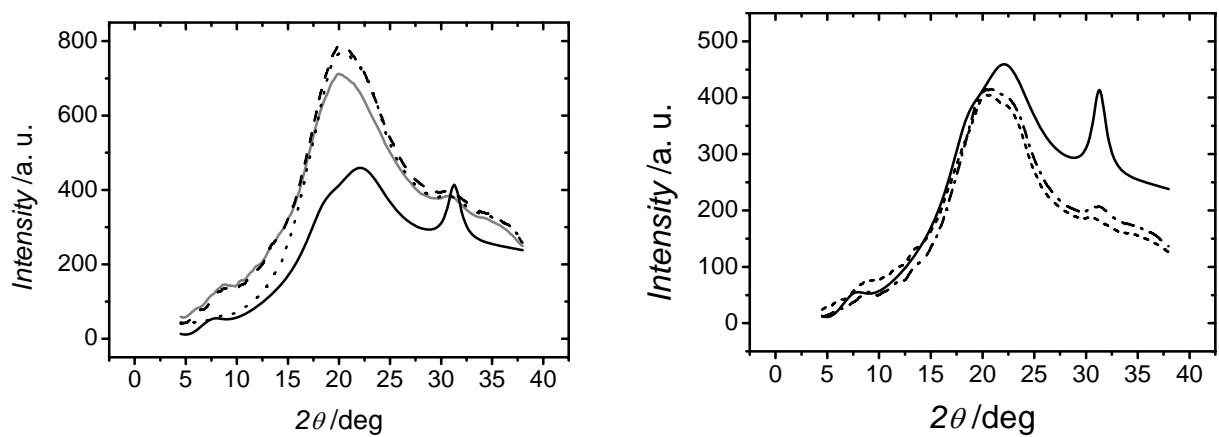

**Figure 5.5. WAXS curves of gelatin scaffolds. (Left): Effect of HDI content: (—): SG10_HNCO3, (——): SG10_HNCO5, (···): SG10_HNCO8, (—): gelatin. (Right): Effect of thermal conditioning: (—·—): SG13_HNCO8 with no thermal conditioning, (——): SG13_HNCO8 with thermal conditioning at 25 °C for 20 days (—): gelatin.**

To identify the reason for the suppressed content of either single- or triple-helices, equilibration with water was applied to a scaffold sample (SG13_HNCO8, 20 days, 25 °C), for potential helical renaturation purposes. Figure 5.5 (right) compares the WAXS curve of a freeze-dried sample after water incubation with the one of a freshly synthesized sample. Only marginal difference between the two spectra can be observed, meaning that both single- and triple-helices were hardly detectable in both samples. The above-described observations highlight that the introduction of chemical crosslinks and grafted side chains was responsible for the permanent suppression of gelatin helices, irrespective of post-synthesis sample thermal history or conditioning. These results suggest that the synthetic route described in this work could allow for defined network structural conformation, on one hand, and reliable network processing, *i. e.* porous scaffolds or bulky films, on the other hand. Those two points are generally considered a challenging task in case of biopolymer-based biomaterials.[133]

# 5.5 Thermal properties determined by TGA and DSC

TGA and DSC analyses were performed on freeze-dried samples in order to characterize the scaffold thermal properties. TGA was performed in a $N_2$ atmosphere in order to identify any thermal event, potentially imputed to either foaming or crosslinking reaction. The TGA plot of pure gelatin normally displays three thermal stages, i. e. loss of water, between 25 and 100 °C, gelatin decomposition, between 250 and 450 °C, and finally combustion of the remaining material, between 450 and 750 °C.[77,134]

The TGA plot of crosslinked gelatin scaffold (SG13_HNCO8) was compared with the one of pure gelatin in a temperature range of 25 – 700 °C (Figure 5.6, left). Thermal phenomena such as water loss, material decomposition and combustion were exhibited by both samples. The gelatin scaffold displayed a 4 wt.-% mass loss, even after 4 days drying, against a 1 wt.-% mass loss of untreated gelatin. At the same time, no additional thermal event was observed in the crosslinked gelatin scaffold compared to untreated gelatin.

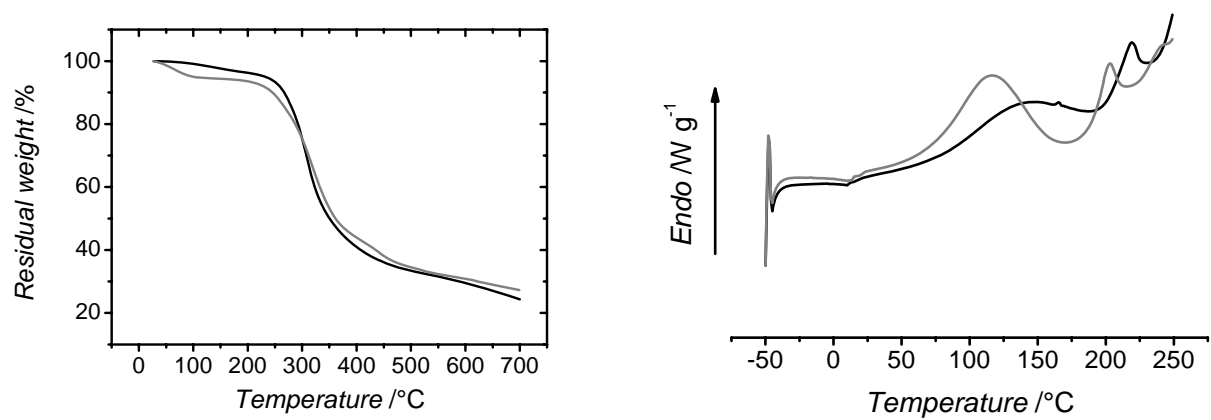

**Figure 5.6. TGA (left) and DSC second heating run (tight) thermograms of gelatin (black) and sample SG13_HNCO8 (gray) in the dry state.**

The above-mentioned results indicate that the presence of the chemical network likely impedes the removal of water from the scaffold. Also, neglectable presence of by-products, e. g. monomers or blends,[77] was observed in the TGA thermograms.

In addition to TGA, DSC analysis was carried out in order to identify any network thermal transition. The thermal characterization of biopolymers, including gelatin, is not relatively straightforward. The DSC spectrum of non-crosslinked gelatin generally displays an endothermic peak associated to the helix-coil transition, which can greatly vary based on the gelatin source and the material thermal history. Therefore, no unique value is provided in the literature.[64,135,136,137] Figure 5.6 (right) described the dry-state DSC curves (2° heating run) of both non-crosslinked and crosslinked gelatin. Two main endothermic peaks are found in the range of 110 – 150 °C and 200 – 220°C. Previous reports that include TGA analysis of gelatin have shown that water loss and gelatin decomposition occurs near these temperature ranges.[81] Determining whether the observed endothermic peak at around 100 °C was related to a helix-coil transition (which was unlikely based on WAXS) or related to the water present in the material was difficult, especially given the high water content (~ 4 wt.-%) present in the material. In order to distinguish these interpretations, DMTA temperature sweeps were carried out, since DMTA is often a much more sensitive method compared to DSC in measuring the thermal transitions of networks. These results will be discussed in section 5.6.1. In the next section, preliminary compression tests on freeze-dried scaffolds will be discussed together with the DMTA compression measurements, in order to have an overview of the thermo-mechanical scaffold properties at room temperature.

# 5.6 Dry-state mechanical properties

In perspective of Regenerative Medicine applications, the scaffold should serve as a physical support, i. e. withstanding the complex loading conditions in the body, conveying the physiological loads to the surrounding tissue, and preserving the porous architecture integrity during or after compression. These three conditions are crucial for a novel biomaterial-based concept for tissue regeneration. In order to address this point, compression tests (50% compression) were carried out on freeze-dried gelatin scaffolds. SEM was performed on compressed samples, in order to elucidate if mechanical load could irreversibly alter the scaffold architecture. Moreover, DMTA was conducted in order to elucidate the physical state of the gelatin scaffold during the compression test with respect to the $T_g$.

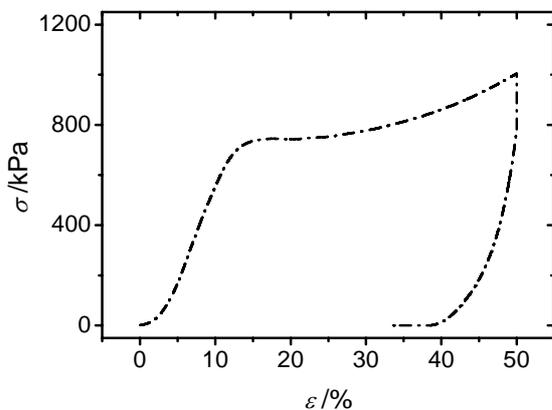

**Figure 5.7. Compressive stress-strain plot of sample SG13_HNCO8 depicting an initial elastic regime, an elastic plateau transition, followed by an ultimate increase of stress during compression.**

Figure 5.7 depicts a representative compression stress-strain plot of the scaffold SG13_HNCO8. Three distinct regions can be observed, namely an elastic regime ($E_c$: 8400 kPa), an elastic transition plateau ($\varepsilon_{el}^{*}$: 10-30 %), and an ultimate region of further increase of compression stress ($\sigma_{max}$: 800-1000 kPa). The collapse plateau and the ultimate region of compression stress regimes have been reported to be caused by pore wall buckling (plateau) and irreversible collapse of pores (densification), respectively, in the case of collagen-glycosaminoglycan scaffolds.[93] Each of these regions was therefore considered for the evaluation of the presented scaffold properties.

The elastic modulus ($E_c$) was determined in the elastic region, while the elastic compression stress ($\sigma_{el}^*$), the elastic compression strain ($\varepsilon_{el}^*$),[iii] and the stress-strain slope ($\Delta\sigma/\Delta\varepsilon$) were calculated in the elastic transition plateau. Finally, the ultimate compression stress ($\sigma_{max}$) was also taken into account for the last compression stage ($\varepsilon$ = 50 %). Table 5.2 gives an overview of the scaffold mechanical properties. Samples displayed a nearly constant $E_c$ (2110 - 2535 kPa) with varied crosslinker content (3-8 NCO/Lys molar excess, 10 wt.-% gelatin), while $\sigma_{max}$ was ranged from 215 to 395 kPa. On the other hand, a clear variation of mechanical properties ($E_c$: 1720 - 8390 kPa; $\sigma_{max}$: 175 - 955 kPa; $\sigma_{el}^*$: 110 - 686 kPa, $\Delta\sigma/\Delta\varepsilon$: 131-558 kPa) was observed in scaffolds with increased gelatin concentration (7 - 13 wt.-%, 8 NCO/Lys molar excess). Despite that, only a slight change of $\varepsilon_{el}^*$ (9-14 %) was observed among the different scaffold compositions.

**Table 5.2. Mechanical properties of gelatin-based scaffolds in the dry state. Values describe the median value and the range of data (four replicas were tested for each sample composition).** [*]

| Sample ID | $E_c$ /kPa | $\sigma_{el}^*$ /kPa | $\varepsilon_{el}^*$ /% | $\Delta\sigma/\Delta\varepsilon$ /kPa | $\sigma_{max}$ /kPa |
|---|---|---|---|---|---|
| SG7_HNCO8 | 1720 (641-1812) | 110 (92-150) | 10 (3-5) | 131 (68-154) | 175 (150-210) |
| SG10_HNCO3 | 2110 (1257-3653) | 155 (140-190) | 9 (8-13) | 123 (94-129) | 260 (230-340) |
| SG10_HNCO5 | 2568 (1432-3120) | 204 (182-258) | 12 (10-17) | 180 (131-196) | 315 (250-360) |
| SG10_HNCO8 | 2535 (2150-3396) | 256 (244-346) | 14 (12-15) | 254 (150-320) | 395 (360-510) |
| SG13_HNCO8 | 8390 (7230-15734) | 686 (650-699) | 10 (6-12) | 558 (468-769) | 955 (910-1030) |

[*] $E_c$: compressive elastic modulus, $\sigma_{max}$: stress at 50% compression, $\Delta\sigma/\Delta\varepsilon$: slope of the plateau region, $\sigma_{el}^*$: compressive plateau stress, $\varepsilon_{el}^*$: compressive plateau strain.

The above-described results supported the original hypothesis that the overall scaffold properties can be varied by changing the network architecture (variation of HDI excess proved to directly influence the network crosslinking density, section 4.2), while keeping

---

[iii] $\sigma_{el}^*$ and $\varepsilon_{el}^*$ were calculated from the intersection of regression curves describing $E_c$ and $\Delta\sigma/\Delta\varepsilon$.

constant the scaffold geometry (only slight variation of pore and interconnection size, porosity and wall thickness, based on SEM and µCT). Compared to presented scaffolds, other porous materials based on the synthesis of a polymer urethane network could not show any adjustment of mechanical properties.[138]

## 5.6.1 Shape recovery of freeze-dried gelatin scaffold

The shape recovery (SR) of freeze-dried gelatin scaffold was determined after mechanical load removal, according to the eq. 5.1:

$$SR = \frac{h_r - h_c}{h_0 - h_c} \times 100 \qquad \text{(eq. 5.1)}$$

where $h_0$ describes the height of the sample before compression, $h_c$ identifies the height of the compressed sample, while $h_r$ corresponds to the height of the sample after removal of loads. Scaffolds did not recover after 50% compression, (SR: 15%), meaning that an elastic transmission of physiological loads could not be achieved between the scaffold and the surrounding tissues *in vivo*. In order to explain this finding, both the scaffold network morphology and the porous architecture were identified as possible reasons. Consequently, the minimal shape recovery could be related to either a glassy- rather than rubbery-like state of the material at the compression temperature, i. e. 25 °C, or to the destruction of the porous scaffold architecture after 50% compression.[93,109]

SEM and DMTA were therefore carried out to identify the reason for the minimal shape recovery of dry compressed scaffolds. SEM investigation on a compressed sample (Figure 5.8, left) showed that the porous structure was not yet entirely compromised. Pores could still be observed after 50% compression, so that the porous architecture alteration was not likely to explain the above-mentioned observation. In contrast, the DMTA thermogram clearly displays a distinct morphological transition with a *tanδ* maximum at 45 ºC (Figure 5.8, right). This peak mostly corresponded to a glass transition temperature ($T_g$), as no melting transitions could be observed on the DMTA thermogram in the temperature range -100–100 °C. This hypothesis is also in agreement with the WAXS spectra, which displayed a permanent suppression of triple helix renaturation (Figure 5.5). The glass transition spanned roughly 70 ºC (0-70 ºC) and caused approximately a 4 MPa decrease in storage modulus ($E''$: 8-12 MPa). Based on the broad glass transition, the material is expected to display a glassy state at room temperature. This is likely the reason why the scaffolds only displayed a limited shape recovery (SR: 0-14%) after compression at 25 °C, resulting in a cold deformation of the material. In contrast to that, complete shape recovery (SR> 95%) was observed by the material when the compression was carried out above the glass transition temperature ($T > 70$ °C). This is likely due to the corresponding rubbery-like state of the material at this temperature. On the other hand, heating of dry compressed scaffolds did not lead to unfolding of the scaffold. This might be caused by kinks in the walls introduced during compression and that cannot be recovered by applying thermal stimulus. Such mechanical flaws could be reversed using water, which overcame the energy barrier for elasticity due to its plasticization effect.

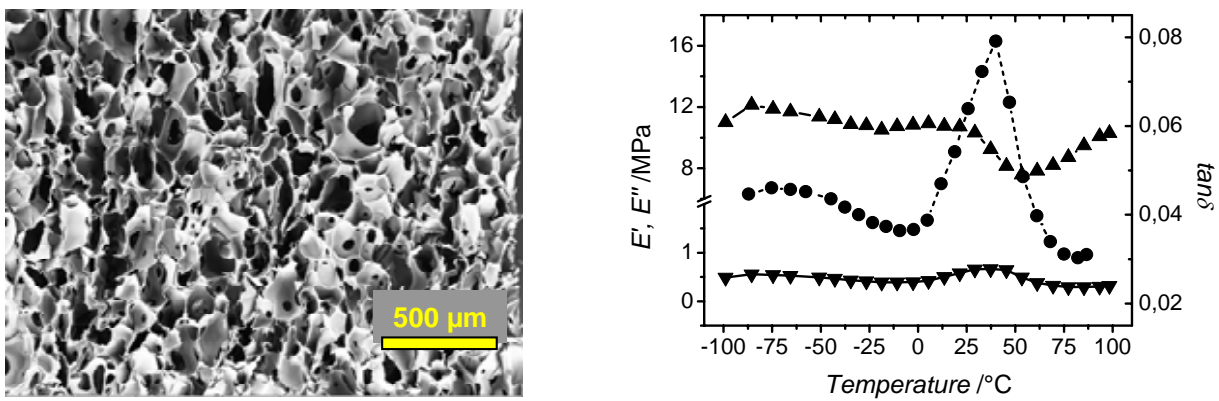

**Figure 5.8. SEM investigation after compression (Left) and dry-state Dynamic Mechanical Analysis at varied temperature (DMTA, right) on sample SG13_HNCO8. *E'*: storage modulus (—▲—), *E''*: loss modulus (—▼—), *tanδ*: loss angle (–●–). Pores are still opened after 50% compression of the scaffold at 25 °C, though the material is still in glassy state, as displayed by a maximum of *tan* δ at 45 °C.**

In view of the morphological observations on freeze-dried scaffolds, chapter 6 will present the material characterization in aqueous environments, in order to have a more closely description of material properties in *in vivo*-like conditions. Therefore, scaffolds were equilibrated in water and macro-compression tests as well as additional wet-state DMTA analyses were performed (section 6.4-6.5). This was also complemented by AFM investigation in order to probe the scaffold mechanics of the pore wall in the wet-state.

# 5.7 Summary

An integrated foaming-crosslinking procedure was developed for the formation of gelatin scaffolds, resulting in the embodiment of the gelatin network into a defined porous shape (section 5.1). A gelatin aqueous solution was foamed with different surfactant type and concentration, enabling the formation of stable foams (> 96 vol.-% foamed volume). The choice of surfactant was based on the extent of foam stability. Indeed, both saponin and Polysorbate 20 led to scaffold with insignificantly different pore size and distribution (section 5.2).

In order to design porous biomaterials with adjustable properties, gelatin was crosslinked with HDI under foaming, aiming at the fixation of the foam by chemical crosslinking (section 5.3) The formulations used to prepare gelatin-based films (chapter 4) were thereby translated into three-dimensional scaffold structures: the HDI excess (3-8 NCO/Lys molar ratio) and the gelatin concentration (7-13 wt.-%) were varied during the synthesis, in order to achieve different bulk material properties. Despite that, the scaffold geometry was kept constant regardless of the sample composition. The scaffolds showed open pores ($D_{dry}$: 117±28–166±32 µm) and pore interconnections ($d_{dry}$: 32±11–50±12 µm), which are crucial to ensure cell growth within the implant. The wall thickness ($Wall_{dry}$: 20±10–24±11 µm) was measured during the SEM investigations and was slightly varied among the different scaffolds.

Once the scaffold geometry was defined and controlled, the analysis on the scaffold network followed (section 5.4). WAXS on the freeze-dried samples confirmed the presence of an amorphous network at the molecular level. Furthermore, sample equilibration in water did not affect the structural organization of gelatin, so that the chain helicity was at least partially suppressed.

Initial thermo-mechanical characterization was carried out on freeze-dried scaffolds. A multistep mechanical response was observed during compression, though the material morphology was glassy at room temperature in the dry state (section 5.5-5.6). The porous architecture was still maintained after 50% compression at room temperature (glassy network morphology). Most importantly, full shape recovery (SR> 95%) was observed when compression was carried out above the glass transition temperature ($T > 70$ °C).

# 6. Characterization of gelatin-based scaffolds in aqueous environments

This chapter deals with the characterization of wet gelatin scaffolds in physiological conditions. Therefore, the form-stability, shape recoverability, macro- and micromechanical characterization, and the degradation behavior will be described.

## 6.1 Swelling behaviour of gelatin-based scaffolds in water

Gelatin scaffolds of known weight were immersed in water and the resulting water uptake ($H$, eq. 4.1) was determined as weight/weight ratio until reaching equilibrium with water. Figure 6.1 depicts the variation of $H$ in scaffolds differing in either HDI excess (left) or gelatin concentration (right). An increase of weight was observed in every sample tested, and $H$ was varied in the range of 630-1680 wt.-% based on the scaffold composition. Samples crosslinked with 3 NCO/Lys molar ratio displayed 1560 wt.-% of $H$, while samples crosslinked with 8 NCO/Lys molar ratio could absorb water for roughly 940 wt.-% of the sample weight. A decrease of $H$ was also observed in scaffolds with increased gelatin concentration, though the constant (8 NCO/Lys molar ratio) HDI excess during crosslinking.

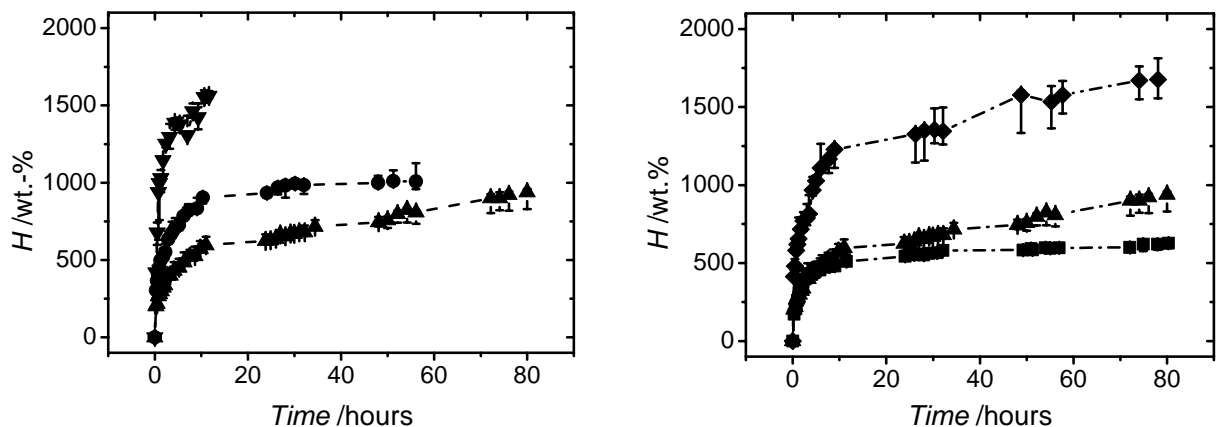

**Figure 6.1. Water uptake ($H$) of scaffolds with different HDI excess (left) and gelatin concentration (right) at 25 °C: (—▼—) SG10_HNCO3, (—●—) SG10_HNCO5, (—▲—) SG10_HNCO8, (—♦—) SG7_HNCO8, (—■—) SG13_HNCO8. Data are expressed in median (n= 3), while the error bars represent the range of data.**

Despite such a high uptake, wet scaffolds showed no change of the outer dimensions, i. e. neither expansion nor shrinkage, in contrast to the homogeneous gelatin films which displayed volumetric swelling. In comparison, clinically-applied collagen sponges were not form-stable in water at room temperature (41 vol.-% reduction upon contact with water and material disintegration due to the mechanical stress).

In order to understand the mechanism of water uptake and form-stability in the wet gelatin scaffolds, the effect of the porous structure was initially excluded. For this purpose, scaffolds were cryogenically milled and swelling studies were performed on homogeneous powdered scaffolds (Appendix, section 9.7.1). The volumetric swelling of powdered scaffolds was larger ($Q$: 1400-1850 vol.-%, eq. 4.2) than in the case of intact samples. An increase of swelling was observed with increasing the gelatin concentration as well as the HDI excess (Figure 9.1, Appendix), which was an opposite observation compared to the results of water uptake. In principle, this observation could be explained since the samples were frozen and mechanically stimulated during cryogenic treatment, so that an alteration of the chemical network is likely. Due the limitation of the above-mentioned observations, Confocal Laser Scanning Microscopy (CLSM) was employed on samples equilibrated in aqueous solutions. With this approach, the swelling was analyzed by considering the change of pore size and wall thickness compared to the dry state. As the porous scaffolds were characterized by a large surface/volume ratio, it was hypothesized that the swelling might occur in the direction of the pores, thereby explaining the minor changes of the scaffold outer dimensions upon water uptake.

# 6.2 Form-stability of gelatin-based scaffolds

Figure 6.2 depicts the light microscopy image of scaffolds wetted in Phosphate Buffer Solution (PBS) showing the intact pores. Here, an overall reduced pore size, i. e. 115±47–130±49 µm (9-22 % pore size reduction),[iv] was observed compared to the dry state. Furthermore, samples showed distinctively lower pore sizes (63±28–89±35 µm) when treated in a 70 vol.-% aq. ethanol solution (EtOH) for disinfection purposes. This decrease in pore volume supported the hypothesis that pore wall swelled inwards during the water uptake, thereby explaining the form-stability of the wet scaffolds compared to the dry samples. By knowing the dry-state wall thickness ($Wall_{dry}$, SEM) as well as the dry ($D_{dry}$, SEM) and wet ($D_{wet}$, CLSM) pore size, the wet wall thickness ($Wall_{wet}$) was calculated according to eq. 6.1, assuming the pores to be spherical and neglecting the presence of pore interconnections:

$$Wall_{wet} = D_{dry} + Wall_{dry} - D_{wet}$$ (eq. 6.1)

Thus, $Wall_{wet}$ was calculated to be in the range of 31±39–57±59 µm (Table 6.1). Consequently, the Wall swelling factor was calculated using eq. 6.2:

---

[iv] Measurements with the confocal light scanning microscope (CLSM) are subject to the non-statistical choice of data points.

*Wall Swelling Factor* $= Wall_{wet} / Wall_{dry}$ **(eq. 6.2)**

**Table 6.1: Pore sizes (*D*) and wall thicknesses (*Wall*) of dry and wet scaffolds, and the wall swelling factor.**

| Sample | $D_{dry}$ (µm) | $Wall_{dry}$ (µm) | $D_{wet}$ (µm) | $Wall_{wet}$ (µm) | Wall swelling factor |
|---|---|---|---|---|---|
| SG7_HNCO8 | 150±35 | 22±11 | 128±40 | 44±54 | 2±2.7 |
| SG10_HNCO3 | 146±25 | 22±10 | 129±45 | 39±52 | 1.8±2.5 |
| SG10_HNCO5 | 127±27 | 20±10 | 116±27 | 31±39 | 1.6±2.1 |
| SG10_HNCO8 | 166±32 | 21±13 | 130±49 | 57±59 | 2.7±3.3 |
| SG13_HNCO8 | 130±38 | 24±11 | 115±47 | 39±61 | 1.6±2.7 |

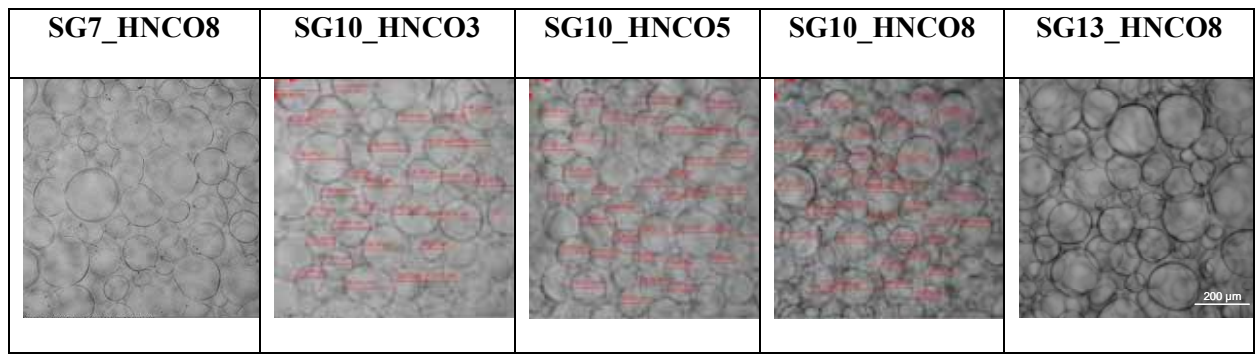

**Figure 6.2. Confocal laser scanning microscopy (CLSM) pictures on scaffolds equilibrated in PBS at 25 °C.**

The pore sizes in the dry and wet state were thereby used to calculate the wall swelling factor. The results range from 1.6±2.1 to 2.7±3.3 µm, so that the values have large errors and no difference was found between different sample compositions. In contrast, corresponding films showed decreasing degree of swelling with increased crosslinker excess or gelatin concentration (section 4.2.3), whereby the absolute swelling factors were much larger than observed for the scaffold walls.

The calculation of such low swelling factors, with respect to the swelling of either films or powdered scaffolds, implied a closer investigation on the mechanism of scaffold form-stability. The low swelling wall factor might imply higher crosslinking densities in the scaffolds than in the corresponding films. However, water uptake was higher in the scaffolds than in the films, contradicting the previous hypothesis. The presence of small pores (i.e. pore size < 100 nm) could also potentially explain the surprisingly high water uptake in the form stable materials. Again, these were not observed in high-resolution SEM. On the other hand, the scaffolds showed composition-dependent densities. For example, SG7_HNCO8 had a scaffold density of 0.038 g·cm$^{-3}$, whereas SG13_HNCO8 had a density of 0.089 g·cm$^{-3}$, which explained why SG7_HNCO8 took up significantly more water ($H$: 1680 wt-%) than SG13_HNCO8 ($H$: 630 wt-%). Hence, SG7_HNCO8 accommodated more water molecules and therefore possessed a higher free volume in the scaffold walls, compared to SG13_HNCO8. This could be related to the fact that SG7_HNCO8 was synthesized from a lower concentrated solution with a lower viscosity.

In this context, the concept of bound and free water is crucial. This refers to water molecules either tightly bound to certain chemical functionalities or atom groups, e.g. by hydrogen bonding, or freely diffusing in the bulk. The bound water molecules generally are thought to require less space, and the gelatin chains can tightly bind water molecules to a high extent.[139] The free volume in the scaffold walls would be directly affected from the solution concentration in the synthetic step, thereby explaining the different water uptake between scaffolds. The free volume in the scaffold walls might also be affected by the freeze-drying step. More likely, the additionally freeze-drying step in the synthesis of scaffolds compared to the films might have resulted in free volume in the scaffold walls, which was concentration-dependent. In this context, the physical crosslinking by oligourea side chains might play an important role. The large free volume could also be a reason for the relatively low porosity of the scaffolds.

# 6.2.1 Water uptake in cell culture medium

As for application in regenerative medicine, the water uptake (eq. 4.1) of freeze-dried scaffolds (SG7_HNCO8 and SG13_HNCO8) was investigated in PBS solutions containing different salt concentrations, and in minimal essential medium (MEM) solution ($T = 37\ °C$). The effect of each MEM component, i. e. Earle's salts, glucose, and amino acids, on the scaffold uptake was considered.

In the current investigation, other methods, besides paper blotting, were considered in order to minimize the contribution of free, non-bounded swelling medium in the uptake. Indeed, in the case of samples blotted with paper, the free water potentially entrapped in the pores could not be removed completely, thereby affecting the determination of the (bounded) water uptake. At the same time, the cryogenic mill treatment was applied to investigate the swelling of the only bulk phase, by powdering of the scaffolds. However, samples were frozen and mechanically stimulated during the cryogenic mill treatment, so that alteration of the bulk material properties was likely after this process.

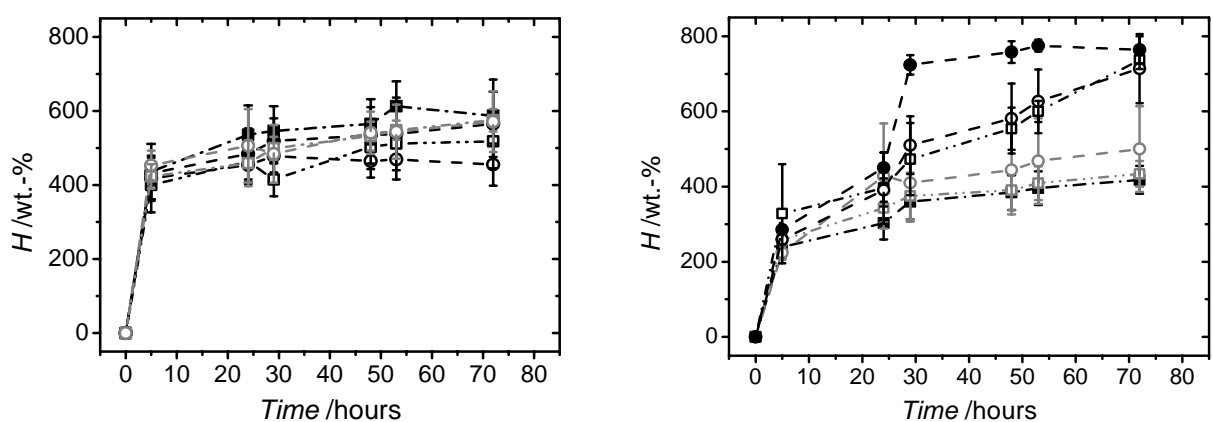

**Figure 6.3. Water Uptake (*H*) of different aqueous solutions in scaffold SG7_HNCO8 (left) and scaffold SG13_HNCO8 (right) at 37 °C.** [*] **(—■—) Water, (··—□—··) PBS, (—○—) PBS (2X), (··—□—··) Earle's salt aqueous solution, (—○—) Earle's salt aqueous solution containing 1 g/L glucose, —●— Minimum Essential Medium (MEM) solution. Data are expressed in average ± standard deviation (n=5).** [*] **Before the weight measurements, samples were placed on a frit and free water was removed by application of vacuum.**

For these reasons, removal of the free water from the porous scaffolds was achieved by placing the wet samples on a frit, under the application of vacuum. Retrieved samples were then weighed and the water uptake in the different medium was calculated by knowing the mass of wet and dry samples. As a result, much lower uptake values (*H*: 400 wt.-% (SG13_HNCO8), 600 wt.-% (SG7_HNCO8)) were obtained compared to the water uptake values obtained by the paper blotting method. This could potentially suggest that not only the free water but also the bounded water was removed upon sample treatment with vacuum. Taking the limitations of the method into account, Figure 6.3 describes the results of uptake in the different mediums. On the one hand, sample SG7_HNCO8 displayed a nearly constant water uptake in the different media (Figure 6.3, left). On the other hand, sample SG13_HNCO8 displayed a higher water uptake in solutions containing Earle's salts, glucose and amino acids, with respect to water (Figure 6.3, right). In these conditions, the water uptake of SG13_HNCO8 was observed to be higher than the uptake of sample SG7_HNCO8. When comparing this to the previous results of water uptake, a somewhat opposite tendency was observed in the composition-dependent variation of water uptake. However, different methods were applied for the medium removal in the above-mentioned tests. Likely, the contribution of the free aqueous solution should be minimal or lower in the current investigation than the one achieved by the paper blotting method, since free, non-bounded medium was here removed by application of vacuum. Following this line of thinking, the difference in the water uptake between samples SG7_HNCO8 and SG13_HNCO8 suggests that the different network architecture in the two samples should play a major role. Considering the occurrence of side reactions due to the competitive reaction of isocyanate with water, a different reaction yield of grafting compared to crosslinking could be identified as the reason for the above-mentioned findings. In case sample SG7_NCO8 was characterized by a higher yield of crosslinking compared to grafting, then only a slight change of uptake would be expected in the case ions/sugars were present in the medium.[140] This hypothesis was also identified to explain the hydrolytic degradation kinetics of the same samples (section 6.7).

Besides the measurements of $H$, the change of scaffold outer dimensions was measured during uptake tests. Scaffolds SG7_HNCO8 and SG13_HNCO8 exhibited no immediate volume reduction under these conditions, in contrast to the collagen control, displaying an immediate 80 vol. % reduction (Figure 6.4).[141] Therefore, the developed gelatin scaffolds could potentially ensure a close contact between the scaffold and the surrounding tissue in case of application *in vivo*.

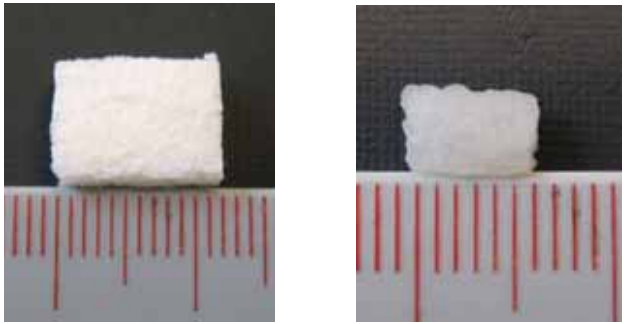

**Figure 6.4. Collagen sponge (Lyostypt[TM]) in the dry (left) and wet state at 37 °C (right). Immediately after contact with water, a shrinking of approximately 80 vol.% compared to the initial volume was observed. In comparison, a gelatin-based scaffold (SG13_HNCO8) sample displayed only 20 vol.-% of volume reduction after equilibrium with water (3 days).**

# 6.3 Macroscopic elasticity

Once the mechanism of form-stability was understood, the macroscopic elasticity of wet scaffolds was investigated, and compression tests were carried out directly after scaffold formation. Figure 6.5 (left) depicts the stress-strain plots of three swollen samples (SG7_HNCO8, SG10_HNCO8, and SG13_HNCO8) compressed at room temperature. In all curves, no elastic transition plateau can be observed, which was typical of the dry-state compression curve. Most importantly, full shape recovery (SR> 95%) was achieved after mechanical load removal. In contrast, dry samples maintained a compressed shape after removal of mechanical loads, due to the partial glassy network state at the compression temperature ($T_g \sim 45$ °C).

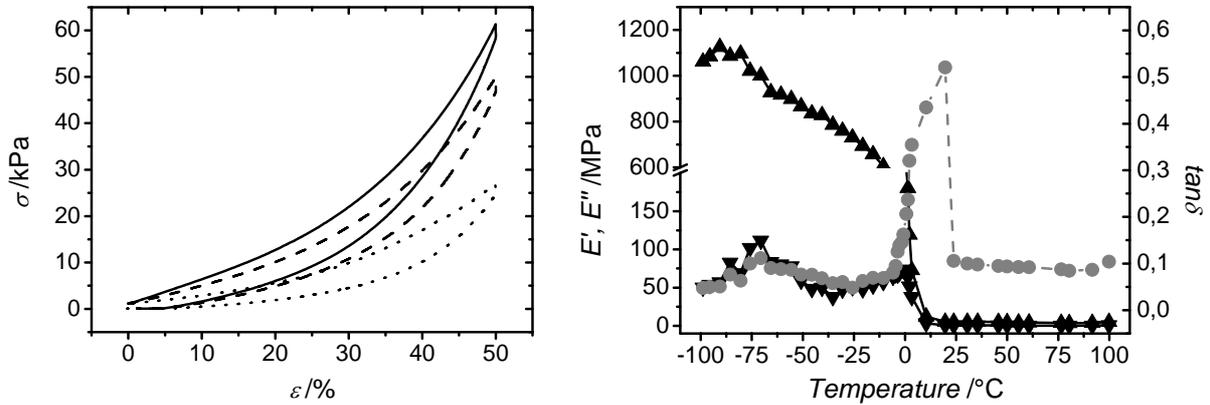

**Figure 6.5. (Left): Compression stress-strain curves of wet gelatin scaffolds tested in air at 25 °C. (···): SG7_HNCO8, (---): SG10_HNCO8, and (—): SG13_HNCO8. (Right): Dynamic Mechanical Analysis at varied temperature (DMTA) on sample SG13_HNCO8 after equilibration with water. (—▲—): storage modulus (*E'*), (—▼—): loss modulus (*E''*), (—●—): loss angle (*tanδ*).**

DMTA was performed on wet samples (Figure 6.5, right), in order to elucidate the elastic behaviour of wet scaffolds in terms of network morphology. A clear morphological transition was observed at around 20 °C, showing a decrease of storage modulus (*E'*: 200 → 5 MPa) and direct increase of *tanδ*. Compared to the dry-state DMTA, a similar transition at 45 °C identified a glass transition, which meant that the presence of the water acted as softener.[142] As a result, $T_g$ is reduced from 45 °C (dry state) to nearly 20 °C (wet state). This corresponds to a rubbery-like network morphology at the compression temperature and at body temperature, explaining the re-expansion of wet compressed scaffolds. Consequently, dry compressed samples also re-adopted their original shape once equilibrated in water, which can be interpreted as a water-induced shape memory effect (SME). This water-induced SME might be helpful for the colonization of the implant with cells. As a result, Figure 6.6 provides a schematic overview of the material behaviour in the wet compared to the dry state. Upon equilibration of water, the scaffold does not change the outer dimension, as due to a reduction of the pore size, i. e. swelling of the pore walls along the pores, from the dry (SEM) to the wet (CLSM) state. At the same time, pictures of wet- and dry-state compression highlight the elastic scaffold compression in the wet state in contrast to the cold deformation in the dry state.

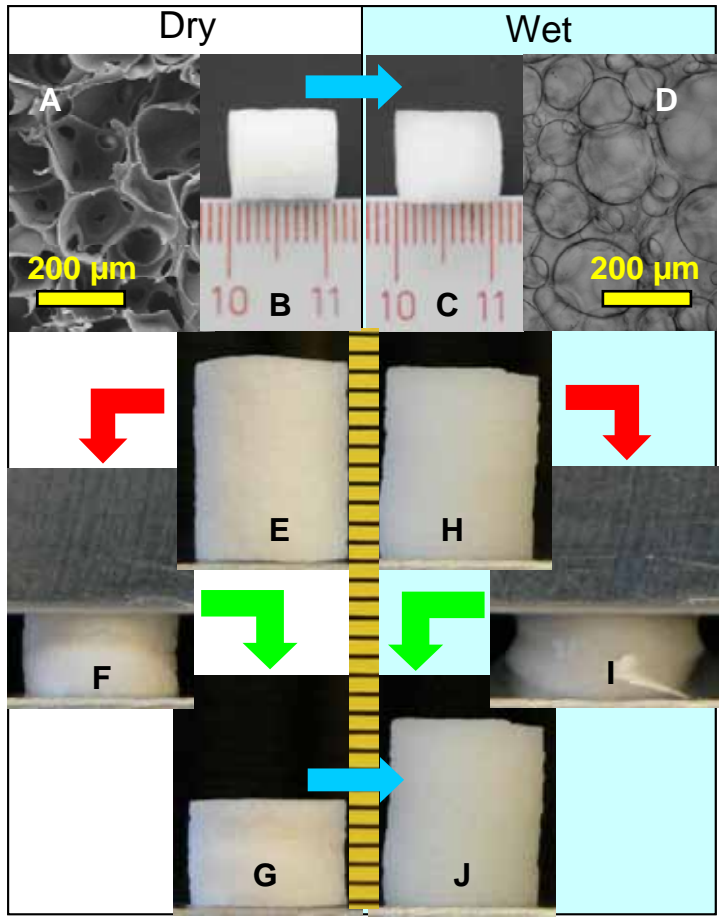

**Figure 6.6. A): SEM picture of the scaffold SG13_HNCO8. B/C) The outer dimensions of the scaffold do not change when being equilibrated in water. D) Light microscopy image of the swollen scaffold. E-G) Non-elastic compression of a dry scaffold. H) SEM of a dry compressed scaffold showing an intact porous structure. I-K) Elastic compression and re-expansion of a scaffold equilibrated in water. The dry compressed scaffold re-expands to the permanent shape when equilibrated in water. →: equilibration in water, →: compression, →: removal of mechanical loads.[13]**

## 6.3.1 Macro-mechanical properties by compression tests

Wet scaffolds displayed tailorable compressive modulus ($E_c$) and ultimate compression stress ($\sigma_{max}$) based on the sample composition. $E_c$ could be successfully adjusted between the elastic modulus of muscle (10 kPa) and the one of bone (100 kPa).[26] Similarly, the ultimate compression stress ($\sigma_{max}$) increased from 18 to 48 kPa by increasing the HDI excess (Table 6.2). Interestingly, $E_c$ (17-48 kPa) and $\sigma_{max}$ (24-60 kPa) could be adjusted not only by variation of the HDI excess, but also by variation of the gelatin concentration in the starting solution, in contrast to the case of homogeneous films.

**Table 6.2. Compressive modulus ($E_c$) and compression stress at 50% compression ($\sigma_{max}$) for wet gelatin scaffolds tested in air at 25 °C. Measurements are expressed as median and range of data.**

| Sample ID | $E_c$ /kPa | $\sigma_{max}$ /kPa |
|---|---|---|
| SG7_HNCO8 | 17 (17-24) | 24 (24-25) |
| SG10_HNCO3 | 9 (7-13) | 18 (17-18) |
| SG10_HNCO5 | 18 (16-23) | 26 (25-26) |
| SG10_HNCO8 | 39 (33-48) | 48 (47-49) |
| SG13_HNCO8 | 48 (44-57) | 60 (57-64) |

This observation is mostly related to the different internal structure of the (porous) scaffolds compared to the one of the homogeneous hydrogels. Indeed, it was already shown that the variation of gelatin concentration resulted in an increase of the pore wall density, which directly affected the overall mechanical properties of the scaffold. Additionally, the $E_c$ and $\sigma_{max}$ values of the wet scaffolds are much lower compared to the dry ones. This is likely explained by the rubbery-like network morphology in the wet state, consequent to the presence of water acting as plasticizer.

## 6.4 Micro-mechanical properties by AFM

The microenvironmental cell-material interactions that are present and are 'felt' by cells through focal adhesion ("microenvironment") are crucial for the success of hydrogels as morphogenetic guide. Tailoring of the micromechanical properties is thereby essential to support specific cell adhesion, migration, growth and differentiation into the scaffolds.

For these reasons, a method to accurately characterize the local stiffness of porous matrices is required. Microspheres probing[143], Atomic Force Microscopy (AFM)[144,145], and Nanoindentation[146,147] techniques were employed for the determination of microscopic elasticity in 2-D hydrogel substrates. On the other hand, few investigations addressed the characterization of the micromechanical properties on 3-D porous scaffolds. Harley et al.[93] performed stiffness measurements on the individual scaffold struts of collagen-glycosaminoglycan scaffolds. Although clear data could be obtained, the analysis was carried out in the dry state, and different mechanical orientation were applied in the macro vs. micro level, *i. e.* bending tests on the scaffold struts were compared to the compression tests on the overall scaffolds. In principle, the removal of scaffold struts also requires long-term preparation of the macroscopic sample, thereby limiting the routine applicability of this method.

As the mechanical behaviour of porous 3-D scaffolds depends on the mechanical properties of the bulk phase as well as on the scaffold geometry and porosity, the evaluation of the micromechanical properties requires the isolation of single pore walls and the determination of the stiffness for each isolated wall. This is a challenging task, firstly, because the presence of heterogeneity in the porous structure, i. e. cut pores resulting from the macroscopic preparation of the sample, might lead to artefacts during the measurements. Secondly, the measurement involves the fixation of the macroscopic sample in wet conditions, meaning that the mechanical properties of the substrate below the sample could affect the measurement. Thirdly, gelatin scaffolds analyzed displayed extremely soft compressive moduli ($E_c$: 10-100 kPa) in the hydrated state. This imposes the demand of relatively low load resolution testing, which is extremely challenging for the conventional mechanical test methods.

For these reasons, micromechanical measurements on wet scaffolds were accomplished at *JPK Instruments AG*, by using a combination of an optical microscope and an AFM. Samples were equilibrated in water at 25 °C and glued, so that the measurements could be performed on regions adjacent to the spots of glue (Figure 6.7).

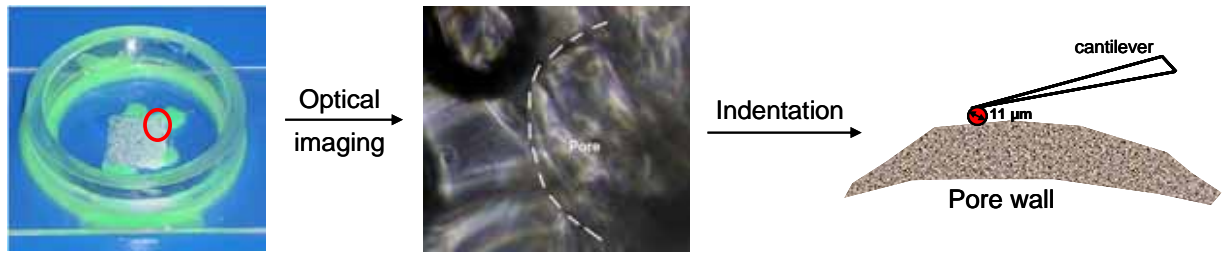

**Figure 6.7. Micromechanical measurements performed on wet gelatin scaffolds. After equilibration in water and fixation at 25 °C, the sample was analyzed with an optical microscope allowing the identification of pore walls. Once that the pore wall is isolated, AFM measurements are carried out with a cantilever bearing a 11 μm sphere.**

Once that the pore wall was optically identified AFM measurements with a cantilever bearing a spherical indenter (Ø =11 μm) followed. Pore walls were indented in form of 5x5 testing site grids (or larger, if possible), whereby the distance between two indentation spots was 2 μm. Each force measurement on immersed samples was taken with a 50 nN maximum applied force and a 2 μm·s⁻¹ velocity, while the height change was continuously read out.

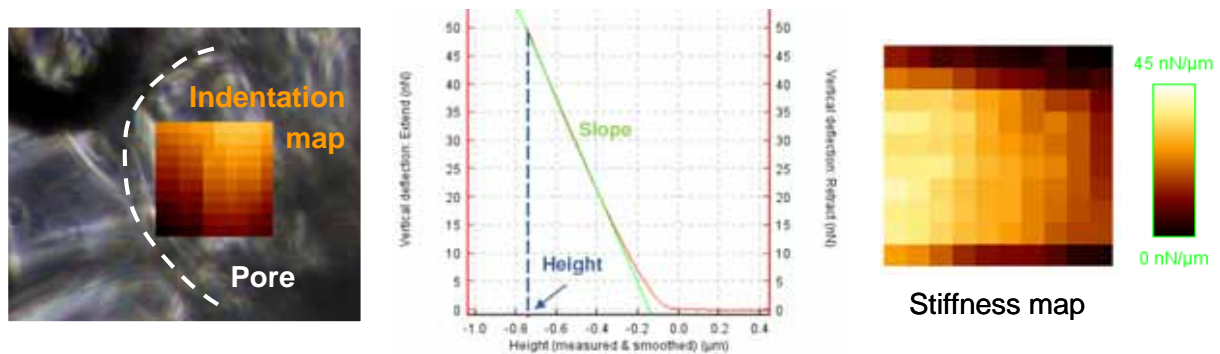

**Figure 6.8. Left: optical image overlaid with an indentation map (right, 10x10 points map) displaying the relative height values of all curves: the darker the color, the lower the height. Middle: the force is plotted over height. The height value of the force map displays the measured height (blue), when the applied force reached the setpoint (50 nN). The Young's modulus is measured in the linear indentation part (green). Right: representation of the stiffness map. The colors display the slope of the indentation (contact) part of the curve: the brighter the color, the steeper the slope, e. g. the stiffer the material.**

The curves were recorded in closed loop mode, which allowed for a constant extend speed of the piezo, and thereby of the probe. Resulting force-indentation curves were fitted with the Hertz model for spherical indenters, allowing for the calculation of the Young's modulus (Figure 6.8).

The results displayed a wide variation of data compared to the macro-compression measurements, which is probably due to either the technical limitation of the instrument or the non-homogeneity of the material. The optical investigation was indeed challenging, since the foam consisted of several layers. Additionally, the macroscopic samples were cut on both sites, so that truncated pores resulted in the upper as well as in the subjacent layer. Hence, it is likely that most of the measurements were taken on the edge of truncated pores (Figure 6.9). The non-homogeneity along with the loosely floating pore material, which derived from the cut pores, often disturbed force measurements, resulting in a variation of the number of usable curves per map and sample.

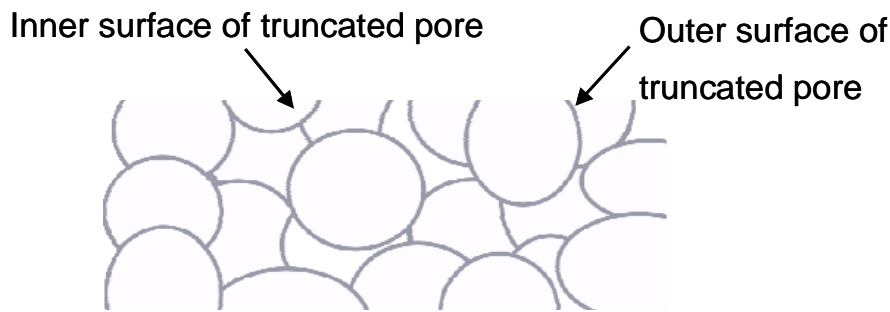

**Figure 6.9. Schematic representation of scaffolds pores during imaging and indentation. Since almost all pores were truncated, most measurements are expected to be taken on the edge of truncated pores, either on the ascending (right arrow) or the descending part (left arrow) of the pore. This resulted in a wide variation of results compared to the macro-compression tests.**

Taking into account the limitation of the measurements, the micro-mechanical Young's moduli are compared with the macro-compressive moduli (Figure 6.10). The results showed that the local $E$ moduli are in the same order of magnitude as the scaffold $E_c$ moduli (10-50 kPa). The local $E$ moduli steadily increased with the HDI excess, likely related to a different crosslinking density of the network. On the other hand, the change of gelatin concentration only had a small effect on the local $E$ modulus, e. g. a local $E$ modulus of 40 kPa could equally be exhibited from a scaffold with a macroscopic $E_c$ modulus of either 30 or 40 kPa. This observation gives further evidence that the change of gelatin concentration mainly affects the density of the pore walls.

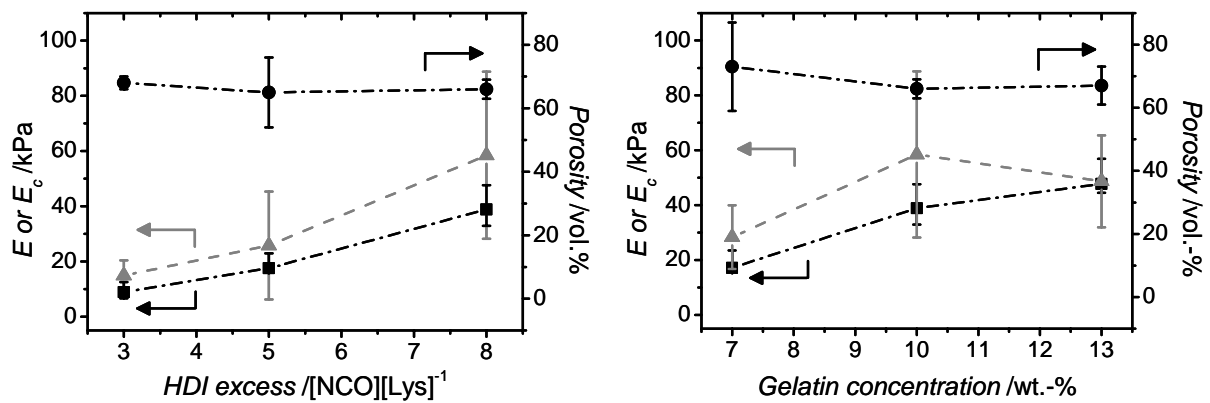

**Figure 6.10. Porosity in dry state and mechanical properties of scaffolds in wet conditions. Compression modulus $E_c$ obtained from compression tests and Young's modulus $E$ determined by AFM, under variation of the HDI excess (left) and gelatin concentration (right). ●: porosity determined by µCT, ▲: $E$, ■: $E_c$.**

Therefore, two mechanisms of tailoring the mechanical properties of the gelatin scaffolds could be highlighted: on the one hand, macro- and micro-mechanical properties can be adjusted by variation of the HDI excess during the crosslinking reaction. On the other hand, the macro-mechanical properties can be tailored independently from the micro-mechanical properties under variation of gelatin concentration. The aforementioned structure-property relationships are crucial in order to adjust the mechanical properties to the in vivo situation and successfully establish a clinical biomaterial-based tissue regeneration approach (Figure 6.11).

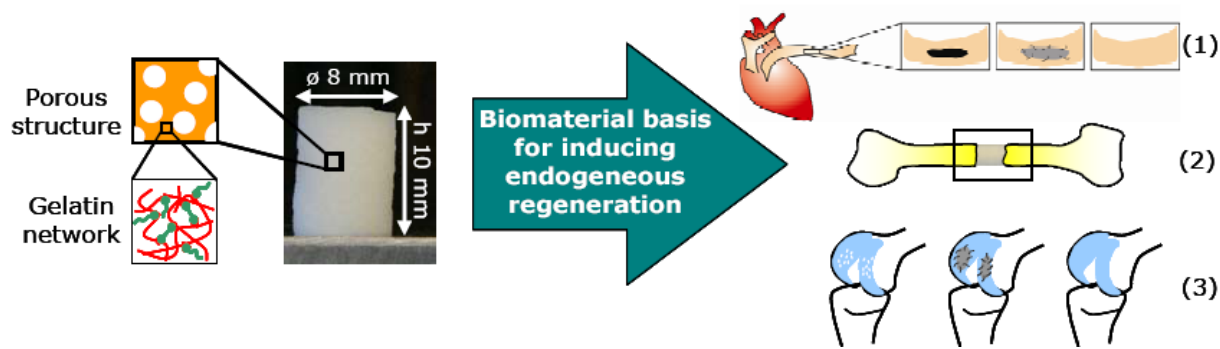

**Figure 6.11. Application of multifunctional gelatin-based scaffolds for inducing the endogenous regeneration of damaged tissues. By controlling the porous geometry, the mechanical properties of the scaffold can be adjusted to the situation *in vivo* under systematic variation of molecular parameters. Potential applications include the reconstruction of blood vessels after surgery (1), healing of critical bone defects (2), or cartilage repair (3).[13]**

# 6.5 Hydrolytic degradation

For successful material-guided tissue regeneration, not only the material properties at the starting point of implantation are of interest, but the course of scaffold biodegradation is of importance.[19] Thus, the change of material properties over the time needs to be studied and adapted to the specific clinical situation.

Two types of gelatin scaffolds were subjected to hydrolytic degradation in PBS buffer (pH 7.4) at 37 °C, and mass loss, water uptake, mechanical properties, chain organization, and porous architecture were monitored over time. Scaffolds SG7_HNCO8 and SG13_HNCO8 were studied because they represented two extreme sample compositions according to the uptake and compression modulus. As the scaffolds were based on a chemical network of gelatin, the hydrolytic cleavage of the network chains could occur along the amide bonds of the gelatin backbone as well as at the urea bonds formed after reaction between with HDI. Thus, all the chemical bonds of the gelatin network could be hydrolyzed, so that a fully degradable material was expected. In order to study the mechanism of material degradation, hydrolysis was chosen as model of degradation.

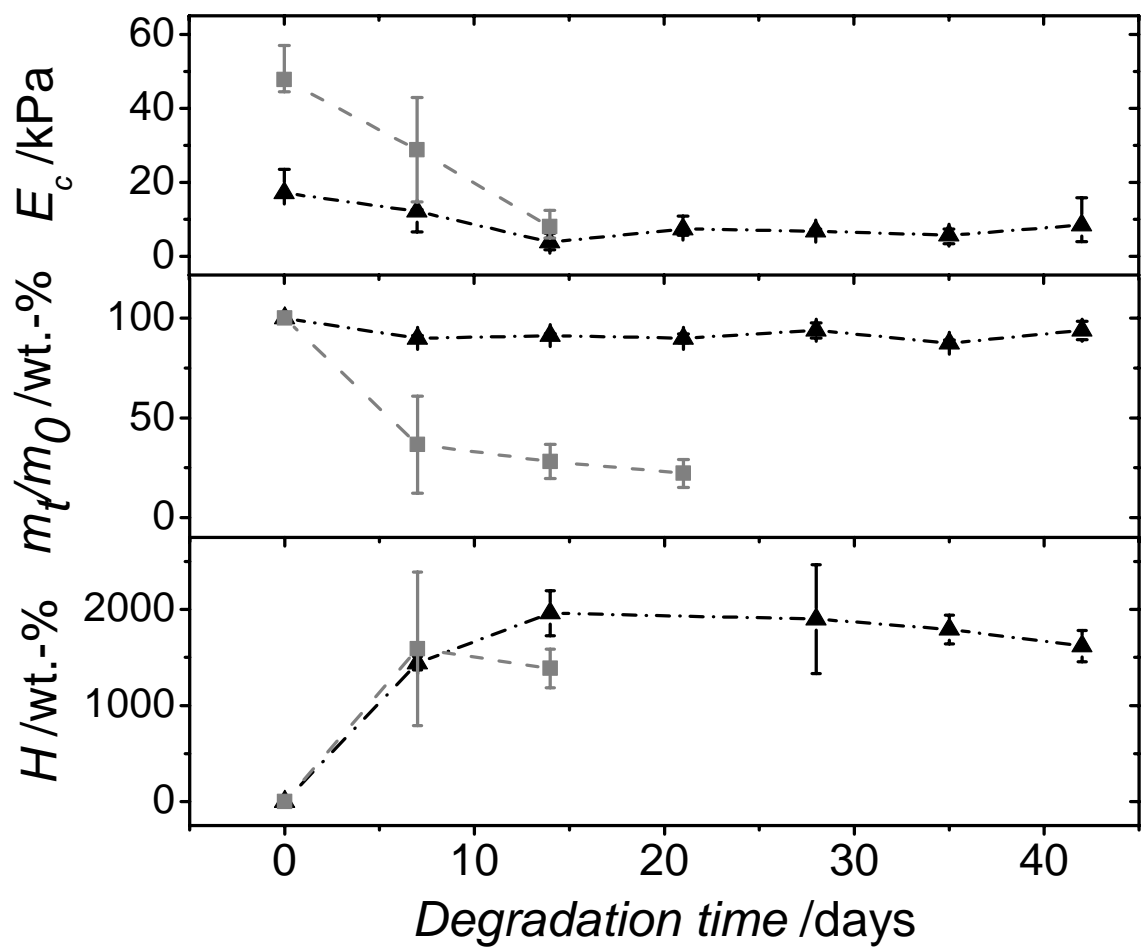

**Figure 6.12. Hydrolytic degradation (PBS, pH 7.4, 37 °C) of two gelatin-based scaffolds (·−▲−· : SG7_HNCO8, −■− : SG13_HNCO8). The compressive modulus ($E_c$, top) was determined by wet-state macro-compression tests, the relative mass ($m_t/m_o$, middle) was measured after freeze-drying of the partially-degraded scaffolds, while the water uptake ($H$, bottom) was measured by paper drying wet samples.**

Figure 6.12 displays the change of material properties during hydrolytic degradation. Macroscopically, the scaffold degradation resulted in a decrease of $E_c$ modulus and sample mass, and in an increase of uptake, though the volume of the partially-degraded samples was slightly changed compared to original size. Samples SG13_HNCO8 degraded much faster, i. e. after roughly 20 days, than samples SG7_HNCO8, showing nearly constant mass up to 40 days degradation. Other than the mass loss, the uptake ($H$) of both samples was increased and then decreased in the last period of degradation, as in accordance with the change in sample mass. The change of $E_c$ followed the same tendency as the mass loss, i. e. it was decreased when the relative sample mass was decreased. A decrease of $E_c$ from 50 to roughly 10 kPa was displayed in sample SG13_HNCO8, while $E_c$ of sample SG7_HNCO8 decreased from 20 to about 10 kPa. Despite the different degradation kinetics, a controlled change of mechanical properties, which is crucial for ensuring the clinical applicability of polymer materials, was observed during the degradation of both samples.

The differences in degradation rate can be explained by differences on the molecular level. Both scaffold compositions have similar bulk properties (Figure 6.10, left) but different densities. Thus, it can be speculated that the ratio of direct covalent crosslinks and physical crosslinks by grafted HDI oligomers varies according to the gelatin concentration in the reaction mixture. If SG13_HNCO8 has a higher degree of grafted chains, this would explain the observed lower degree of swelling and higher material density. Consequently, already a small number of hydrolyzed bonds would lead to large changes in material properties during degradation.

In order to follow the change of molecular network structure, WAXS was carried out on the partially-degraded samples. While the formation of single- and triple helices in the freshly-synthesized gelatin scaffolds was inhibited by the crosslinking step, a step-by-step self-organization of the peptide chains into single helices was observed during the time of degradation (Figure 6.13).

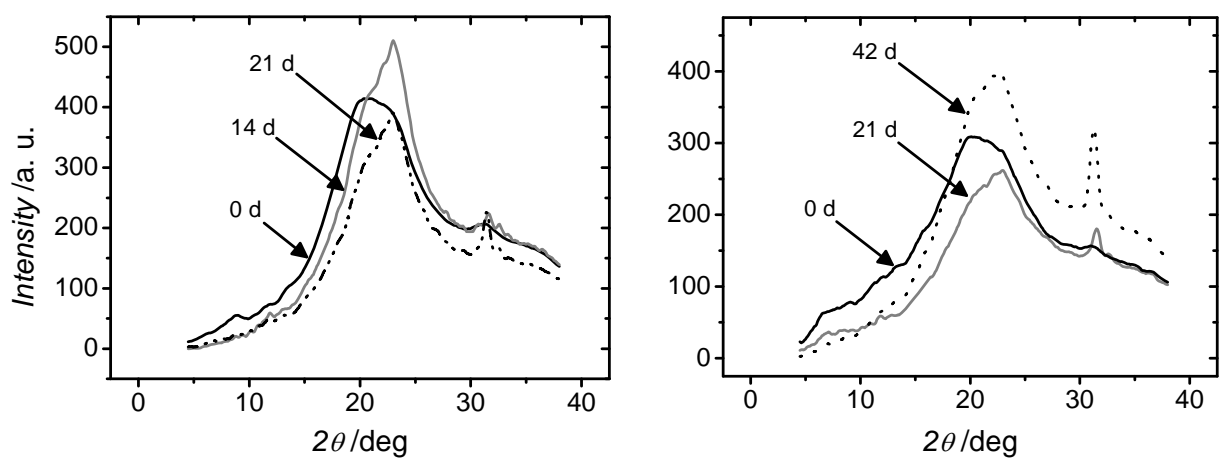

**Figure 6.13. WAXS analysis of freeze dried gelatin-based scaffolds during hydrolytic degradation. (Left): scaffold SG13_HNCO8: (—) freshly-synthesized sample (0 d), (—) sample analyzed after 14 days of hydrolytic degradation (14 d), and (----) samples analyzed after 21 days of hydrolytic degradation (21 d). (Right): scaffold SG7_HNCO8: (—) freshly-synthesized sample (0 d), (—) sample analyzed after 21 days of hydrolytic degradation (21 d), and (---) samples analyzed after 42 days of hydrolytic degradation (42 d).**

The observed chain helicity might counteract an undesired increase of swelling during degradation. By comparing WAXS spectra of the two samples, the partially-degraded scaffold SG13_HNCO8 showed helicity during mass loss. In contrast, scaffold SG7_HNCO8 revealed a change of network structure with only a partial decrease of either mass or compressive modulus. Therefore, the formation of dangling chains might be expected in the sample SG7_HNCO8, in view of the uptake increase, and the negligible mass loss.

Synthetically, the surfactant employed promoted the stabilization of air bubbles in the foaming solution of gelatin. Here, air bubbles were fixed as pores in the resulting scaffold through chemical crosslinking of gelatin, leading to a narrow size distribution of the pores. A slight reduction of scaffold size was observed in the partially-degraded scaffolds, which is likely due to the wall degradation on the sample surface, resulting in an increase of the average pore size (Figure 6.14).

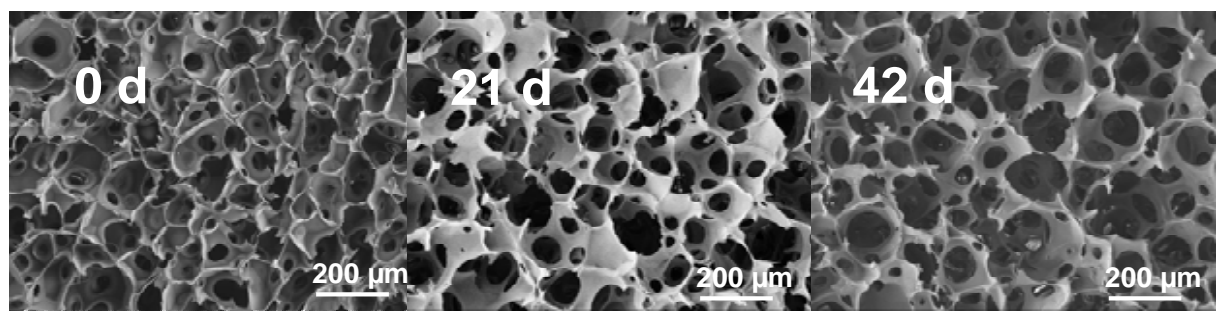

**Figure 6.14. Porous architecture of a gelatin-based scaffold (SG7_HNCO8) at different time points of hydrolytic degradation (PBS, pH 7.4, 37 °C).**

After 6 weeks of degradation, the porous architecture was still observed in the material and an increase of either pore size (113±23 μm → 216±51 μm) or interconnection size (36±8 μm → 82±22 μm) was detected. The destruction of pore walls at the micro-level might explain the translation from the freshly-synthesized scaffolds with homogeneous pore distribution to the partially-degraded scaffolds with increased size of pores and interconnections. Thus, the hydrolytic cleavage of the network chains at the molecular level resulted in the wall destruction in the scaffold. Thus, it is speculated that pores behave like merging air bubbles, i.e. when a wall separating two pores is degraded, the surface tension of the merged pore results in the formation of a new, larger, round pore rather than in tunnels (Figure 6.15). This corresponds to the demands for scaffolds in regenerative therapies in which smaller pores leads to better adhesion of cells,[148,149] while cell in-growth and tissue formation is profiting from larger pores, higher surface areas, and greater interconnectivity.[150]

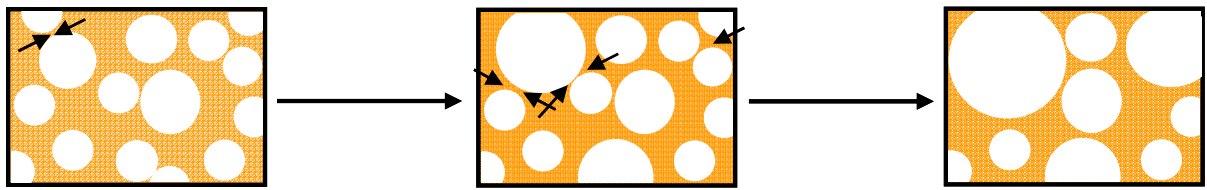

**Figure 6.15: Schematic representation of the scaffold degradation mechanism, leading to increased pore size with the increase of degradation time. Pores previously stabilized through crosslinking of the gelatin foam (left), are connected through the degradation of the pore walls (middle), so that in increase of pore size in the partially-degraded scaffold is observed (right).**

# 6.6 Summary

A gelatin-based scaffold system was successfully characterized in aqueous environments to fulfill the complex biomaterial requirements for regenerative medicine applications. The scaffolds increased their weight once equilibrated in the aqueous solution (section 6.1) while a pore size decrease was observed with respect to the dry state. In this way, pore walls swelled inwards rather than outwards, ensuring only minimal change of the scaffold outer dimensions (section 6.2). Other than the form-stability in physiological conditions, scaffolds were elastic and full shape recovery (SR> 95%) was observed after mechanical load removal (section 6.3). Thanks to the hierarchical structure, the macroscopic (scaffold) and microscopic (pore walls) mechanical properties could be adjusted in their modulus (10-50 kPa) to values near the ones of soft tissue (section 6.4). The scaffold bulk phase was directly ruled at the molecular level by the synthesis of an entropy-elastic network. Likewise, the macroscopic compression behavior was also controlled by varying the crosslinking density, since the scaffold geometry was kept constant. The hydrolytic degradation study highlighted a controlled decrease of compression modulus ($E_c$: 50 $\rightarrow$ 10 kPa, sample SG13_HNCO8), while the degradation kinetics was successfully adjusted based on the sample composition (section 6.5). Over the time of degradation (0-42 days), the hydrolysis of gelatin chains on the molecular level resulted in growing pore size (113±23 $\rightarrow$ 216±51 µm) on the porous architecture level. Therefore, it is speculated that pore walls are partially disintegrated, resulting in the formation of bigger round pores, which is crucial for accomplishing cell growth into the scaffold. Due to the material behavior in physiological environments, these structured gelatin-based scaffolds are very promising candidates to be applied in biomaterial-induced endogenous regeneration.

# 7. Development of medical-grade hydrogels

In the following sections, the development of fully cytocompatible, medical-grade scaffolds is presented. The biological evaluation includes direct and indirect cell tests as well as the investigation of the biomaterial endotoxin contamination (*cooperation with Prof. Volk group – Charitè, Berlin, Germany*).

## 7.1 First results of direct cell tests on scaffolds

In biomaterial-assisted tissue regeneration, the adjustment of material properties to the *in vivo* situation is of paramount importance, in order to restore the hierarchical organization and ensure the functionality of neo-tissue. At the same time, the scaffold implanted *in vivo* must display a porous architecture enabling cell proliferation in the implant. Also the scaffold must cause as few complications to the surrounding biological environment as possible. Therefore, leaching of toxic chemicals into the body, immunological responses, tissue inflammation, and blood clotting must all be minimized.

A pilot cell culture study was initially carried out, in order to ensure that the scaffold pore size and porosity enabled cell proliferation in the implant. Thus, HFIB-D fibroblasts were seeded onto the scaffold sections (1 mm thickness), and confocal laser scanning microscopy was carried out at different depth section levels. The presence of cells was homogeneous on the material, and vital cells were found in the inner part of the scaffold, up to approximately 90 µm of depth (Figure 7.1).

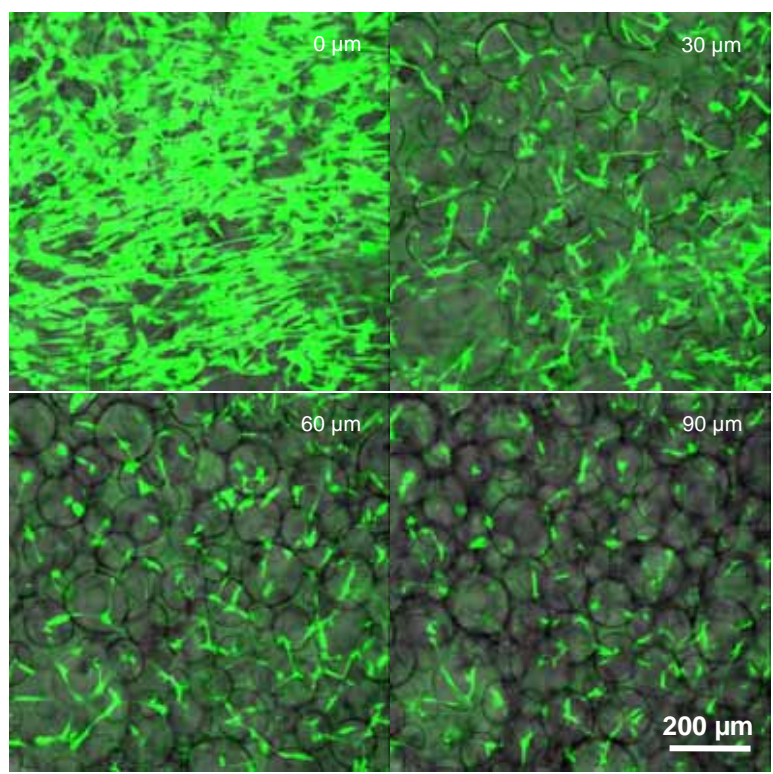

**Figure 7.1. Visualization of human fibroblasts by vital staining with fluorescein diacetate. Vital cells could be observed by CLSM either on the top (top-left) or at different depth level of the sample, i. e. at 30 (top-right), 60 (bottom-left), and 90 (bottom right) µm underneath. The size of the scale bar is 200 µm.**

The lower amount of vital cells deep within the scaffold can be attributed to the closed colonization of the surface, since the transport of nutrients and waste is difficult in static cell cultures. Most importantly, cells were still vital after 21 days of cell culture on the scaffolds, and the growth patterns were the same regardless of scaffold composition. Therefore, it was possible to tailor the mechanical properties of the scaffolds without affecting their biocompatibility.

# 7.2 Decreasing the endotoxin contamination by optimizing scaffold preparation

Endotoxins are well known to stimulate a wide range of inflammatory mediators and can therefore compromise the implant biocompatibility *in vivo*.[151] Especially for scaffolds based on biopolymers, the material purity requires careful consideration. This involves the effective removal of antigenic epitopes associated with cell membranes and intracellular components of tissues.[152]

The endotoxin contamination was evaluated on either the pure gelatin or the synthesized scaffolds. The human whole blood TNF-α induction and the endotoxin content (LAL assay) were quantified. In a preliminary experiment, a sample of aqueous gelatin solution displayed high TNF-α induction (124 ng/mL, i. e. 620 EU/mL, in case of 1:51 gelatin:whole blood sample), much higher than the FDA accepted limit (100 pg/mL or 0.5 EU/mL, Figure 7.2). In contrast, a sample of doubly distilled water showed negligibly TNF-α induction, well below the control. As both samples were prepared in the same conditions, i. e. the water and the equipment used during the synthesis, i. e. glassware, potential contamination during the sample preparation was unlikely to be the reason for the TNF-α induction of the gelatin sample. Therefore, three different situations were hypothesized to explain this finding: (i) either the material was contaminated from low purity gelatin source (ii) or the contamination occurred during a synthetic step, handling and work-up; (iii) a combination of (i) and (ii) cases.

In order to exclude the first case (i), a certified low-endotoxin gelatin was applied for the synthesis of gelatin-based scaffolds. LAL assay proved an endotoxin material contamination (21 pg/mL) below the FDA limit. However, the TNF-α induction (820 pg/mL) was still above the accepted threshold.

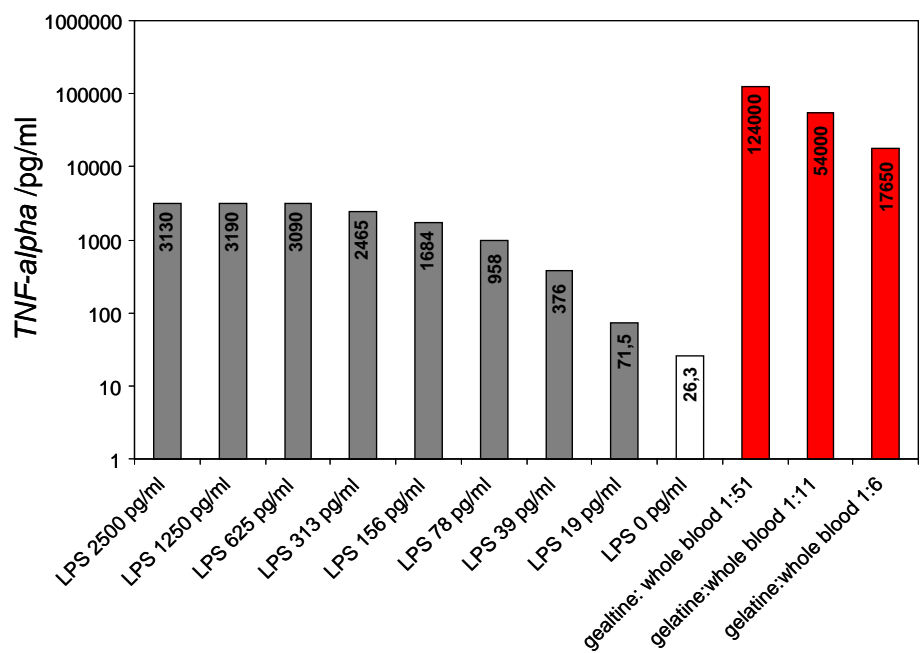

**Figure 7.2. TNF-α induction on the whole blood upon contact with a gelatin (chemical grade) sample eluate. Solutions with a known content of endotoxin (here stated as LPS) have been applied as control.**

It was therefore likely that the TNF-α induction was not imputed to an endotoxin contamination. In this case, the synthetic step (ii) was identified as a possible reason for this finding, likely related to the presence of non-reacted moieties in the final material. In order to validate this hypothesis, scaffolds were washed with water for two more days, aiming at the removal of HDI oligomers from the material. Noteworthy, resulting samples successfully displayed a 85 wt/vol.-% reduction of TNF-α induction compared to the non-washed samples, suggesting that longer washing was needed to achieve pure networks.

Consequently, synthesis and washing were carried out under a laminar flow cabinet, instead of the chemical fume hood, and all lab equipment was (steam) sterilized before use, in order to minimize the material contamination. Samples were synthesized and washed using filtered (Ø 0.2 μm), sterilized water. During washing, samples were also rinsed with a 70 vol.-% ethanol aqueous solution, aiming at the extraction of hardly-water soluble moieties, i. e. HDI oligomers/rings. After washing, samples were sealed in sterilization bags, in order to avoid environmental contamination, before being dried and EtO sterilized. At this stage, EtO sterilization was carried out as a compulsory step before any *in vitro* and/or *in vivo* tests.

Resulting samples highlighted a minimal TNF-α induction (~ 50 pg/mL), which was far below the FDA limit (100 pg/mL). Besides the low TNF-α induction, LAL assay confirmed an endotoxin contamination (0.16 EU/mL) within the accepted FDA limit (0.5 EU/mL). Overall, these results proved that the previously-observed TNF-α induction was related to soluble factors entrapped in the material after the synthesis. Pure and nearly-endotoxin free materials were therefore successfully accomplished by establishing the synthesis and washing in cleaned conditions.

# 7.3 Optimizing the cell-compatibility determined by indirect cell tests

Preliminary direct tests proved the growth of vital fibroblasts at different depth level within the scaffolds (section 7.1). This finding suggested that the scaffold porous morphology, i. e. pore size and porosity, was suitable to enable the migration of cells within the implant. Once the biomaterial endotoxin contamination was suppressed, indirect eluate assays were performed according to EN DIN ISO 10993-5 standards, in order to investigate the cytotoxicity of the gelatin-based hydrogels. Thus, the material was equilibrated in cell culture medium and cells were cultured on the material extract as well as on the pure cell culture medium, as control. Figure 7.3 displays the cell morphology on the control (spread cells) and on the sample extract (nearly complete destruction of cell layers).

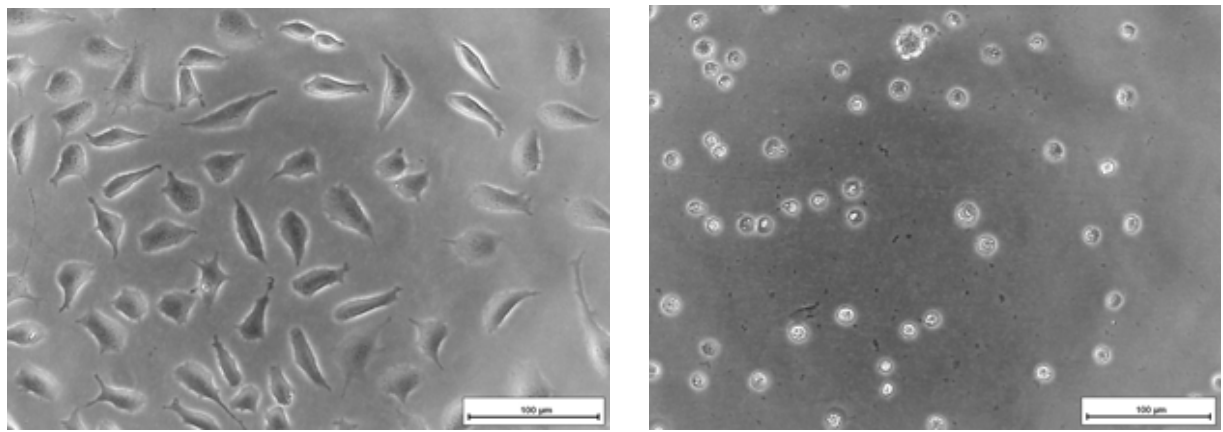

**Figure 7.3. Morphology of L929 cells 24 hours after culturing with a pure culture medium (left) and a 3-days (undiluted) extract (right) of a crosslinked hydrogel (G10_HNCO5).**

As a result, a severe cytotoxic response was observed when L929 mouse fibroblasts were cultured on a 3-days material extract in contrast to the first cell culture study performed on the scaffolds. In order to identify the reason for this finding, the overall synthesis of the material was closely inspected, and the influence of the crosslinker and surfactant was investigated in order to optimize the cell-compatibility of gelatin scaffolds.

# 7.3.1 Influence of parameters varied during scaffold formation on material properties

Due to the high reactivity of decarboxylated amino functions, gelatin-based scaffolds were synthesized through crosslinking with LDI. Crosslinking of gelatin with LDI is of advantage compared to HDI: in view of the chemical structure of LDI, non-toxic lysine-derivatives products can be expected in the case of gelatin grafting as well as in the partially-degrading network (Figure 7.4).[94]

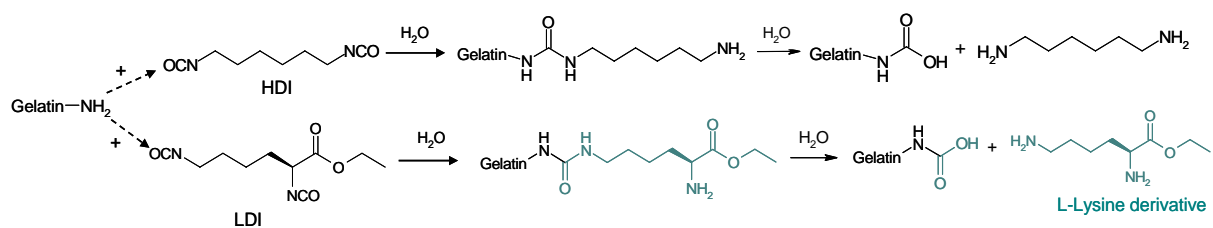

**Figure 7.4. Chemical reaction of gelatin with diisocyanates. In the case of reaction with HDI, hexamethylene diamine could be released during degradation, which might be toxic. Other than HDI, LDI potentially leads to the formation of non-toxic lysine-derivatives products.**

Additionally, crosslinking of gelatin with LDI was already accomplished, and led to the formation of tailorable entropy-elastic hydrogels (Chapter 4). The reaction between gelatin (chemical grade) and LDI under foaming resulted in the formation of porous gelatin-based scaffolds, whose internal architecture was comparable to the one of previously-synthesized HDI-crosslinked scaffolds (Figure 7.5). Tests in water at 50 °C for three days also confirmed the material stability in aqueous environment.

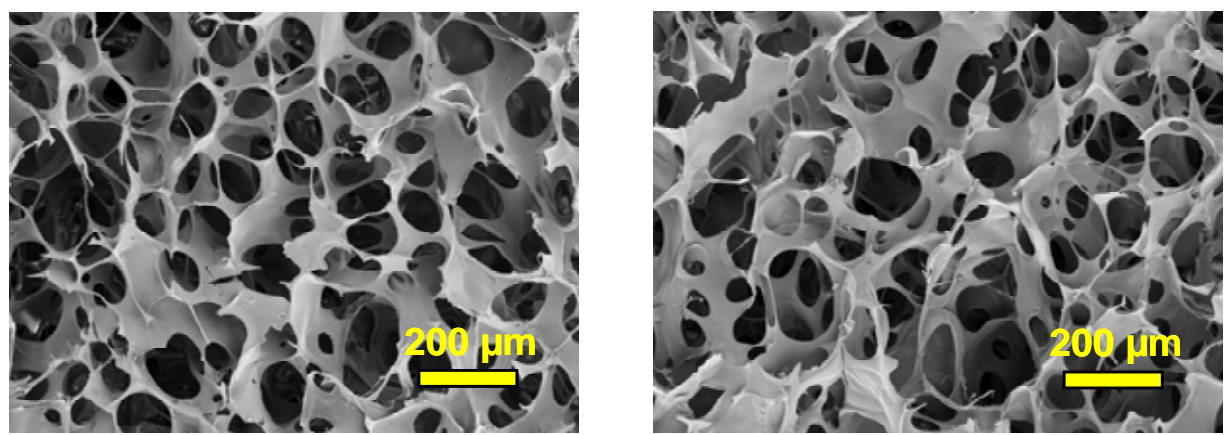

**Figure 7.5. SEM pictures depicting the vertical (left) and horizontal (right) section of an LDI-crosslinked gelatin-based scaffold (SG7_LNCO8).**

Besides the crosslinker, the effect of the surfactant was also considered in view of optimizing the scaffold cell-compatibility. In principle, the optimal surfactant for scaffold formation was selected considering the foam stability as well as the pore size distribution of the resulting gelatin scaffolds (chapter 5). Besides the HLB value, the potential cell-compatibility was another crucial point for the surfactant selection at this stage of the work.

Polymeric surfactants such as poly(ethylene oxide)-poly(propylene oxide)-poly(ethylene oxide) (PEO-PPO-PEO, Pluronic® F-108) have recently gained great attention for the design of compatible biomaterials,[153] since they are FDA approved for *in vivo* injection,[154,155,156] and, as such, widely established in clinical applications. Among these, $PEO_{40}$-$PPO_{20}$-$PEO_{40}$ (20 mol.-% PPO, 80 mol.-% PEO, $M_n$ = 14600 Da) was considered for the formation of stable and porous foams (Figure 7.6). Although its HLB value (~ 16) was lower than the one of saponin (9-13), its high molecular weight could potentially enable an increase of solution viscosity, which is of advantage to ensure foam stability.[97] Also, the high molecular weight of $PEO_{40}$-$PPO_{20}$-$PEO_{40}$ suggests a decreased concentration of hydroxyl terminations of the polymer, which makes the potential reaction of these with the crosslinker isocyanate groups improbable.

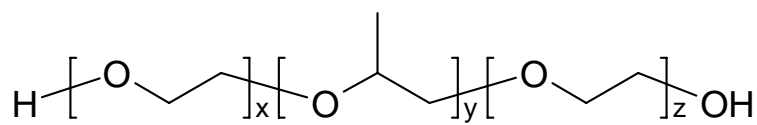

**Figure 7.6. Chemical structure of $PEO_{40}$-$PPO_{20}$-$PEO_{40}$ (HLB ~ 16), as block copolymer of poly(ethylene oxide)-poly(propylene oxide)-poly(ethylene oxide. y= 20 mol.-%, x= z= 40 mol.-%.**

Initial experiments were carried out in order to elucidate the foam stability as well as the porous architecture of resulting LDI-crosslinked scaffolds. Foaming and crosslinking with 10 wt.-% aqueous gelatin (chemical grade) solutions revealed the formation of stable foam, with only 3 vol.-% of non foamed phase. Additionally, SEM investigation (SG10_LNCO8) proved the existence of a porous structure (Figure 7.7) in resulting samples. In another trial, solutions with 7 wt.-% gelatin were also foamed. However, higher phase separation was observed (20 vol.-% of non-foamed solution), which was expected based on the HLB value of $PEO_{40}$-$PPO_{20}$-$PEO_{40}$, which was slightly higher than the HLB value of saponin. For these reasons, only samples synthesized from 10 wt.-% gelatin solution were further considered for EtO sterilization and eluate cytotoxicity tests.

Once a preliminary material characterization was performed on scaffolds synthesized with LDI and saponin, on the one hand, and LDI and polaxomer, on the other hand, the evaluation of the cytocompatibility followed.

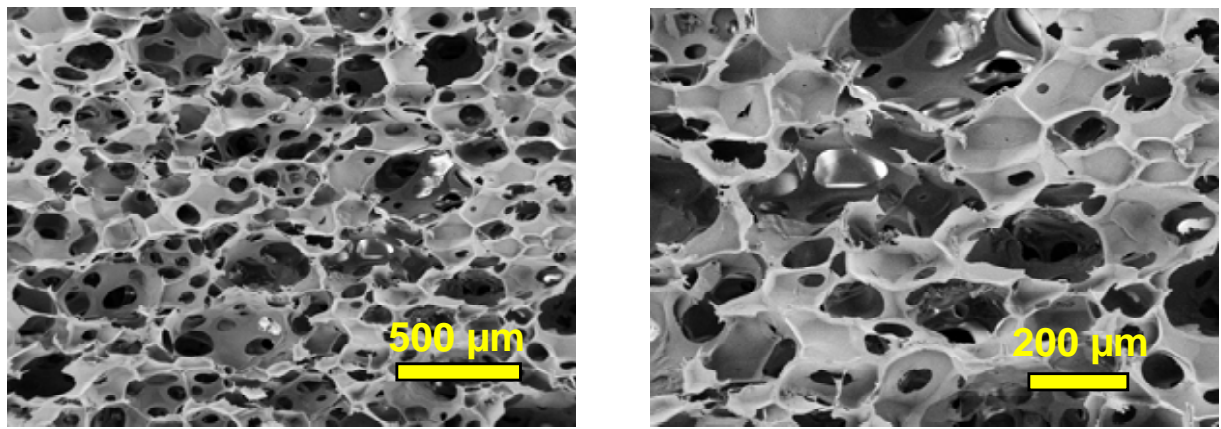

**Figure 7.7. SEM picture depicting the porous architecture of an LDI-crosslinked gelatin-based scaffold (SG10_LNCO8) synthesized by using PEO$_{40}$-PPO$_{20}$-PEO$_{40}$ as foaming agent.**

# 7.3.2 Influence of parameters of scaffold formation process on the cell-toxicity

Once the influence of LDI and PEO$_{40}$-PPO$_{20}$-PEO$_{40}$ was elucidated on the material properties, the cell-compatibility of resulting scaffolds was investigated. EtO sterilization was applied for the scaffold sterilization. After EtO sterilization, sample synthesized with LDI as chemical crosslinker and saponin as surfactant highlighted a moderate cytotoxicity, i. e. reduced cytotoxic effect compared to the HDI-crosslinked materials (Figure 7.8). In comparison to the control (Figure 7.8, left), cells cultured on the sample extract were round and loosely attached (< 70 %), so that many empty areas characterized the cell layer. In contrast to HDI-crosslinked materials, cell lysis counted for less than 70 %, though many intracellular vacuoles were observed. Other than EtO sterilized samples, EtOH disinfected samples showed comparable results, giving evidence that EtO sterilization did not affect the material cell response. Overall, the above-mentioned results supported the strategy of using

LDI rather than HDI in order to suppress the cytotoxic material response. The resulting materials were still moderately cytotoxic, which might be caused by the employment of saponin for the foaming process. For these reasons, $PEO_{40}$-$PPO_{20}$-$PEO_{40}$ was employed as surfactant, while LDI was applied as chemical crosslinker.

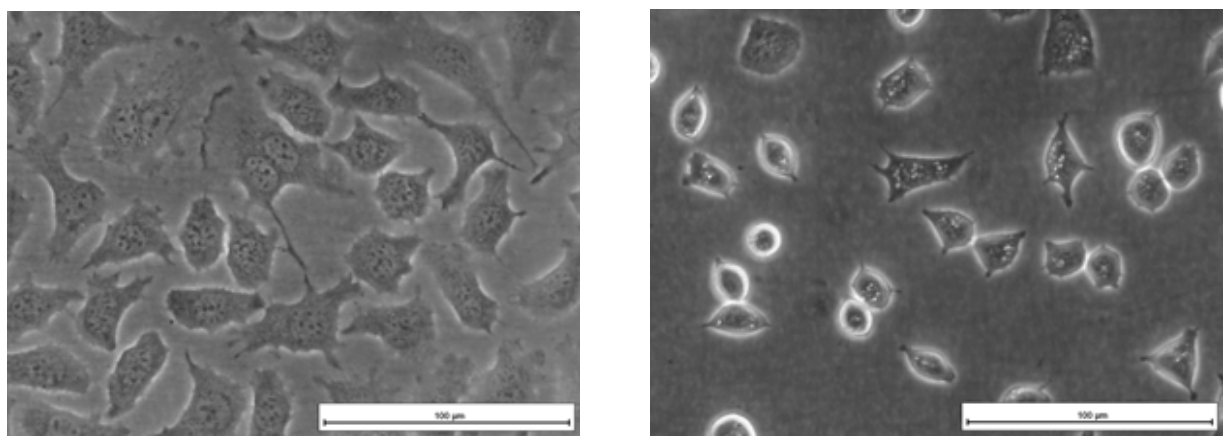

**Figure 7.8. Morphology of L929 cells 24 hours after culturing with a pure culture medium (left) and a 3-days (undiluted) extract (right) of an LDI-crosslinked gelatin-based scaffold (SG7_LNCO8).**

Cytotoxicity tests on resulting EtO sterilized samples displayed a different morphology of L929 cells cultured for 48 hours with an extract obtained by stirring the sample in cell culture medium at 37°C for 72 hours (72h-extract) compared to the morphology of cells cultured for 48 hours with pure culture medium (Figure 7.9). Higher cell proliferation was observed in sample extract cell culture rather than in cell culture medium, resulting in a tight cell layer and completely absence of cell lysis. The mitochondrial dehydrogenases (MTS) activity was significantly ($p<0.01$) higher than the activity of these enzymes seen after 48 hours of L929 cell culture in pure cell culture medium (Figure 7.10, right). Also, the lactate dehydrogenase (LDH) activity in the extracellular fluid was significantly ($p<0.05$) different from the extracellular LDH-activity seen 48 hours after culturing L929 with pure cell culture medium (Figure 7.10, left).

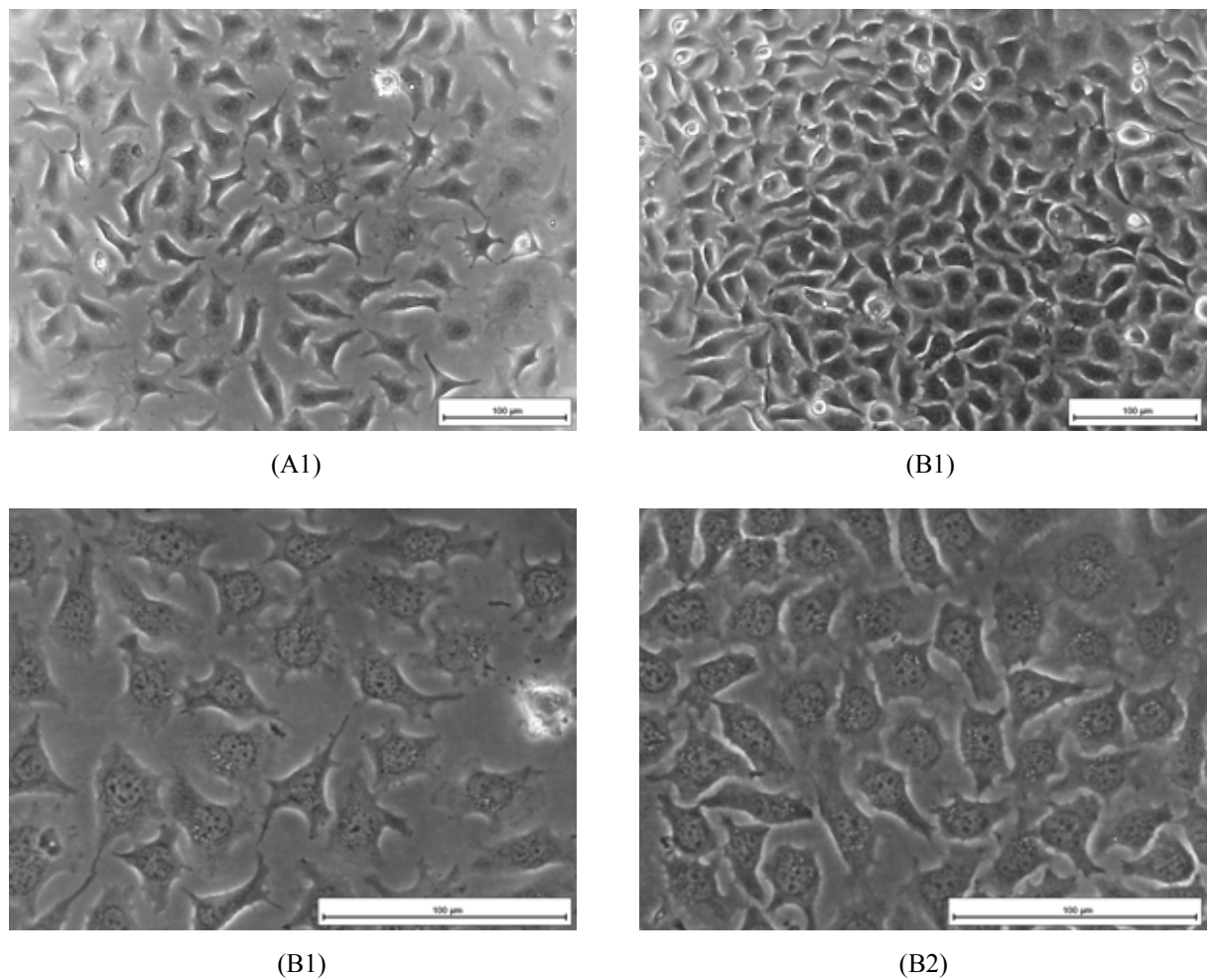

(A1)

(B1)

(B1)

(B2)

**Figure 7.9. (A1, A2): L929 cell morphology observed after 48 hours of cell culture in pure cell culture medium. (B1, B2): L929 cell morphology observed after 72 hours of cell culture in the sample extract. Image obtained with a transmitted light microscopy (phase contrast mode), primary magnification 20× (A1, B1) and 40× (A2, B2).**

This indicated that the sample had no negative influence on the activity of the mitochondrial dehydrogenases as well as on the functional integrity of the outer cell membrane. Therefore, the above-mentioned results highlighted no cytotoxic effect of the gelatin scaffold, while promoting an increased cell proliferation with respect to the control.

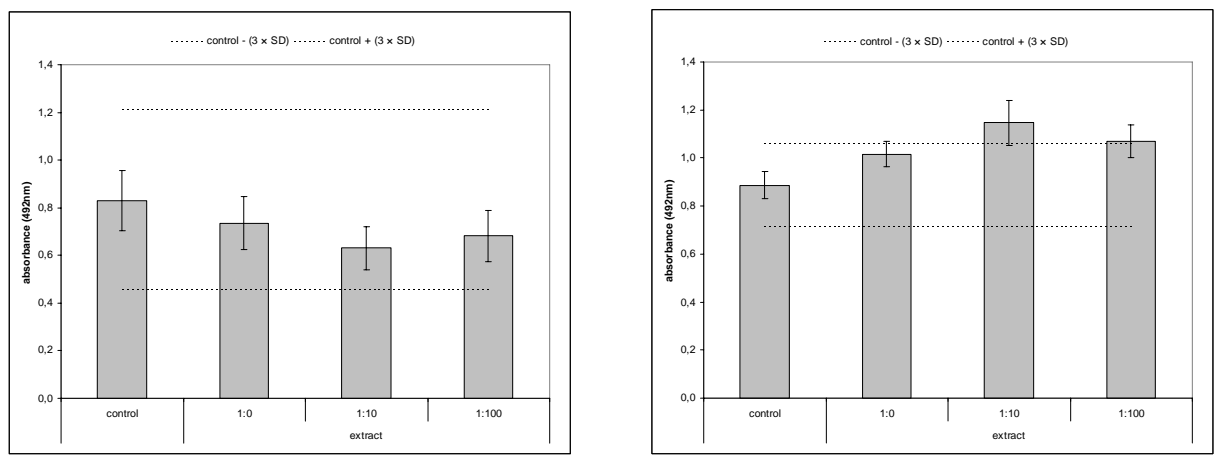

**Figure 7.10. Lactate dehydrogenase activity (LDH) (left) and mitochondrial dehydrogenases activity (right) of L929 cells after 48 hours of cell culture in pure cell culture medium (control) and after 72 hours of cell culture in the sample extract.**

# 7.4 Summary

Gelatin-based scaffolds were characterized including the suppression of the both the endotoxin contamination and the cytotoxic material response. Pure, nearly-endotoxin-free materials were accomplished, while the complete cell compatibility was demonstrated by indirect cell tests with L929 mouse fibroblasts. A pilot cell culture study demonstrated the proliferation of vital human dermal fibroblasts up 90 µm depth (section 7.1). This gave evidence that the scaffold pore size and porosity enabled the cell growth in the implant. Afterwards, the quantification (and suppression) of the biomaterial endotoxin contamination was investigated, as the purity source represents a crucial aspect for biopolymer-based material (section 7.2). Therefore, scaffolds based on low endotoxin gelatin were successfully synthesized, showing a minimal endotoxin content. At the same time, the high induction of TNF-α was completely suppressed by establishing the synthesis in cleaned condition, in order to minimize the risk of contamination. Also, a systematic washing/sterilization procedure enabled the removal of non-reacted moieties, so that pure endotoxin-free materials were accomplished. Once the biomaterial endotoxin contamination was solved, quantification and suppression of any cytotoxic material response was the goal (section 7.3). Eluate tests according to EN DIN ISO 10993-5 standards revealed a severe cytotoxic scaffold effect. In contrast to this, no cytotoxic response was observed in the pilot cell culture study, when testing saponin-containing scaffolds. The different result between direct and indirect tests might be explained as due to the heterogeneous saponin composition, and the occurrence of toxic and non-toxic saponin in the extract. Following this finding, a systematic study was conducted connecting the network architecture/reactivity with the cellular response. The effect of crosslinker and surfactant was investigated. Therefore, LDI was employed instead of HDI, as resulting in potentially non-toxic lysine-derivative degradation products. Also the use of $PEO_{40}$-$PPO_{20}$-$PEO_{40}$ during foaming was based on its well established clinical application. Resulting cytotoxicity tests highlighted higher cell proliferation for cells cultured on the sample extract compared to cells cultured on the pure cell culture medium. In agreement with the increase in cell proliferation, a consequent increase was observed in the MTS activity. Therefore, fully compatible gelatin-based scaffolds were successfully synthesized when LDI and $PEO_{40}$-$PPO_{20}$-$PEO_{40}$ as surfactant were used as chemical crosslinker, respectively.

# 8. Summary and outlook

Biomaterials have to fulfil complex clinical requirements of regenerative medicine applications.[13,157] This situation motivated this work, which aimed at the development of cytocompatible, gelatin-based scaffolds for potential applications in the induced autoregeneration of tissues, such as bone tissue. An interdisciplinary approach was pursued, which included the synthesis of entropy-elastic gelatin hydrogel films (Chapter 4), the translation from homogeneous (2-D) films to structured (3-D) scaffolds (Chapter 5), their characterization in physiological conditions (Chapter 6), and the optimization of the scaffolds in terms of endotoxin contamination and cell-toxicity (Chapter 7).

The crosslinking reaction of gelatin was carried out with either HDI or LDI above the gelatin gel-sol transition temperature ($T > 37\,°C$), which led to the fixation of gelatin chains in the randomly coiled state. Consequently, the gelatin chain helicity was suppressed, which was confirmed by WAXS, and the formation of entropy-elastic gelatin-based polymer network was accomplished. The crosslinking reaction was mechanistically complex: an excess of crosslinker (3-8 NCO/Lys molar ration) was necessary to achieve direct crosslinks (3-13 mol.-%) and overcome the competitive reaction of isocyanates with water. Despite the complex cascade of simultaneous reactions, defined structure-property relationships, similar to what is observed in the case of classical polymer networks, could be identified by varying the crosslinker excess.

An integrated foaming-crosslinking process was developed for the formation of 3-D porous scaffolds. The surfactant employed during foaming, saponin, proved to stabilize the foam (> 96 vol.-% foamed volume) and mediate the crosslinker solubility in the aqueous environment better than Polysorbate 20, which was also tested as a possible surfactant. As a result, scaffolds highlighted a porous architecture (pore size > 100 μm, porosity: 65-73 vol.-%), while the inherent gelatin network still displayed an amorphous morphology, as was observed in the case of homogeneous films.

In addition to being characterized in the dry state, scaffolds were characterized in the wet state, which revealed an interesting behaviour in physiological conditions. The scaffolds took up high amount of water ($H$: 630-1680 wt.-%), and the pore size decreased (9-22 % pore size reduction) due to the swelling of the pore walls along the pores. However, only minimal change of the outer dimensions was observed, meaning that the scaffolds were form-stable. The wet scaffolds were elastic, highlighting full shape recovery on the macro- and microlevel, as explained by the rubbery-like network morphology ($T_g \sim 20$ °C) in wet conditions at 25 °C and at 37 °C. The macroscopic (compression tests on the scaffold) and microscopic (AFM on the pore wall) mechanical properties could be adjusted in modulus ($E_c$: 10-50 kPa) to values near those of soft tissue.[26,145] The scaffolds showed different degradation kinetics in physiological conditions, depending on their network architectures. A controlled change of macroscopic properties was observed ($E_c$: 50 → 10 kPa) together with an increase of pore size (113±23 → 216±51 μm) during degradation.

Gelatin-based scaffolds were demonstrated to enable the proliferation of human dermal fibroblasts at different depth level (up to 90 μm scaffold depth), and the suppression of the biomaterial endotoxin contamination to appropriate levels (endotoxin content <0.5 EU/mL). Furthermore, complete suppression of the cytotoxic response was observed by indirect eluate tests (according to the US Pharmacopeial Convention). The employment of LDI and PEO$_{40}$-PPO$_{20}$-PEO$_{40}$ was thereby crucial to accomplish fully cell-compatible scaffolds, as higher cell (L929 mouse fibroblasts) proliferation and activity was observed when cells were cultured on the sample extracts with respect to cells cultured in pure cell culture medium. The fulfilment of these criteria successfully enabled the formation of medical-grade gelatin-based scaffolds.

The flexible and tailorable nature of the presented scaffold formation procedure could pave the way to an improved biomimetic design of the bone structure. A further step could include the formation of gelatin composites that contain calcium phosphate nanoparticles as the bone inorganic phase.[11,81] The presence of the inorganic phase should play a role in the scaffold bioactivity and potentially offer a new dimension for tailoring of mechanical properties and swelling behaviour.

One of the most promising applications is the potential of such scaffolds to induce regeneration within critical bone sized defects.[1] With this perspective, next steps could include direct tests of the gelatin scaffolds with specific and primary cells, e. g. osteoblasts or co-culture studies. In view of both the full recoverability and material form-stability, it would be also interesting to investigate the scaffold performance under dynamic cell culture conditions or in an *in vitro* bioreactor. Bioreactors provide technological instruments to perform controlled studies aimed at understanding the effects of specific biological, chemical, or physical cues on basic cell functions in a 3-D spatial arrangement.[158] The employment of the bioreactor would be of high importance to screen which types of scaffold are more suitable for a specific tissue. This would be mainly important for determining suitable *in vivo* studies, but could also drive the development of structurally defined and functionally effective 3D engineered constructs.

# 9. Appendix

## 9.1 Materials

Gelatin (type A), saponin, and $PEO_{40}$-$PPO_{20}$-$PEO_{40}$ (Pluronic$^{®}$ F-108), and Polysorbate 20 (Tween-20) were purchased from Fluka. Hexamethylene diisocyanate (HDI, 98%), and Cyclohexyl Isocyanate (98%) were purchased from Aldrich. Ethyl L-lysine diisocyanate was received from Shanghai Infine Chemical Co., Ltd. Glycine (> 99%) was obtained from Merck. Minimal Essential Medium with Earle's salts was purchased from Biochrom AG.

## 9.2 Synthesis of gelatin-based films in DMSO

The synthesis of gelatin-based films was carried out in a round bottom flask under magnetic stirring, whereby gelatin was dissolved in DMSO at 60 °C and cooled to 45 °C following complete dissolution. An appropriate amount of HDI in DMSO was cast in a Petri dish followed by fast addition of the gelatin solution, which crosslinked immediately. The DMSO was removed by drying the films at 60 °C for 2 d followed by washing with methylene chloride and extracting with water for 1 d at room temperature. The films were again dried at 60 °C overnight before further analysis. For mechanical testing, the films were equilibrated in water.

# 9.3 Synthesis of gelatin-based films in water

The synthesis of gelatin films was carried out in a round bottom flask under magnetic stirring, whereby gelatin was dissolved in water at 45 °C. A clear solution with a pH of 5.2 was obtained after roughly 30 min. Saponin was introduced to the solution while continuous stirring for 5 min, prior to adding the crosslinker. After 8 or 20 min, in case of crosslinking reaction with HDI or LDI, respectively, 15 mL of crosslinking mixture was cast in Petri dish, covered and placed at room temperature overnight. The end point of the reaction was qualitatively determined as gelation of the films. Investigation by FTIR spectroscopy demonstrated that no more unreacted isocyanate groups were present. Resulting films were washed with doubly distilled water (DDW) for several steps, in order to remove water soluble side products, and dried at temperature as low as 30-40 °C, in order to avoid bubble formation in gelatin films during solvent evaporation. In order to achieve the formation of bulky samples with cylindrical shape, the crosslinking mixture (30 mL) was cast into a centrifuge tube, instead of a Petri dish, and washed and dried as described above. For mechanical testing, the films were equilibrated in water.

## 9.4 Synthesis of gelatin-based scaffolds

An aqueous solution of gelatin (7-13 wt.-%) was obtained at 45 °C in a flat flange cylindrical jacketed vessel with bottom outlet valve (HWS Labortechnik, Germany) by mixing for 10 min at 1500 rpm with a basket type stirrer according to Hoesch (HWS labortechnik, Germany). Further stirring took place for 5 min once that the surfactant (1 g) was added and the foam was formed. Afterwards HDI was applied to the foam for gelatin crosslinking by 1 min reaction with stirring. Crosslinked foams were casted in cylindrical beakers of polypropylene (50 mm height x 72 mm diameter) and put in a refrigerator at 0 °C for 10 min. Following complete washing in double distilled water (DDW), foams were lyophilized. Further experimental details can be found in the supporting information.

## 9.5 Investigation of the crosslinking mechanism

Grafting was investigated by reaction of gelatin (10 wt.-% aqueous solution) with Cyclohexyl Isocyanate (CHI) with 8 molar excess of isocyanate groups to amino functions. Stability of resulting materials was checked at 25 °C by washing with water.

A solution of 223 mg glycine (2.97 mmol) in 100 mL of water at 45 °C was stirred at 1500 rpm for 20 min. 1 g of saponin was added to the solution, while stirring for further 5 min. Afterwards, 2.5 g of HDI (14.85 mmol) were added to the solution and reacted for 1 min. The reaction mixture was then transferred into a round bottom flask, cooled (no precipitate formed) and freeze dried. For LDI crosslinking, the method has been used accordingly. Crosslinking in DMSO has been investigated by reacting HDI (2.5 molar excess) with glycine for <1 min in the presence of 100 mL DMSO at 45 °C. A small aliquot was removed from the reaction mixture, from which DMSO was removed under reduced pressure, leaving a solid residue for ESI-MS analysis.

## 9.5.1 ESI-Mass Spectrometry

0.5 mg of the residue were dissolved in 1 mL of water containing 0.2% formic acid, and subjected to ESI-Q-ToF analysis on a Micromass Q-ToF Ultima with a source temperature of 120 °C, desolvation temperature of 150 °C, and capillary voltage of 2.5 kV (positive ion mode, *Dr. Jana Falkenhagen, BAM, Berlin*).

## 9.6 Wide angle X-ray scattering (WAXS)

Wide angle X-ray scattering (WAXS) measurements were carried out using the X-ray diffraction system Bruker D8 Discover with a two-dimensional detector from Bruker AXS (Karlsruhe, Germany). The X-ray generator was operated at a voltage of 40 kV and a current of 40 mA, producing Cu Kα-radiation with a wavelength λ = 0.154 nm. WAXS images were collected from either gelatin films or scaffolds (up to 850 μm sample thickness, 9 min exposure time) in transmission geometry with a collimator-opening of 0.8 mm at a sample-to-detector distance of 15 cm. Integration of the two-dimensional scattering data gave the intensity as a function of the scattering angle *2θ*.

# 9.7 Swelling tests

Swelling tests were performed in distilled water as well as Minimal Essential Medium (MEM). Dry films of known weight and standard dimensions (ISO 527-2/1BB) were individually placed in centrifuge tubes containing 5 mL of swelling medium and located in a water bath at 25 °C. Until equilibrium with water was reached, samples were removed, plotted with tissue paper and weighed, at specific time points. Swelling medium was daily changed during the experiment. The volumetric swelling ($Q$) was calculated according to eq. (1).

## 9.7.1 Swelling tests on powdered scaffolds

As for the swelling measurements on powdered scaffolds, samples were initially cryogenically milled. 0.01 g of the powder was collected in a 2 mL Eppendorf tube and 1.5 mL distilled water was added. The tubes were put in a water bath at 25 °C for 16 hours and then centrifuged. The supernatant was weighed and the swelling degree was calculated according to eq. 4.2.

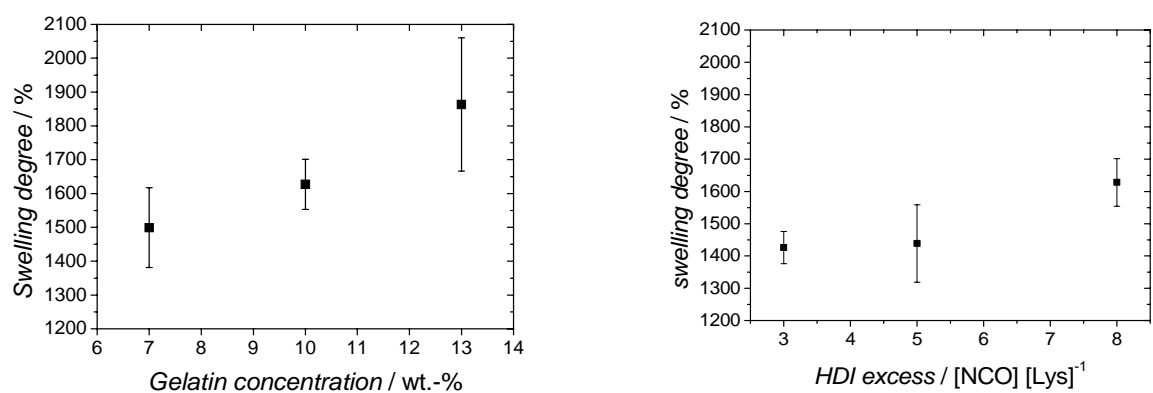

**Figure 9.1. Swelling experiments on powdered scaffolds in water at 25 °C: (left) effect of the gelatin concentration (8 NCO/Lys molar ratio); (right): effect of the HDI excess (10 wt.-% gelatin).**

The powder density of was measured by using a micro ultrapycnometer (P/N 02112-1, Quantachrome, Germany). Each experiment was carried out with six samples. The results are expressed as the mean ± standard deviation.

# 9.8 Uptake tests

Lyophilized scaffold samples (1 cm x 1.5 cm x 0.5 cm) were immersed in the aqueous solution and the uptake ($H$) was measured according to eq. 4.2. $H$ was determined with three replicas for each sample composition. Data points described the median, while error bars the range of data. During equilibration, the change of sample outer dimensions was also measured and compared to the one of a commercially available, clinically applied collagen sponge (Lyostypt™).

# 9.9 Mechanical tests

Bulky samples and porous scaffolds were analyzed by a mechanical tester (Zwick Z2.5, Zwick GmbH, Germany) equipped with a 100 N cell load, either in tension or compression mode. Film samples of standard dimensions (ISO 527-2/1BB) and equilibrated in water were tested with 0.02 N pre-force and 5 mm·min$^{-1}$ cross-head speed, until break. Young's modulus of gelatin films was calculated from the linear region, generally around 4-8% strain of the tensile stress-strain plot.

Compression tests were performed on swollen bulky cylinders of 3 cm diameter and 1.5 cm height. Samples were compressed up to 50% of their original height, with 0.3 N pre-force and 2 mm·min$^{-1}$ cross-head speed. For either tensile or compression tests, measurements were conducted at room temperature and each sample composition was tested with three replicas. Data were expressed in median, while bars described the range of data.

Specimens prepared by foaming were punched (2 cm diameter) in cylindrical shapes and then freeze dried. Afterwards they were cut with razor blades by 1.2 cm height, paying special attention in order to obtain parallel surfaces. Compression tests up to 50% of the initial sample height followed, either on the freeze-dried or water equilibrated (24 h) samples. A cross-head speed of 2 mm·min$^{-1}$ was employed during the measurement and each sample composition was tested with either three (wet-state) or four (dry-state) replicas. Data are expressed in median and the bars represent the range of data.

Dynamic mechanical analysis at varied temperature (DMTA) was performed in compression mode by Eplexor 25N (Gabo) equipped with a 25 N load cell. Measurements were conducted with 2 K·min$^{-1}$ heating rate and 10 Hz frequency. Samples swollen in $H_2O$ with cylindrical shape of 3 cm diameter and 1.5 cm height were used for measurements conducted in air.

# 9.10 Atomic force microscopy (AFM)

AFM measurements were carried out at *JPK Instruments AG, Berlin*. The foam was immobilized with a two component repro rubber. The measurements were performed on regions adjacent to the spots of glue, to allow for plenary optics and to ensure tight adhesion of the sample to the glass substrate. In two replicas of each scaffold composition, measurements in several pores in form of 5x5 testing site grids (or larger, if possible) within each of the investigated pores have been measured to get reliable statistics. The distance between two indentation spots was 2 μm. Optical calibration (DirectOverlay™), i.e. correlation of the AFM space to the optical space, was performed directly adjacent to the sample. Then an appropriate sample position was determined and an optical image was taken. Finally one or several force maps were taken of one, or, if possible, of a second pore, orientating on the optical image. Force measurements on immersed samples were all taken with a maximum applied force of 50 nN and a velocity of 2 μm·s$^{-1}$ and the height change was continuously read out. The curves were recorded in closed loop mode, which allowed for a constant extend speed of the piezo and therewith of the probe.

# 9.11 Scanning Electron Microscopy (SEM)

Lyophilized scaffolds were cut by using a razor blade, fixed on holders with conductive adhesive and then spattered with Gold-Palladium (Au 80% - Pd 20%) with a Polaron SC7640 sputter coater (Quorum Technologies Ltd, UK). The prepared samples were investigated using a SUPRA 40 VP electron microscope with a Schottky emitter at an acceleration voltage of 3 kV (Carl Zeiss NTS GmbH, Germany). Quantitative characterization of scaffold porous morphology was achieved by analyzing high resolution images.

# 9.12 Micro-computed Tomography (μCT)

Three dimensional structures of scaffold sample were characterized by using an X-ray micro computer tomography (ProCon X-ray GmbH). During the measurement, the sample rotates around a stable vertical axis while the X-ray source and the detector are fixed. X-ray radiographs were recorded at different angles during stepwise rotation of the sample between 0° and 360° around the vertical axis. The attenuation of the X-rays passing through the sample depends on the density of the material. The density of the material is directly related to the resulting contrast of the image. Starting by the two-dimensional cross-sections, the three dimensional architecture of the sample was reconstructed by mathematical algorithms,[5] with a maximal resolution of 1 μm. The modular constructed software "MAVI" (ITWM – Fraunhofer Institut Techno- und Wirtschaftsmathematik) was then used to determine the pore size and porosity of samples. Measurements were carried out at a specific distance between x-ray source and manipulator, i.e. the place where the sample was located. Samples were characterized by using different distances among the measurements, in order to obtain different resolutions.

# 9.13 Cell culture/imaging

## 9.13.1 Scaffold preparation for cell culture experiments

To get uniform open porous samples for cell culture experiments, cylinders were punched out from the core structure of the molded gelatine-scaffold. The cylindrical punches with a size of 10 mm approximately and 13 mm diameter were completely soaked in tissue freezing medium™ from Leica Microsystems GmbH (Nussloch, Germany) over night (*Dr. G. Boese, GKSS*). After freezing, exact uniform discs with a size of 1 mm were generated by cryosectioning with Leica CM3050S (Leica Microsystems GmbH, Nussloch, Germany). The freezing agent was removed by three gentle washing steps in phosphate buffered saline (PBS).

---

[5] Utilization of several modular algorithm: Binarizing by „Otsu", double edge filter, segmentation by preflooded watershed model, adaptation of frequency parameters to model of logarithm normal distribution.

## 9.13.2 Cell culture

Human dermal fibroblasts from passage 5-9 used for our experiments were purchased from Cell Lining GmbH (Berlin, Germany). The cells were grown in a solution containing Dulbecco's modified Eagle's medium (DMEM), supplemented with N-(2-hydroxyethly)piperazine-N'-ethanesulphonic acid (HEPES) buffer (Biochrom KG, Berlin, Germany), antibiotic/antimycotic solution (Sigma-Aldrich, Taufkirchen, Germany) and 10 % fetal boving serum (FBS) (Biochrom KG, Berlin, Germany) in a humidified incubator at 37 °C and 5% carbon dioxide. Cells were harvested with 0.05% trypsin/0.025% ethylenediaminetatraacetic acid (EDTA) (Sigma-Aldrich).

## 9.13.3 Cell growth on gelatin scaffolds

Cells were plated at a density of 2.0 x 105 cells per well, using 24-well plates (Corning Inc., NY, USA), containing 13 mm-diameter disks of gelatin scaffolds, which were disinfected with 70 % ethanol. The medium was exchanged every second day during cultivation of cells.

## 9.13.4 Morphology of cells

Cells were stained with fluorescein diacetate (FDA) to visualize living cells. The gelatin scaffolds with attached cells were incubated with 5 µl of the FDA stock solution (5 mg·ml$^{-1}$ in acetone) in 1 ml culture medium for 5 min. Then, the scaffolds were transferred into wells containing new culture medium. The investigation was started immediately with confocal laser scanning microscope LSM 510 (Carl Zeiss, Jena, Germany).

# 9.14 Cytotoxicity eluate tests

Cytotoxicity eluate tests were carried out according to EN DIN ISO 10993-5 norm (*Dr. B. Hiebl, S. Reinhold, GKSS*). 0.7 g of freeze-dried samples was sterilized with ethylene oxide (1.7 bar, 54 °C, 24 h, 3 h exposure time). Resulting samples were equilibrated with cell culture medium (MEM containing 7.5 wt.-% NaHCO$_3$ solution and 1 vol.-% solution of Penicillin-Streptomycin-Glutamine) for three days and the three-days sample extract was applied for culture of L929 cells. Cell morphology-based cytotoxicity was determined according to USP23-NF18, US Pharmacopoeial Convention categorization (Table 9.1).

**Table 9.1. Cell-morphology-based categorization of cytotoxicity (USP23-NF18, US Pharmacopoeial Convention).**

| cytotoxicity level | interpretation | changes in cell morphology |
|---|---|---|
| 0 | no cytotoxicity | • no morphological changes<br>• cell layer: tight<br>• cell lysis: none<br>• intracellular vacuoles: only discrete<br>• diameter ☑ small □middle □ big |
| 1 | slight cytotoxicity | • round and loosely attached cells: < 20%<br>• cell-layer: almost tight<br>• cell lysis: occasional (< 5 %)<br>• intracellular vacuoles: only discrete<br>• diameter ☑ small □ middle □ big |
| 2 | mild cytotoxicity | • round and loosely attached cells : < 50%<br>• cell layer: few empty areas<br>• cell lysis: extensive (< 50 %)<br>• intracellular vacuoles: few<br>• diameter ☑ small ☑ middle □ big |
| 3 | moderate cytotoxicity | • round and loosely attached cells : < 70 %<br>• cell layer: many empty areas<br>• cell lysis:  extensive (< 70 %)<br>• intracellular vacuoles: many<br>• diameter ☑ small ☑ middle ☑ big |
| 4 | severe cytotoxicity | • Nearly complete destruction of the cell layer |

Lactate dehydrogenase (LDH, Roche) and mitochondrial dehydrogenase (MTS, Promega) activities were determined. Data were reported as mean value ± standard deviation ($n$=8) for continuous variables, and analyzed by Student's t-test or the Chi square test. A $p$ value of less than 0.05 was considered significant.